\documentclass[11pt, a4paper]{article}
\usepackage{jheppub}
\pdfoutput=1

\usepackage{amssymb}
\usepackage{makeidx}
\usepackage{graphicx}
\usepackage{ccaption}
\usepackage{subfigure}
\usepackage{rotating}
\usepackage{lscape}
\usepackage{verbatim}
\usepackage{amsmath, amsthm,amssymb}
\usepackage{mathrsfs}
\usepackage{hypernat}
\usepackage{hyperref}
\usepackage{amsfonts}
\usepackage{dsfont}
\usepackage{bbm}
\usepackage{slashed}
\usepackage{hyperref}
\usepackage{booktabs}

\newcommand {\cB}{{\cal B}}
\newcommand {\cC}{{\cal C}}
\newcommand {\cD}{{\cal D}}

\newcommand {\cF}{{\cal F}}
\newcommand {\cG}{{\cal G}}
\newcommand {\cH}{{\cal H}}

\newcommand {\cK}{{\cal K}}
\newcommand {\cL}{{\cal L}}
\newcommand {\cM}{{\cal M}}
\newcommand {\cN}{{\cal N}}
\newcommand {\cO}{{\cal O}}

\newcommand {\cR}{{\cal R}}

\newcommand {\cU}{{\cal U}}
\newcommand {\cV}{{\cal V}}

\def\a{\alpha}
\def\b{\beta}
\def\c{\chi}
\def\d{\delta}
\def\e{\epsilon}
\def\f{\phi}
\def\g{\gamma}
\def\G{\Gamma}

\def\k{\kappa}
\def\l{\lambda}
\def\m{\mu}
\def\n{\nu}
\def\o{\omega}

\def\q{\theta}
\def\r{\rho}
\def\s{\sigma}
\def\t{\tau}

\def\x{\xi}

\def\D{\Delta}
\def\F{\Phi}

\def\L{\Lambda}
\def\O{\Omega}

\def\S{\Sigma}

\def\ri{{\rm i}}
\def\re{{\rm e}}

\newcommand{\ve}{\varepsilon}

\newcommand{\pa}{\partial}
\newcommand{\hf}{\frac12}

\newcommand{\vf}{\varphi}
\newcommand{\be}{\begin{equation}}
\newcommand{\ee}{\end{equation}}
\newcommand{\bea}{\begin{eqnarray}}
\newcommand{\eea}{\end{eqnarray}}
\newcommand{\non}{\nonumber}
\newcommand{\ba}{\begin{array}}
\newcommand{\ea}{\end{array}}

\newcommand{\bm}[1]{\mbox{\boldmath$#1$}}

\def\double #1{#1{\hbox{\kern-2pt $#1$}}}

\newcommand{\hm}{{\hat{m}}}
\newcommand{\hn}{{\hat{n}}}
\newcommand{\hp}{{\hat{p}}}
\newcommand{\hq}{{\hat{q}}}

\newcommand{\de}{{\nabla}}

\newcommand{\bbD}{{\mathbb {D}}}

\newcommand{\bsubeq}{\begin{subequations}}
\newcommand{\esubeq}{\end{subequations}}

\newcommand{\eps}{{\ve}}

\newcommand{\rd}{\mathrm d}
\newcommand{\gD}{{\mathbb D}}

\newcommand{\veps}{\varepsilon}

\newcommand{\hmu}{{{\hat{ \mu}}}}

\DeclareMathOperator{\Tr}{Tr}

\newcommand{\loco}{\vert}
\newcommand{\doubar}{{{\loco}\!{\loco}}}

\newcommand{\tgamma}{\tilde\gamma}

\makeatletter
\g@addto@macro\bfseries{\boldmath}
\makeatother

\newcommand{\ft}[2]{{\textstyle\frac{#1}{#2}}}
\newcommand{\hnu}{{\hat{\nu}}}
\newcommand{\hr}{{\hat{\r}}}

\newcommand{\hs}{{\hat{\s}}}

\title{Curvature squared invariants in six-dimensional $\cN = (1,0)$ supergravity}

\author[a]{Daniel Butter,}
\author[b]{Joseph Novak,}
\author[c]{Mehmet Ozkan,}
\author[b,d]{Yi Pang}
\author[e,f]{and Gabriele Tartaglino-Mazzucchelli}

\affiliation[a]{George and Cynthia Woods Mitchell Institute for Fundamental Physics and Astronomy,
Texas A\&M University, College Station, TX 77843, USA}
\emailAdd{dbutter@tamu.edu}

\affiliation[b]{Max-Planck-Institut f\"ur Gravitationsphysik, Albert-Einstein-Institut,
Am M\"uhlenberg 1, D-14476 Golm, Germany}
\emailAdd{joseph.novak@aei.mpg.de}

\affiliation[c]{Department of Physics, Istanbul Technical University,
Maslak 34469 Istanbul, Turkey}
\emailAdd{ozkanmehm@itu.edu.tr}

\affiliation[d]{Mathematical Institute, University of Oxford,
Woodstock Road, Oxford, OX2 6GG, UK}

\emailAdd{yi.pang@maths.ox.ac.uk}

\affiliation[e]{Instituut voor Theoretische Fysica, KU Leuven,
Celestijnenlaan 200D, B-3001 Leuven, Belgium}
\affiliation[f]{Albert Einstein Center for Fundamental Physics,
Institute for Theoretical Physics,\\
University of Bern,
Sidlerstrasse 5, CH-3012 Bern, Switzerland}

\emailAdd{gtm@itp.unibe.ch}

\arxivnumber{1808.00459}

\abstract{
We describe the supersymmetric completion of several curvature-squared invariants for ${\cal N}=(1,0)$ supergravity 
in six dimensions. The construction of the invariants is based on a close interplay between superconformal tensor 
calculus and recently developed superspace techniques to study general off-shell supergravity-matter couplings. 
In the case of minimal off-shell Poincar\'e supergravity based on the dilaton-Weyl multiplet 
coupled to a linear multiplet as a conformal compensator, 
we describe off-shell supersymmetric completions for all the three possible purely 
gravitational curvature-squared terms in six dimensions: Riemann, Ricci, and scalar 
curvature squared. A linear combination of these invariants describes the off-shell completion of the 
Gauss-Bonnet term, recently presented in arXiv:1706.09330. We study properties of the Einstein-Gauss-Bonnet 
supergravity, which plays a central role in the effective low-energy description of $\alpha^\prime$-corrected 
string theory compactified to six dimensions, including a detailed analysis of the spectrum about the 
${\rm AdS}_3\times {\rm S}^3$ solution.
We also present a novel locally superconformal invariant based on a higher-derivative action
for the linear multiplet. This invariant, which includes gravitational curvature-squared  terms,
can be defined both coupled to the standard-Weyl or dilaton-Weyl multiplet for conformal supergravity.
In the first case, we show how the addition of this invariant to the supersymmetric Einstein-Hilbert term 
leads to a dynamically generated cosmological constant and non-supersymmetric (A)dS$_6$ solutions.
In the dilaton-Weyl multiplet, the new off-shell invariant includes Ricci and scalar curvature-squared terms 
and possesses a nontrivial dependence on the dilaton field.
}

\numberwithin{equation}{section}
\allowdisplaybreaks

\begin{document}
\maketitle


\section{Introduction}

Over the years, six-dimensional (6D) $\cN=(1,0)$ supergravity theories
\cite{Nishino:1984gk,Nishino:1986dc,Nishino:1997ff,Sagnotti:1992qw,
Ferrara:1996wv,Ferrara:1997gh,Riccioni:2001bg}
have been a fertile ground of studies in various contexts
due to their relationship with string theories and 10D supergravities, and
the role they have played in various phenomenological scenarios.
It was already realized in the 80s by Salam and Sezgin that
a prototypical scenario in six-dimensional  $\cN=(1,0)$ gauged supergravity did not have
${\rm Minkowski}_6$ (nor ${\rm(A)dS}_6$) backgrounds
as solutions and the equations of motion led to a spontaneous compactification to
lower-dimensional spaces \cite{Salam:1984cj}.
In the absence of a background three-form flux \cite{Gueven:2003uw},
smooth symmetric solutions of the Salam--Sezgin model
take the form of a half-BPS ${\rm Minkowski}_4\times {\rm S}^2$ background
\cite{Salam:1984cj,Gibbons:2003di}, which are phenomenologically relevant
and can also be embedded in a String/M-theory framework \cite{Cvetic:2003xr}.
In the context of warped braneworld scenarios,
6D supergravities have also been investigated in the past to propose possible scenarios to solve
the cosmological constant problem and
build models possessing ${\rm dS}_4$ vacua, see, e.g.,
\cite{Aghababaie:2003wz,Aghababaie:2003ar,Burgess:2004dh}.

Being based on a 6D  theory possessing  chiral fermions, all the previously mentioned
models are generically anomalous.
As part of the effort to embed these theories in a consistent quantum theory of gravity,
anomaly free 6D $\cN=(1,0)$ supergravities have been constructed
in the ungauged and gauged cases
\cite{AlvarezGaume:1983ig,Green:1984bx,Ketov:1988nn,Ketov:1990jr,Erler:1993zy,Schwarz:1995zw,Seiberg:1996vs,RandjbarDaemi:1985wc,Salam:1985mi,Bergshoeff:1986hv,Avramis:2005qt,Avramis:2005hc,Suzuki:2005vu}.
In the ungauged case various  anomaly-free models were originally constructed by means of compactification
of the heterotic string and Green-Schwarz anomaly cancellation.
Recently, the classification of 6D $\cN=(1,0)$ supergravity theories consistent with quantum gravity
have also been systematically approached in the context of F-theory compactifications
on elliptically fibered Calabi-Yau threefolds.
We refer the reader to 
\cite{Vafa:1996xn,Morrison:1996na,Morrison:1996pp,Kumar:2009ac,Seiberg:2011dr,Taylor:2011wt,Park:2011wv,Grimm:2012yq,Bonetti:2014tqm,Monnier:2017oqd,Taylor:2018khc,Weigand:2018rez} 
and references therein for some of the literature on the subject.

The ungauged six-dimensional supergravity based on the dilaton-Weyl multiplet admits a unique supersymmetric 
${\rm AdS}_3 \times {\rm S}^3$ solution. 
The solution is supported by a self-dual 3-form flux and possesses vanishing Weyl tensor reminiscent of
the ${\rm AdS}_5\times {\rm S}^5$ solution in IIB supergravity. Moreover, there have been arguments indicating 
that the supersymmetric ${\rm AdS}_5\times {\rm S}^5$ 
is an exact solution in the full string theory based on its vanishing 
Weyl tensor. By analogy, it is tempting to conjecture that the supersymmetric ${\rm AdS}_3\times {\rm S}^3$ solution 
is an exact solution in the six-dimensional compactified string theory. 
We have verified this up to the inclusion of the Gauss-Bonnet 
and Riemann squared super-invariants \cite{NOPT-M17}.\footnote{One can also show that it is true with the inclusion 
of the Ricci scalar squared invariant given in this paper.} 
We recall that the ${\rm AdS}_3 \times {\rm S}^3$ geometry is 
locally ${\rm BTZ}\times {\rm S}^3$ 
and it is also the near horizon geometry of a black string. Thus, the 6D supergravity models 
can also be a useful arena for studying the black hole/string entropy correspondence. Although the supersymmetric 
${\rm AdS}_3\times {\rm S}^3$ 
solution is not affected by the 
curvature-squared corrections, 
$\alpha^\prime$ corrections in the action do 
modify the macroscopic entropy via Wald's entropy formula and thus can be used to compare with future microscopic 
computations at the same order. 
Curvature-squared terms are particularly important for computing the entropy of 
small black holes as the ``would-be'' leading Bekenstein-Hawking piece vanishes and therefore the 
curvature-squared corrections serve as the first non-vanishing contribution. 
The inclusion of the curvature-squared invariants 
also modifies the spectrum of fluctuations around the supersymmetric ${\rm AdS}_3\times {\rm S}^3$ solution. 
On top of the short multiplets of ${\rm SU}(1,1|2)$, 
there are also infinite towers of long ${\rm SU}(1,1|2)$ multiplets with mass proportional to 
the inverse of the $\alpha^\prime$ parameter. 
Expanding on the results presented in \cite{NOPT-M17}, 
in this paper we give a detailed analysis of the spectrum around the 
supersymmetric ${\rm AdS}_3\times {\rm S}^3$ solution.
Note also that the spectrum of fluctuations provides logarithmic corrections to the effective action, 
which has been proposed as a new probe to the possible UV completion of a low-energy effective 
theory of quantum gravity \cite{Sen:2012dw}.


It is worth stressing that there is a sharp difference between the spectrum of the short and long multiplets 
mentioned above. The spectrum of short multiplets is protected and can in principle be inferred by using
group theoretical considerations along the analysis of \cite{deBoer:1998kjm}. 
Contrarily, the AdS energies of the long-multiplet states are not protected by shortening conditions and depend on
the detail of the supergravity theory (in a low-energy approximation). 
Once the model is fixed, we can adopt the same method as \cite{Deger:1998nm} 
to compute the spectrum of long multiplets
by solving the linearized equations for the supegravity fields about the supersymmetric AdS vacuum.
In this regard, our results based on the new off-shell Einstein-Gauss-Bonnet supergravity
presented in \cite{NOPT-M17} and, in more detail, in the present paper are new and overcome the limitations of 
an algebraic approach. 


Higher order curvature terms are of importance in string theory where
the leading supergravity actions are necessarily corrected by an infinite series of quantum corrections
parametrized by the string tension $\alpha'$ and the string coupling $g_s$.
The purely gravitational higher-curvature terms are also related by supersymmetry to
contributions depending on $p$-forms. These terms, that have not yet been systematically analyzed in the literature,
play an important role in understanding the moduli in compactified string theory and the low-energy description of
string dualities, see, e.\,g., \cite{Antoniadis:1997eg,Antoniadis:2003sw,Liu:2013dna}.

Since multiplets of extended supersymmetry can be decomposed into $(1,0)$ multiplets, in principle, 
all supergravity models in ${\rm 6D}$ can be formulated in the $(1,0)$ framework.
Among all the supergravity models in ${\rm 6D}$, the most interesting ones are those coming from
Calabi-Yau compactifications of string theory/F-theory which are anomaly free and possess the duality properties
inherited from the parent theories.
For instance, the two $(1,1)$ supergravities descending from the Heterotic string on ${\rm T}^4$ and IIA-string on
${\rm K}3$ are expected to be related to each other by the six-dimensional Heterotic/IIA duality transformation
to all order in $\alpha'$, see the discussion in \cite{Liu:2013dna}.
So far the ${\rm 6D}$ duality has been verified rigorously at the two-derivative level
and connects the two $(1,1)$ supergravity theories introduced by Romans in \cite{RomansF4}. 
Beyond the leading order, examining the duality becomes a difficult task as it requires the knowledge of the
fully-fledged supersymmetric higher-derivative corrections involving delicate couplings between gravity and matter
fields.
Some recent progress was made to the first order in $\alpha'$ in \cite{Liu:2013dna}.
Furthermore, ${\rm 6D}$ $\cN=(1,0)$ supergravity admits an off-shell formulation which significantly
facilitates the construction of higher-derivative super-invariants.
Thus the off-shell $(1,0)$ supergravity should provide a useful framework for testing string duality at higher order.
We also notice that under the proposed six-dimensional duality transformation \cite{Liu:2013dna},
 fields belonging to the dilaton-Weyl multiplet, which is one of the two variant multiplets of ${\rm 6D}$ $(1,0)$
 conformal supergravity \cite{BSVanP},  form a closed structure.
Consequently, testing the duality can be further simplified by focusing on the subsector consisting of only 
the dilaton-Weyl multiplet, which  will be a main player in our paper.

In compactified string theory
the leading corrections in the  higher-derivative series
come from curvature-squared terms given by  the Gauss-Bonnet (GB) combination
which has the form
$\cR^{ab}{}^{cd}\cR_{ab}{}_{cd} - 4\cR^{ab} \cR_{ab} + \cR^2=6\cR_{[ab}{}^{ab} \cR_{cd]}{}^{cd}$,
where $\cR_{abcd}$ is the Riemann tensor, $\cR_{ab}$ is the Ricci tensor, and $\cR$ is the Ricci scalar.
This is the simplest example of  Lovelock gravity theory and as such it is singled out since
it is ghost free, and its equations of motion are second order in
derivatives \cite{Zwiebach:1985uq,Deser:1986xr}.
In the case of string theory compactified to ${\rm D}\leq 6$, the construction of
higher-derivative supergravity invariants can be
simplified thanks to the fact that off-shell supersymmetric techniques can be efficiently used.
In particular, the construction of the GB supergravity invariant by using off-shell techniques
was achieved in ${\rm 4D}$ in \cite{N=1-GB,Butter:2013lta},
in ${\rm 5D}$ in \cite{OP131,OP132, Butter:2014xxa}, while for the 6D
case only partial results were obtained more than thirty years ago \cite{BSS1,BSS2,Nishino:1986da,BR}.
Before continuing our discussion it is worth underlining what was the state of the art in the description of 
curvature squared invariants in off-shell 6D $\cN=(1,0)$ supergravity and what is new in our paper.
Note that in 6D, up to total derivatives, the general non-supersymmetric 
curvature squared Lagrangian has the form $\alpha \cR^{abcd}\cR_{abcd}+\b \cR^{ab}\cR_{ab}+\g \cR^2$.

A supersymmetric extension of the Riemann curvature squared term
was constructed more than 30 years ago \cite{BSS1,BSS2,Nishino:1986da,BR}. 
This invariant has been studied in some detail in the literature.
For instance, it has been coupled to
the gauged chiral supergravity in six dimensions extending the Salam-Sezgin model with curvature squared 
corrections \cite{Bergshoeff:2012ax} and the exact spectrum of this model
around the half-BPS Minkowski$_4\times {\rm S}^2$ background was analyzed  in \cite{Pang:2012xs}. 
A feature of this model that is worth underlining is that, 
compared to the 6D $\cN=(1,0)$ Einstein-Hilbert Poincar\'e supergravity, 
the addition of the Riemann squared invariant modifies the spectrum of the theory. 
In fact, as we will also review in section \ref{EH} of our paper, off-shell $\cN=(1,0)$ Poincar\'e 
supergravity (we focus for simplicity on the ungauged case) can be constructed by coupling the dilaton-Weyl multiplet
of conformal supergravity \cite{BSVanP} to a linear multiplet.
The independent field content of the first off-shell multiplet comprises the vielbein and chiral gravitini 
$(e_m{}^a,\,\psi_m{}_i)$, the dilatation $(b_m)$ and SU(2)$_R$ $(\cV_m{}^{kl})$ gauge connections together with a
real scalar field $\s$, a chiral fermion $\psi^i$, and a gauge two form $b_{mn}$.
The off-shell linear multiplet is described by an SU(2)$_R$ triplet of scalar fields $L^{ij}$ together
with a gauge $4$-form $b_{mnpq}$ and a chiral fermion $\varphi_\a^i$. The off-shell 
Einstein-Hilbert (EH) invariant that can be constructed with the previous field content \cite{BSVanP} 
is such that $\{e_m{}^a \,,\psi_{m}{}_i \,, b_{mn} \,,L=\sqrt{L_{ij}L^{ij}/2} \,, \vf^i \}$ are the on-shell physical fields
while $\{\s,\,b_m,\,b_{mnpq},\,\cV_m{}^{kl},\,\psi^i\}$ are either pure gauge degrees of freedom or auxiliary fields.
Once the Riemann tensor squared invariant is added to the supersymmetric EH term
some of the latter fields start to propagate and are not auxiliary any longer \cite{Bergshoeff:2012ax,Pang:2012xs}. 
In principle, their elimination can be performed in a nontrivial expansion in $\alpha^\prime$, 
see \cite{Bergshoeff:2012ax} for the analysis at first order.
Moreover, the gravitational spectrum is also modified due to the higher-derivative term that in this case 
generate ghost modes.
A similar pattern emerges when one consider a linear combination of the supersymmetric EH term and
a supersymmetric completion of the scalar curvature squared which was constructed in 2013
in \cite{OzkanThesis}.

Recently, in \cite{NOPT-M17} we presented a new independent curvature squared invariant completing 
the classification of the supersymmetric extension of the three curvature squared terms
$\alpha \cR^{abcd}\cR_{abcd}+\b \cR^{ab}\cR_{ab}+\g \cR^2$.
By taking a linear combination of the Riemann squared invariant and the new invariant, 
in \cite{NOPT-M17} we described for the first time 
the full bosonic sector of the off-shell $\cN=(1,0)$ Gauss-Bonnet term.
Besides being free of ghosts in the gravitational sector, the supersymmetric Gauss-Bonnet invariant
stands apart being the only off-shell curvature squared invariant that
remarkably preserves the same structure of physical and auxiliary degrees of freedom of the EH term.
It is exemplary to note that a general combination of Riemann squared, Ricci squared and scalar curvature 
squared, contains a dynamical terms $\cR_{ab}{}^{kl}(\cV)\cR^{ab}{}_{kl}(\cV)$ for the SU(2)$_R$ 
connection $\cV_m{}^{kl}$ which exactly cancels out only for the GB invariant.
One of the main aims of this paper is to describe in detail the construction of the three curvature squared invariant
with the techniques of \cite{BSVanP,BKNT16, BNT-M17}.
Though these invariants were already known in the literature, some of the detail of their construction were missing.
In particular, in \cite{NOPT-M17} we presented the Gauss-Bonnet invariant but this paper
gives a comprehensive description of its construction and the analysis of the spectrum for the
Einstein-Gauss-Bonnet supergravity.

Another main purpose of this paper is to present 
two other new curvature-squared invariants.
They both arise from a four-derivative action for a compensating linear multiplet coupled to
off-shell $\cN=(1,0)$ conformal supergravity.
If the linear multiplet is coupled to the dilaton-Weyl multiplet, as for all the curvature squared invariants discussed 
above, one obtains a new invariant
that contains a linear combination of Ricci tensor and scalar curvature squared terms multiplied by a prefactor for the 
dilaton field $L=\re^{-v}$ which also appears in higher-derivative self-interactions.
Alternatively, by coupling the new higher-derivative action for the linear multiplet to the standard-Weyl 
multiplet of off-shell conformal supergravity\footnote{As we will review in detail later, the standard-Weyl 
 multiplet  of $\cN=(1,0)$ conformal supergravity includes 
 $(e_m{}^a,\,\psi_m{}_i,\,b_m,\,\cV_m{}^{kl})$  independent gauge fields, 
as for the dilaton-Weyl  multiplet, 
but comprises an anti-self-dual tensor $T_{abc}^-$, a real scalar field $D$, and  a chiral fermion $\c^i$
as covariant matter fields \cite{BSVanP}.}
one can obtain another supersymmetric extension of a linear combination of Ricci tensor and scalar 
curvature squared terms. 
This latest invariant differs considerably from the previous four since it is constructed by using a variant
off-shell supergravity multiplet. Interestingly, as an application,
we will show that the coupling of this latest invariant to the supersymmetric Einstein-Hilbert term 
in a standard-Weyl multiplet
leads to a dynamically generated cosmological constant and non-supersymmetric (A)dS$_6$ solutions.

Before turning to the main technical sections  of our paper,
it is worth emphasizing some important features and recent
developments of off-shell supergravity that might not be familiar to all the readers and will play an important
role in our paper.
A main feature of off-shell techniques is that the local supersymmetry transformations
 close without the use of the equations of motion.
This feature makes off-shell formalisms extremely powerful to describe general supergravity-matter systems
without having to worry about the dependence of the supersymmetry transformations upon specific models.
In particular, this is a clear advantage if one is interested in constructing higher-derivatives interactions
that on-shell introduce highly nontrivial modifications to the supersymmetry transformations.
As such, a description of the supergravity effective action of superstrings where supersymmetry closes off-shell
would eliminate the 
complexity of an infinite series of $\alpha^\prime$ and $g_s$ corrections that in an on-shell setting
not only appear in the action but also in the supersymmetry transformations that are both corrected order by order.
Furthermore, an off-shell description naturally solves the problem that
higher-derivative terms in the on-shell string theory effective action
possess ambiguities arising from curvature dependent field redefinitions of the metric, such as
$g_{mn}' = g_{mn} + a g_{mn} \cR  + b \cR_{mn} + \dots$ with $\cR_{mn}$ the Ricci tensor and $\cR$
the scalar curvature. This conceptual issue, which plagues the organization of the effective action, simply disappears
if supersymmetry is implemented off-shell since no such redefinition leaves the supersymmetry algebra invariant.
It is important to stress that such an off-shell organization of the low-energy string effective action remains
in general an open problem. One of its very nontrivial issues  
is to understand how the off-shell supergravity ``auxiliary'' fields  
get integrated out\footnote{We refer the reader to \cite{Chemissany:2012pf} for some references on the subject
and a very interesting example
of how non-trivial and counterintuitive the integration of auxiliary fields can be.
In \cite{Chemissany:2012pf} it was shown how the infinite series of higher derivative interactions 
of DBI theories coupled to supergravity can remarkably arise by integrating out the auxiliary fields of 4D $\cN=2$
conformal supergravity in an off-shell $\cR^4$ action.}
 and if and how they alter the low-energy spectrum.
 Though still a longstanding open problem, in this paper
we will show how it is possible to deal with these issues in the restrictive truncation to six dimensional $\cN=(1,0)$ 
and at first order in $\alpha^\prime$.

Formalisms to describe off-shell supergravity-matter systems 
make use of component field techniques within the superconformal tensor calculus or superspace approaches.
The literature on these subjects is vast and we only refer to standard reviews for the 4D case, respectively
\cite{FVP} and \cite{GGRS,WB,BK}. In the case of 6D $\cN=(1,0)$ supergravity, the superconformal tensor
calculus was first applied in \cite{BSVanP} and further developed in \cite{CVanP,Bergshoeff:2012ax}
where the complete off-shell action for minimal Poincar\'e supergravity was presented  \cite{CVanP}
as well as that of the  gauged minimal 6D supergravity \cite{Bergshoeff:2012ax}.
For standard superspace techniques applied to supergravity in six dimensions
see \cite{BreitenlohnerRQ,GatesQV,GatesJV,SmithWAA,AwadaER,BergshoeffSU,BR}
and \cite{LT-M12}. 
Note that these references employ
a ``Wess-Zumino'' superspace approach analogue to the one used to study 4D $\cN=1$
 (see, e.g., \cite{GGRS,WB,BK})
and $\cN=2$ supergravity (see, e.g., \cite{Grimm,Howe1,Howe2,KLRT-M2008})
where the structure group of the superspace geometry is chosen to be the Lorentz group times (a subgroup of) the 
$R$-symmetry group.

It turns out that the superconformal tensor calculus and the standard superspace approach to off-shell supergravity 
are naturally related through the so-called \emph{conformal superspace}.
In this formalism the entire superconformal algebra is manifestly gauged in superspace
combining the main advantages of both approaches and providing a streamlined approach for component reduction
of superspace results.
As such, this approach is a bridge between the superconformal tensor calculus, where
 the superconformal group is gauged manifestly in standard space-time, and standard superspace approaches,
 where typically part of the superconformal transformations are non-linearly realized as super-Weyl transformations
 \cite{Howe-Tucker,GGRS,BK}.
Conformal superspace was first introduced for 4D $\cN=1,2$ supergravity in \cite{Butter:2009cp,Butter:2011sr}
(see also the seminal work by Kugo and Uehara \cite{Kugo:1983mv})
and it was developed and extended to 3D $\cN$-extended supergravity \cite{Butter:2013goa},
5D $\cN=1$ supergravity \cite{Butter:2014xxa},
and recently to 6D $\cN=(1,0)$ supergravity \cite{BKNT16},
see also \cite{BNT-M17}.
In the last few years, this approach has proven to be efficient to construct higher-derivative supergravity invariants.
These include the construction of:
 the three-dimensional $3\leq \cN \leq 6$ conformal supergravity
actions \cite{BKNT-M13,KNT-M13}; the 4D $\cN=2$  Gauss-Bonnet invariant  \cite{Butter:2013lta};
curvature squared invariants in 5D $\cN=1$ supergravity \cite{Butter:2014xxa};
 the 6D $\cN=(1,0)$ conformal supergravity actions related to the Type-B conformal anomalies
\cite{BKNT16,BNT-M17};
and recently the 6D $\cN=(1,0)$ Gauss-Bonnet invariant \cite{NOPT-M17}.
This paper will then show how this formalism can be efficiently employed to construct
curvature-squared invariants
for six-dimensional $(1,0)$ supergravity.

This paper is organized as follows.
In section \ref{section-2} we describe the 6D $\cN=(1,0)$ locally superconformal multiplets that will be used 
in our paper. In particular, we will describe the standard-Weyl multiplet, the non-abelian vector multiplet,
the linear multiplet, the gauge 3-form multiplet, the tensor multiplet and the dilaton-Weyl multiplet. 
We will describe the structure of each of the multiplets both in superspace and in components.
In section \ref{section-3},
by using the superform approach to the construction of supersymmetric invariants
\cite{Hasler, Ectoplasm, GGKS, Castellani},
we describe how to construct various locally superconformal invariants that will play the role of action principles. 
In section \ref{EH} we review the construction of the minimal off-shell $\cN=(1,0)$ 
two-derivative Poincar\'e supergravity theory of \cite{BSVanP} within our approach.
The off-shell extension of the Einstein-Hilbert term arises by using a linear multiplet compensator coupled to 
a dilaton-Weyl conformal supergravity multiplet.
Section \ref{main-curvature-squared} contains some of the main results of our paper:
the locally $\cN=(1,0)$ supersymmetric extension of all the
curvature squared terms for the minimal Poincar\'e supergravity of section \ref{EH} which is based on a dilaton-Weyl
multiplet of conformal supergravity.
In particular, two of these invariants, the Riemann squared and an invariant
that was first constructed in \cite{NOPT-M17}, are locally superconformal
and do not need the coupling to a matter compensator in contrast to a supersymmetric extension
of the scalar curvature squared term.
We also present in detail the off-shell Gauss-Bonnet 
invariant which is relevant to describing $\alpha^\prime$ corrections
in string theory compactified to six dimensions.
Section \ref{Einstein-Gauss-Bonnet} is devoted to defining the 
6D $\cN=(1,0)$ Einstein-Gauss-Bonnet supergravity theory which arises by
adding the off-shell Gauss-Bonnet to the Einstein-Hilbert invariant.
By integrating out the auxiliary fields, 
whose on-shell values can be set to zero as in the two-derivative 
Poincar\'e theory of section \ref{EH}, we derive the on-shell Einstein-Gauss-Bonnet supergravity
which was first obtained in \cite{Liu:2013dna} by using string theory arguments.
In section \ref{new-R2} we present new curvature squared invariants based on an higher-derivative action
for the linear multiplet compensator which, depending on the choice of the dilaton-Weyl or the standard-Weyl
as conformal supergravity multiplets, proves to describe two new curvature-squared terms compared
to the ones of section \ref{main-curvature-squared}.
In section \ref{spectrum} we turn to an application of the results of section \ref{Einstein-Gauss-Bonnet}
and describe in detail the computation of the  
spectrum of the Einstein-Gauss-Bonnet supergravity theory around 
the supersymmetric ${\rm AdS}_3\times {\rm S}^3$ vacuum.
In section \ref{conclusion} we conclude by discussing our results and possible future lines of research based on our
findings.
The paper includes four appendices. Appendix \ref{conformal-identities} summarises useful results
of  \cite{BKNT16,BNT-M17}
regarding 6D $\cN=(1,0)$ conformal superspace that are necessary for our paper.
In appendix \ref{descendants-C2gauge3form} we collect relevant descendant components
for the composite gauge 3-form multiplet based on the primary  \eqref{C2gauge3form}, 
which are necessary to derive the complete result for the invariant \eqref{newInvariant}.
Appendix \ref{full-nasty-new-invariant} includes the full bosonic part of the new locally superconformal
invariant constructed in section \ref{new-R2}.
In appendix \ref{notation} we describe how to 
map our notation and conventions, based on \cite{BKNT16,BNT-M17},
to the ones of \cite{BSVanP} and \cite{CVanP}.


\section{$\cN = (1, 0)$ superconformal multiplets}
\label{section-2}

In this section, we describe several superconformal multiplets that will serve as building blocks
for the various curvature squared invariants presented in this paper.
We will first discuss the standard-Weyl multiplet of conformal supergravity before moving on to the description
of various matter multiplets, 
including the non-abelian vector,
linear, and  gauge 3-form multiplets, and concluding with the description of the tensor and dilaton-Weyl multiplets.
For each of the multiplets,
we first elaborate on a superspace description  and then provide  their structure in terms of component fields.
The readers only interested in the component results can
direct their attention to the second half of each subsection.
For 6D $\cN=(1,0)$ superspace we make use of the formulation and results of \cite{BKNT16,BNT-M17}.

Before turning to the technical presentation of this section it is worth commenting about the fact that 
the non-abelian vector multiplet is described by a closed super $2$-form, 
the tensor multiplet is described by a closed super $3$-form,  
the gauge 3-form multiplet is described by a closed super $4$-form, 
and the linear  multiplet is described by a closed super $5$-form.
As was shown in \cite{ALR14}, these multiplets are in fact part of a tensor hierarchy of superforms
that also contains as top-form the closed super $6$-form used in \cite{BKNT16,BNT-M17}.
Note that, despite the natural organization of these multiplet in a tensor hierarchy, we will organize this section
by following a more traditional order, as for instance similar to the one used in \cite{BSVanP},
and present these multiplets by increasing complexity of their structure and the physical role they will play in 
the applications studied in our paper. 
For instance, we leave the description of the tensor multiplet to the end of this section. 
Compared to the other matter 
multiplets, the tensor multiplet stands apart since
in the flat limit it is on-shell and in the curved case
it is directly linked to the description of the off-shell dilaton-Weyl  multiplet of conformal supergravity.


\subsection{The standard-Weyl multiplet}

The standard-Weyl (or type I) multiplet  of $\cN=(1,0)$ conformal supergravity
is associated with the local off-shell gauging of the
superconformal group OSp$(8^*|1)$ \cite{BSVanP}.
The multiplet contains $40+40$ physical components described by a set of independent gauge fields:
the vielbein $e_m{}^a$ and a dilatation connection $b_m$;
the gravitino $\psi_m{}^\alpha_i$, associated with the gauging of $Q$-supersymmetry;
and SU(2)$_R$ gauge fields $\cV_m{}^{ij}$.
The other gauge fields associated with the  remaining generators of OSp$(8^*|1)$  are composite fields.
In addition to the independent gauge connections, the standard-Weyl multiplet comprises
a set of  covariant matter fields: an anti-self-dual tensor $T_{abc}^-$; a real scalar field $D$;
and  a chiral fermion $\c^i$.
We start by reviewing how to embed this in conformal superspace
\cite{BKNT16} and then, following \cite{BNT-M17},
we will show how to derive the component  structure of the multiplet.

\subsubsection{The standard-Weyl multiplet in superspace}

We begin with a curved six-dimensional $\cN=(1,0)$ superspace
 $\cM^{6|8}$ parametrized by
local bosonic $(x^m)$ and fermionic $(\theta_i)$ coordinates:
\be z^M = (x^m,\q^\mu_i) \ ,
\ee
where $m = 0, 1, \cdots, 5$, $\mu = 1, \cdots , 4$ and $i = 1, 2$.
By gauging the
full 6D $\cN = (1, 0)$ superconformal algebra we introduce
covariant derivatives $ {\nabla}_A = (\nabla_a , \nabla_\a^i)$
that have the form
\be
\nabla_A
= E_A - \hf  {\Omega}_A{}^{ab} M_{ab} - \Phi_A{}^{kl} J_{kl} - B_A \mathbb D
	-  {\mathfrak{F}}_A{}_B K^B \ .
\ee
Here $E_A = E_A{}^M \partial_M$ is the inverse super-vielbein,
$M_{ab}$ are the Lorentz generators, $J^{ij}$ are generators of the
${\rm SU(2)}_R$ $R$-symmetry group,
$\mathbb D$ is the dilatation generator and $K^A = (K^a, S^\a_i)$ are the special superconformal
generators.
 The supervielbein one form is $E^A=\rd z^ME_M{}^A$ with $E_M{}^AE_A{}^N=\d_M^N$,
 $E_A{}^ME_M{}^B=\d_A^B$.
The Lorentz ${\Omega}_A{}^{ab}=-{\Omega}_A{}^{ba}$, 
SU(2)$_R$ $\Phi_A{}^{kl}=\Phi_A{}^{lk}$, 
dilatation $B_A$ and
special conformal ${\frak{F}}_{AB}$
connections are associated with their respective structure group
generators $(M_{ab} , J^{ij} , \mathbb D , K^a, S^\a_i )$.
The super one-form connections are
${\Omega}^{ab}:=\rd z^M{\Omega}_M{}^{ab}
=E^A{\Omega}_A{}^{ab}$,
$\Phi^{kl}=\rd z^M\Phi_M{}^{kl}=E^A\Phi_A{}^{kl}$,
$B:=\rd z^MB_M=E^AB_A$, and
${\frak{F}}_{B}:=\rd z^M{\frak{F}}_{M}{}_{B}=E^A{\frak{F}}_{AB}$.

To describe the standard 6D $(1,0)$ Weyl multiplet in conformal superspace,
one constrains the algebra of covariant derivatives
\begin{align}
[  {\nabla}_A ,  {\nabla}_B \}
&= - {T}_{AB}{}^C  {\nabla}_C
- \frac{1}{2}  {R}(M)_{AB}{}^{cd} M_{cd}
-  {R}(J)_{AB}{}^{kl} J_{kl}
\non \\ & \quad
-  {R}(\mathbb D)_{AB} \mathbb D
-  {R}(S)_{AB}{}_\g^k S^\g_k
-  {R}(K)_{AB}{}_c K^c~,
\label{nablanabla}
\end{align}
to  be   completely determined   in terms of the super-Weyl tensor
superfield
 $W^{\a\b}$ \cite{Gates6D,LT-M12,BKNT16}
 satisfying
\be
W^{\a\b} = W^{\b\a} \ , \quad
K^AW^{\a\b} = 0 \ ,
 \quad \mathbb D W^{\a\b} = W^{\a\b} \ ,
\ee
and the Bianchi identities
\bsubeq\label{WBI}
\bea
\nabla_\a^{(i} \nabla_{\b}^{j)} W^{\g\d} &=& - \d^{(\g}_{[\a} \nabla_{\b]}^{(i} \nabla_{\r}^{j)} W^{\d) \r} \ , \\
\nabla_\a^k \nabla_{\g k} W^{\b\g} - \frac{1}{4} \d^\b_\a \nabla_\g^k \nabla_{\d k} W^{\g\d}
 &=& 8 \ri \nabla_{\a \g} W^{\g \b} \ .
\eea
\esubeq
Due to the relation  $W^{\a\b}=1/6(\tilde{\g}^{abc})^{\a\b}W_{abc}$,
the super-Weyl tensor is equivalent to an anti-self-dual rank-3 tensor superfield $W_{abc}$.
In  \eqref{nablanabla}
$ {T}_{AB}{}^C$ is the torsion, and $ {R}(M)_{AB}{}^{cd}$,
$ {R}(J)_{AB}{}^{kl}$, $ {R}(\mathbb D)_{AB}$, $ {R}(S)_{AB}{}_\g^K$ and $ {R}(K)_{AB}{}_c$
are the curvatures corresponding to the Lorentz, ${\rm SU(2)}_R$,
dilatation, $S$-supersymmetry and special conformal
boosts, respectively. Their expressions in terms of $W^{\a\b}$ and its descendant superfields of
dimension 3/2
\begin{align}
X^{\a i} := - \frac{\ri}{10} \nabla_{\b}^i W^{\a\b}
~,\quad
X_\g^k{}^{\a\b} :=
- \frac{\ri}{4} \nabla_\g^k W^{\a\b} - \d^{(\a}_{\g} X^{\b) k}
~,
\label{Xfields}
\end{align}
and of dimension 2
\bsubeq \label{YfielDs}
\begin{align}
Y_\a{}^\b{}^{ij} &:= - \frac{5}{2} \Big( \nabla_\a^{(i} X^{\b j)} - \frac{1}{4} \d^\b_\a \nabla_\g^{(i} X^{\g j)} \Big)
= - \frac{5}{2} \nabla_\a^{(i} X^{\b j)} \ , \\
Y &:= \frac{1}{4} \nabla_\g^k X^\g_k \ , \\
Y_{\a\b}{}^{\g\d} &:=
\nabla_{(\a}^k X_{\b) k}{}^{\g\d}
- \frac{1}{6} \d_\b^{(\g} \nabla_\r^k X_{\a k}{}^{\d) \r}
- \frac{1}{6} \d_\a^{(\g} \nabla_\r^k X_{\b k}{}^{\d) \r} \ ,
\end{align}
\esubeq
are collected in appendix \ref{conformal-identities}.
There we also collect the (anti-)commutators among the structure group generators and with
the covariant derivatives.
Note that, compared to \cite{BKNT16}, in this paper we will make use of conformal superspace
with a redefined vector covariant derivative which corresponds
to choosing the ``traceless'' frame conventional constraints
employed for the first time
in \cite{BNT-M17}.\footnote{In \cite{BNT-M17} we denoted the covariant derivatives of
\cite{BKNT16} as $\de_A=(\de_a,\de_\a^i)$ while
derivatives in different  frames were denoted by
$\hat{\de}_A=(\hat{\de}_a,{\de}_\a^i)$.
Since in this paper we will always use the traceless frame of \cite{BNT-M17} we will remove everywhere
the hats but the reader
should not confuse $\de_A=(\de_a,\de_\a^i)$ with the ones of \cite{BKNT16}. Equation \eqref{new-vector-derivatives}
explains the relation between the two vector derivatives.}
The superspace and component structures corresponding to this choice are summarized
below and in appendix \ref{conformal-identities}.

The superfields
$X^{\a i}$,
$X_\g^k{}^{\a\b}$,
$Y_\a{}^\b{}^{ij}$,
$Y$ and $Y_{\a\b}{}^{\g\d}$
 satisfy the nontrivial Bianchi identities \eqref{deaBI} \cite{BKNT16}
 that are consequences of  \eqref{WBI}.
These imply that the previous five superfields are the only independent
descendants obtained by acting with spinor derivatives on $W^{\a\b}$.
At higher mass dimension all the descendants
are vector derivatives of the previous five, see \eqref{eq:Wdervs}.
See also \eqref{S-on-X_Y} for the action of the $S$-generators on
$X^{\a i}$,
$X_\g^k{}^{\a\b}$,
$Y_\a{}^\b{}^{ij}$,
$Y$ and $Y_{\a\b}{}^{\g\d}$
that prove to be all annihilated by $K^a$.

In conformal superspace,
the gauge group  of conformal supergravity, $\cG$,
is generated by
{\it covariant general coordinate transformations},
$\delta_{\rm cgct}$, associated with a local superdiffeomorphism parameter $\xi^A$ and
{\it standard superconformal transformations},
$\delta_{\cH}$, associated with the following local superfield parameters:
the dilatation $\s$, Lorentz $\L^{ab}=-\L^{ba}$,  ${\rm SU(2)}_R$ $\L^{ij}=\L^{ji}$,
 and special conformal (bosonic and fermionic) transformations $\L_A=(\L_a,\L_\a^i)$.
The covariant derivatives transform as
\bea
\d_\cG \nabla_A &=& [\cK , \nabla_A] \ ,
\label{TransCD}
\eea
where $\cK$ denotes the first-order differential operator
\bea
\cK = \xi^C  {\nabla}_C + \hf  {\L}^{ab} M_{ab} +  {\L}^{ij} J_{ij} +  \s \mathbb D +  {\L}_A K^A ~.
\eea
A covariant (or tensor) superfield $U$ transforms as
\be
\d_{\cG} U =
(\d_{\rm cgct}
+\d_{\cH}) U =
 \cK U
 ~.
\ee
The superfield $U$ is said to
be \emph{primary} and of dimension $\D$ if $K_A U = 0$ and $\mathbb D U = \D U$.
The super-Weyl tensor $W^{\a\b}$ is a primary dimension 1 covariant superfield.

\subsubsection{The standard-Weyl multiplet in components}

Following the analysis of \cite{BNT-M17},
 let us now describe how to obtain the component description of the Weyl multiplet
from the previous superspace geometry.

The vielbein ($e_m{}^a$) and gravitini ($\psi_m{}^\a_i$) appear as the $\q=0$ projections of the coefficients of
$\rd x^m$ in the supervielbein $E^A$ one-form,
\begin{align}
e{}^a = \rd x^m e_m{}^a = E^a\doubar~,~~~~~~
\psi^\a_i = \rd x^m \psi_m{}^\a_i =  2 E^\a_i \doubar ~,
\end{align}
where the double bar denotes setting $\q = \rd \q = 0$ \cite{Baulieu:1986dp,BGG01,BNT-M17}.
The remaining fundamental and composite one-forms correspond to
double-bar projections of superspace one-forms,
\begin{align}
	\cV^{kl} := \Phi{}^{kl} \doubar~, \quad
	b := B\doubar ~, \quad
 \omega^{cd} :=  \Omega{}^{cd} \doubar ~, \quad
{ \phi}_\g^k := 2 \,{ {\frak F}}{}_\g^k\doubar~, \quad
 {\frak{f}}{}_c :=  {\frak{F}}{}_c\doubar~.
\end{align}

The covariant matter fields are contained within the super-Weyl tensor $W_{abc}$
and its independent descendants as follows:
\bsubeq
\bea
T^-_{abc} &:=& -2 W_{abc}\loco \ ,
\\
\chi^{\a i} &:=& \frac{15}{2}X^{\a i} \loco=-\frac{3\ri}{4}\de^i_\b W^{\a\b}\loco~,
\\
D&:=&
\frac{15}{2}Y \loco=
-\frac{3\ri}{16}\de_\a^k\de_{\b k}W^{\a\b}\loco
~,
\eea
\esubeq
where a single line next to a superfield denotes setting $\q=0$.
The lowest components of the
other nontrivial descendants of $W^{\a\b}$, specifically $X_{\a}^i{}^{\b\g}|$, $Y_{\a}{}^{\b}{}^{kl}|$
and $Y_{\a\b}{}^{\g\d}|$,
prove to be directly related to component curvatures and then they
are composite fields.

By taking the double bar projection of $\de=E^A\de_A$,
the component vector covariant derivative ${\de}_a$ is defined to coincide with
the projection of the superspace derivative ${\nabla}_a \loco$\footnote{Depending on the context it should 
be clear to the reader whether ${\nabla}_a$ denotes the superspace or the component vector derivatives.}
\begin{align}
e_m{}^a  {\nabla}_a = \pa_m
	- \frac{1}{2} \psi_m{}^\a_i \nabla_{\a}^i \loco
	- \frac{1}{2}  {\omega}_m{}^{cd} M_{cd}
	-  {b}_m \mathbb D
	-  {\cV}_m{}^{kl} J_{kl}
	- \frac{1}{2}  {\phi}_m{}_{\a}^{ i} S^{\a}_{ i}
	-  {{\frak f}}_m{}_{a} K^a
	~.
\end{align}
In this framework, the projected spinor covariant derivative $\nabla_{\a}^i\loco$
corresponds to the generator of $Q$-supersymmetry, and is defined so that
if $\cU = U\vert$, then $Q_\a^i\cU:=\nabla_\a^i| \cU := (\nabla_\a^i U) \loco$.
For the other generators, as e.g.~$M_{c d}\cU =(M_{c d} U)\loco$,
there is no ambiguity in identifying the bar projection
and then the local diffeomorphisms,
$Q$-supersymmetry transformations, and so on descend naturally
from their corresponding rule in superspace.

The component supercovariant curvature tensors,
arising from the commutator of two $ \de_a$ derivatives,
are defined as
$ {R}(P)_{a b}{}^{c} =  {T}_{a b}{}^c \loco$ and
$ {R}(Q)_{a b}{}^\g_k =  {T}_{ab}{}^\g_k \loco$,
and with
$ {R}(M)_{ab}{}^{cd}$, $ {R}(J)_{ab}{}^{kl}$, $ {R}(\gD)_{ab}$, $ {R}(S)_{ab}{}_{\g}^{k}$ and $ {R}(K)_{ab}{}_c$
coinciding with the lowest components of the corresponding superspace curvatures
that are given in appendix \ref{conformal-identities}.
Note that \eqref{new-frame_ALL-T-c}, \eqref{new-frame_ALL-R-J} and
\eqref{new-frame_ALL-R-i} imply that
$X_{\a}^i{}^{\b\g}|$,
$Y_{\a\b}{}^{\g\d}|$
and
$Y_{\a}{}^{\b}{}^{kl}|$  are identified with the
$ {R}(Q)_{a b}{}^\g_k$, $ {R}(M)_{ab}{}^{cd}$ and $ {R}(J)_{ab}{}^{kl}$ component curvatures, respectively.

The constraints on the superspace curvatures determine how to supercovariantize
a given component curvature by simply taking the double bar projection of the
superspace torsion and each of the superspace curvature two forms. Upon doing so one
finds \cite{BNT-M17}
\bsubeq\label{comp-curv_2-1}
\bea
 {R}(P)_{ab}{}^c
&=&
0
~,
\label{comp-curv_2-1-a}
\\
 {R}(Q)_{ab}{}_k
&=&
\frac{1}{2} {\Psi}_{ab}{}_k
+\ri \tilde{\g}_{[a} {\phi}_{b]}{}_{k}
+\frac{1}{24}T^-_{cde}\tilde{\g}^{cde}\g_{[a}\psi_{b]}{}_k
~,
\label{comp-curv_2-1-b}
\\
 {R}(\mathbb D)_{ab}
&=&
2e_a{}^me_b{}^n\pa_{[m}b_{n]}
 +4  {\mathfrak{f}}_{[a}{}_{b]}
+\psi_{[a}{}^{ i}  {\phi}_{b]}{}_{ i}
+\frac{\ri}{15} \psi_{[a}{}^j\g_{b]} \chi_{ j}
~,
\label{comp-curv_2-1-c}
\\
 {R}(M)_{ab}{}^{cd}
&=&
\cR_{ab}{}^{cd}( {\o})
+8 \d_{[a}^{[c}  {\mathfrak{f}}_{b]}{}^{d]}
+\ri\psi_{[a}{}_j\g_{b]} {R}(Q)^{cd}{}^j
+2\ri\psi_{[a}{}_j\g^{[c} {R}(Q)_{b]}{}^{d]}{}^{ j}
\non\\
&&
-\psi_{[a}{}_j  \g^{cd} {\f}_{b]}{}^j
-\frac{2\ri}{15} \d_{[a}^{[c}\psi_{b]}{}_j \g^{d]}\chi^{j}
+\frac{\ri}{2} \psi_{[a}{}^{ j}\g^e\psi_{b]}{}_j\,T^-_{e}{}^{cd}
~,
\label{comp-curv_2-1-d}
\\
 {R}(J)_{ab}{}^{kl}
&=&
\cR_{ab}{}^{kl}(\cV)
+4\psi_{[a}{}^{ (k}  {\f}_{b]}{}^{l)}
+\frac{4\ri}{15} \psi_{[a}{}^{ (k}\g_{b]} \chi^{ l)}
~,
\label{comp-curv_2-1-e}
\eea
\esubeq
where we have introduced the derivatives
\be \label{hatDder}
 {\cD}_m = \partial_m
- \hf  \omega_m{}^{bc} M_{bc}
- b_m \bbD
- \cV_m{}^{ij} J_{ij}
\ , \quad
 {\cD}_a = e_a{}^m  {\cD}_m \ ,
\ee
together with the curvature and field strengths
\bsubeq
\bea
 {\Psi}_{ab}{}^\g_k
&:=&
2e_a{}^me_b{}^n {\cD}_{[m}\psi_{n]}{}^\g_k
~,
\\
 {\cR}_{ab}{}^{cd}
&:=&
 {\cR}_{ab}{}^{cd}(\o)
=
e_a{}^me_b{}^n\Big(
2\pa_{[m}\o_{n]}{}^{cd}
-2\o_{[m}{}^{ce} \o_{n]}{}_e{}^d
\Big)
~,
\label{Romega}
\\
 {\cR}_{ab}{}^{kl}
&:=&
 {\cR}_{ab}{}^{kl}(\cV)
=
e_a{}^me_b{}^n\Big(
2\pa_{[m}\cV_{n]}{}^{kl}
+2\cV_{[m}{}^{p(k} \cV_{n]}{}_p{}^{l)}
\Big)
~.
\eea
\esubeq
On the other hand, due to
the superspace curvature constraints
$T_{ab}{}^c=0$,
 \eqref{new-frame_ALL-T-c} and \eqref{new-frame_ALL-R-i},
the   ``traceless'' conventional constraints for the component curvatures follow
\cite{BNT-M17}
\bsubeq \label{componentConstTRv2}
\bea
  R(P)_{ab}{}^c &=& 0~,
\label{componentConstTRv2-a}
\\
\g^b {R}(Q)_{ab}{}_k
&=&
0
~,
\label{componentConstTRv2-b}
\\
{  R}(M)_{ac}{}^{bc} &=& 0
~.
\label{componentConstTRv2-c}
\eea
\esubeq
The
conditions \eqref{componentConstTRv2} allow one to solve for the
composite connections as follows:
\bsubeq
\bea
 {\o}_{a}{}_{bc}
&=&
\omega(e)_{a}{}_{bc}
-2\eta_{a[b}b_{c]}
-\frac{\ri}{4}\psi_b{}^k\g_a\psi_c{}_k
-\frac{\ri}{2}\psi_a{}^k\g_{[b}\psi_{c]}{}_k
~,
\\
 {\phi}_m{}^{k}
&=&
\frac{\ri}{16}\Big(
\g^{bc}\g_m-\frac{3}{5}\g_m\tilde{\g}^{bc}
\Big)
\Big(
 {\Psi}_{bc}{}^{k}
+\frac{1}{12}T^-_{def}\tilde{\g}^{def}\g_{[b}\psi_{c]}{}^{k}
\Big)
~,~~~~~~
\label{eq:SConn-new}
\\
 {\mathfrak{f}}_{a}{}^{b}
&=&
-\frac{1}{8}\cR_{a}{}^{b}( {\o})
+\frac{1}{80}\d_{a}^{b}\cR( {\o})
+\frac{1}{8}\psi_{[a}{}_j  \g^{bc} {\f}_{c]}{}^j
-\frac{1}{80}\d_{a}^{b}\psi_{c}{}_j  \g^{cd} {\f}_{d}{}^j
\non\\
&&
+\frac{\ri}{16}\psi_{c}{}_j\g_{a} {R}(Q)^{bc}{}^j
+\frac{\ri}{8}\psi_{c}{}_j\g^{[b} {R}(Q)_{a}{}^{c]}{}^{ j}
+\frac{\ri}{60} \psi_{a}{}_j\g^b\chi^j
\non\\
&&
+\frac{\ri}{16}\psi_{a}{}^{ j}\g_c\psi_{d}{}_j\,T^-{}^{bcd}
-\frac{\ri}{160}\d_{a}^{b} \psi_{c}{}^{ j}\g_d\psi_{e}{}_j\,T^-{}^{cde}
~,
\label{eq:KConn-new}
\eea
\esubeq
where $\omega(e)_{a}{}_{bc} = -\frac{1}{2} (\cC_{abc} + \cC_{cab} - \cC_{bca})$
is the usual torsion-free spin connection given
in terms of the anholonomy coefficient
$\cC_{mn}{}^a := 2 \,\partial_{[m} e_{n]}{}^a$.

The supersymmetry transformations of the independent fields in the 
standard-Weyl multiplet may also
be read off from the superspace results 
\bsubeq\label{eq:WeylSUSY2}
\bea
\delta e_m{}^a &= &
-\ri\xi_{k}\g^a\psi_m{}^{k}
~,
\label{eq:WeylSUSY2-a}
\\
\delta \psi_m{}_i
&= &
2 {\cD}_m \xi_i
+\frac{1}{12}T^-{}^{abc}\tilde{\g}_{abc}\g_m\xi_i
+2\ri\tilde{\g}_m\eta_{ i}
~,
\label{eq:WeylSUSY2-b}
\\
\delta \cV_m{}^{kl}
&=&
-4 \xi^{(k} {\f}_m{}^{l)}
-\frac{4\ri}{15} \xi^{(k}\g_m\chi^{l)}
+4 \psi_m{}^{(k}\eta^{l)}
~,
\label{eq:WeylSUSY2-c}
\\
\delta b_m &=&
\xi_i  {\f}_m{}^i
+\frac{\ri}{15} \xi_i\g_m\chi^i
+\psi_m{}^i \eta_i
-2e_m{}^a  \l_a
~,
\label{eq:WeylSUSY2-d}
\\
\d T^-_{abc}
&=&
-\frac{\ri}{8}\x^{ k}\g^{de}\g_{abc} {R}(Q)_{de}{}_k
-\frac{2 \ri}{15} \x_k\g_{abc}\chi^{ k} ~,
\label{eq:WeylSUSY2-e}
\\
\d \chi^{ i}
&=&
\hf  D\x^{ i}
-\frac{3}{4}  {R}(J)_{ab}{}^{ij}\tilde{\g}^{ab}\x_j
+\frac{1}{4} {\de}_a T^-_{bcd}\tilde{\g}^{bcd}\g^a\x^{i}
-\ri T^-_{abc}\tilde{\g}^{abc}\eta^i
~,
\label{eq:WeylSUSY2-f}
\\
\d D
&=&
-2 \ri \,\xi_j  {\slashed{\de}}\chi^{ j}
-4\chi^k\eta_k
~.
\label{eq:WeylSUSY2-g}
\eea
\esubeq
Here
\bsubeq
\bea
\nabla_d T^-_{abc}
&=&
\cD_d T^-_{abc}
+\frac{\ri}{15}(\g_{abc})_{\a\b}\psi_d{}^{\a}_k\chi^{\b k}
+\frac{\ri}{2}(\g_{abc})_{\a\b} \psi_d{}^{\g}_kX_\g^k{}^{\a\b}
~,
\label{deW}
\\
{\slashed{\de}}\chi^{ j}
&=&
{\slashed{\cD}}\chi^{ j}
+\frac{\ri}{15}T^-_{bcd}\tilde{\g}^{bcd}\g^a\f_a{}^j
~,
\eea
\esubeq
and we have restricted to the  $Q$, $S$ and $K$ transformations,
$\d=\d_Q+\d_S+\d_K$, whose local component parameter are
 given by $\x_i$, $\eta^i$ and $\l_a$
respectively defined as the $\q=0$ components of the corresponding superfield parameters,
$\x^\a_i|$, $\eta_\a^i:=\L_\a^i|$ and $\l_a:=\L_a|$.


\subsection{The non-abelian vector multiplet} \label{YMmultiplet}

By following the discussion and conventions of \cite{BKNT16,BNT-M17},
let us turn to the description of a non-abelian vector multiplet.

\subsubsection{The non-abelian vector multiplet in superspace}

To describe the non-abelian vector multiplet in superspace,
we introduce the gauge-covariant derivatives
\be
{\bm \nabla} = E^A {\bm \nabla}_A \ , \quad
{{\bm\nabla}}_A := {\nabla}_A  - \ri \bm \cV_A \ ,
\ee
where the gauge connection one-form $\bm \cV_1 = E^A \bm \cV_A$
takes its values in the Lie algebra
of the (unitary) Yang-Mills gauge group, $G_{\rm YM}$, with its (Hermitian) generators
commuting with all the generators of the superconformal algebra.
The gauge-covariant derivatives satisfy the algebra
\bea
[{{\bm \nabla}}_A, {{\bm \nabla}}_B \}
&=&
 -{T}_{AB}{}^C {{\bm \nabla}}_C
-\hf {R}(M)_{AB}{}^{cd} M_{cd}
- {R}(J)_{AB}{}^{kl} J_{kl}
- {R}(\mathbb D)_{AB} \mathbb D
\non\\
&&
 -{R}(S)_{AB}{}^\g_k S_\g^k
-{R}(K)_{AB}{}^c K_c
- \ri \bm \cF_{AB} \ ,
\eea
where the torsion and curvatures are those of conformal superspace
while  $\bm \cF_{AB}$ corresponds
to the gauge covariant field strength two-form
$\bm \cF_2 = {\bm \nabla} \bm \cV_1 = \hf E^B \wedge E^A \bm \cF_{AB}$.
The field strength $\bm \cF_2$
satisfies the Bianchi identity
\be
{\bm \nabla} \bm \cF_2 = 0  \quad \Longleftrightarrow \quad
{\bm \nabla}_{[A} \bm \cF_{BC)}
+ {T}_{[AB}{}^D \bm \cF_{|D| C)} = 0
~. \label{FBI45}
\ee
The Yang-Mills gauge transformation acts on the gauge-covariant
derivatives ${\bm \nabla}_A$ and a matter  superfield $U$ (transforming
in some representation of the gauge group)
as
\be
{\bm \nabla}_A ~\rightarrow~ \re^{\ri  \bm  \t} {\bm \nabla}_A \re^{- \ri \bm \t } ,
\qquad  U~\rightarrow~ U' = \re^{\ri  \bm  \t} U~,
\qquad \bm \t^\dag = \bm \t \ ,
\label{2.2}
\ee
where the Hermitian gauge parameter ${\bm \t} (z)$ takes its values in the Lie algebra
of $G_{\rm YM}$.
This implies that the gauge one-form and the field strength transform as follows:
\bea
\bm \cV_1 ~\rightarrow ~\re^{\ri \bm \t} \bm \cV_1 \re^{-\ri \bm \t}
+ \ri \,\re^{\ri \bm \t} \rd \, \re^{- \ri \bm \t} \ , \qquad
\bm \cF_2 ~\rightarrow ~ \re^{\ri \bm \t} \bm \cF_2 \re^{- \ri \bm \t} \ .
\eea

Some components of the field strength have to be constrained
in order to describe an irreducible multiplet.
In conformal superspace
the right constraints are \cite{Siegel:1978yi,LT-M12,BKNT16}
\bsubeq\label{Vector-2-form}
\bea
\bm \cF_\a^i{}_\b^j = 0 \ , \quad \bm \cF_a{}_\b^j = (\g_a)_{\a\b} \bm \L^{\b i} \ , \label{YMsuperformFa}
\eea
where $\bm \L^{\a i}$ is a conformal primary of dimension 3/2,
 $S^\g_k \bm \L^{\a i} =0$ and
 $\bbD \bm \L^{\a i}= \frac{3}{2} \bm \L^{\a i}$. The Bianchi identity \eqref{FBI45}
  together with the constraints \eqref{YMsuperformFa}
 fix the remaining component of the two-form field strength to be
\bea
\bm \cF_{ab} = - \frac{\ri}{8} (\g_{ab})_\b{}^\a \bm \nabla_\a^k \bm \L^\b_k
\eea
\esubeq
and fix $\bm \L^{\a i}$ to obey the differential constraints
\cite{Siegel:1978yi,LT-M12,BKNT16}:
\bea
\bm \nabla_\g^k \bm \L^\g_k = 0 \ , \quad \bm \nabla_\a^{(i} \bm \L^{\b j)}
= \frac{1}{4} \d_\a^\b \bm \nabla_\g^{(i} \bm \L^{\g j)}
\label{vector-Bianchies} \ .
\eea

It is useful to list some identities for spinor covariant derivatives acting on the primary $\bm \L^{\a i}$
and the descendant superfields
${\bm \cF}_\a{}^\b=-\frac{1}{4}(\g^{ab})_\a{}^\b{\bm\cF_{ab}}$
and 
\begin{align}
\bm X^{ij} := \frac{\ri}{4} \bm\nabla^{(i}_\g \bm \L^{\g j)} \ .
\label{C.8}
\end{align}
These are
\bsubeq \label{VMIdentities}
\bea
\bm\nabla_\a^i \bm \L^{\b j}
&=&
- \ri \d_\a^\b \bm \L^{ij} - 2 \ri \eps^{ij} \bm \cF_\a{}^\b \ , \\
\bm \nabla_\a^i \bm \cF_\b{}^\g
&=&
- {\bm \nabla}_{\a\b} \bm \L^{\g i}
- \d^\g_\a {\bm \nabla}_{\b\d} \bm \L^{\d i}
+ \frac{1}{2} \d_\b^\g {\bm \nabla}_{\a \d} \bm \L^{\d i}
-\ve_{\a\b\r\t}W^{\g\r}  {\bm \L}^{\t i}
\ , \\
\bm\nabla_\a^i \bm X^{jk}
&=&
2 \eps^{i(j} {\bm \nabla}_{\a\b} \bm \L^{\b k)}
\ .
\eea
\esubeq
It is also useful to note that the $S$-supersymmetry
generator acts on the descendants of the superfield $\bm \L^{\a i}$ as follows:
\begin{align} \label{VMIdentitiesS}
S^\g_k \bm \cF_\a{}^\b = - 4 \ri \d^\g_\a \bm \L^\b_k
+ \ri \d^\b_\a \bm \L^\g_k \ , \qquad
S^\g_k \bm X^{ij} = - 4 \ri \d_k^{(i} \bm \L^{\g j)} \ .
\end{align}
Note that the two-form field strength is invariant under special superconformal transformations 
and is hence a primary superform, $K^C{\bm\cF}_2=0$.

\subsubsection{The non-abelian vector multiplet in components}

The component structure  of the vector multiplet
can be readily extracted from the superfield description above,
see \cite{BNT-M17} for more details.
The gaugino of the vector multiplet is given  by the projection ${\bm \L}^{\a i}|$.
The component one-form $\bm v_m$ and its field strength $\bm f_{mn}$ are given by
$\bm \cV_m\loco$ and $\bm \cF_{mn}\loco$, respectively. The \emph{supercovariant}
field strength ${\bm F}_{ab}$ is simply given by $\bm \cF_{ab}\loco$
and is related to $\bm f_{mn}=2\pa_{[m}\bm v_{n]}$ as
\begin{align}
{\bm F}_{ab}
= e_a{}^m e_b{}^n \bm f_{mn} +\psi_{[a}{}_k  \g_{b]} \bm \L^{k}
~.
\label{VM-field-strengths}
\end{align}
The last physical field of the vector multiplet is simply the bar-projection of $\bm X^{ij}$.

In what follows,  to avoid awkward notation, when  the correct interpretation is clear from the context,
we will associate the same symbol for the covariant component fields
and the associated superfields. The superfields
$\bm \L^{\a i}$, $\bm X^{ij}$, together with
$\bm \cF_\a{}^\b$,
are all annihilated by $K^a$
hence all their bar-projections are $K$-primary fields.
Using this fact  together with \eqref{VMIdentities} and \eqref{VMIdentitiesS} the
$\d=\d_Q+\d_S+\d_K$ transformations of the component fields follow:
\bsubeq
\bea
\d   \bm \L^{j}
&=&
- \ri \,\x_i   \bm X^{ij}
+\frac{\ri}{2}\tilde{\g}^{ab}\x^{j}  {\bm \cF}_{ab}
~,
\\
\d  \bm X^{ij}
&=&
-2\,\x^{(i}{ {{\bm {\slashed{\nabla}}}}}   \bm \L^{j)}
- 4 \ri\,\eta^{(i}   \bm \L^{j)}
~,
\\
\d  {\bm \cF}_{ab}
&=&
2\,\x_i{\g}_{[a} { {{\bm \nabla}}}_{b]} {\bm \L}^{i}
+\,T^-_{abc} \,\x_i\g^c {\bm \L}^{ i}
+ 2\ri \,\eta^i \tilde{\g}_{ab}  \bm \L_i
~,
\eea
\esubeq
where
\bsubeq\bea
&{\bm \de}_a \bm \L^{i}
=
{\bm \cD}_a \bm \L^{i}
+\frac{\ri}{2} \bm X^{ij} \psi_a{}_j
-\frac{\ri}{4} \tilde{\g}^{bc}\psi_a{}^{i} {\bm \cF}_{bc}
~,
\\
&{\bm \cD}_a:=
e_a{}^m\big( \pa_m
- \frac{1}{2}\omega_m{}^{cd} M_{cd}
- b_m \mathbb D
- \cV_m{}^{kl} J_{kl}
-\ri {\bm v}_m
\big)
~.
\eea
\esubeq
The transformation rule of the component connection $\bm v_m$ can be computed by first noticing that
$K^A \cF_2=0$ and then by taking the double bar projection of the supergravity gauge transformation
of $\bm\cV_1$, $\delta_\cG \bm\cV_1 = E^A \xi^B \bm\cF_{BA}$,
which leads to
\bea
\delta \bm v_m
= e_m{}^a \xi^\g_k \bm \cF_\g^k{}_{a}|
= - \xi_k \g_m \bm\L^{k}
~.
\eea


\subsection{The linear multiplet}
\label{linear-multiplet-section}

Let us turn to describing the 6D off-shell linear multiplet coupled to conformal supergravity.

\subsubsection{The linear multiplet in superspace}

The linear multiplet, or $\cO(2)$ multiplet, can be described using a 4-form gauge
potential $B_4 = \frac{1}{4!} E^D \wedge E^C \wedge E^B \wedge E^A B_{ABCD}$
possessing the gauge transformation
\be \d_{\r_3} B_4 = \rd \r_3 \ ,
\label{dB4}
\ee
where the gauge parameter $\r_3$ is an arbitrary super  3-form.
The corresponding 5-form field strength is
\be E_5 = \rd B_4 = \frac{1}{5!} E^E \wedge E^{D} \wedge E^C \wedge E^B \wedge E^A E_{A B C D E} \ ,
\label{BI_H5}
\ee
where
\be
E_{A B C D E} = 5 {\nabla}_{[A} B_{BCDE)} + 10 {T}_{[A B}{}^{F} B_{|F|CDE)} \ .
\ee
The field strength satisfies the Bianchi identity
\be
\rd E_5 = 0  \quad \Longleftrightarrow \quad
{\nabla}_{[A} E_{BCDEF )} + \frac{5}{2} {T}_{[AB}{}^{G} E_{|G|CDEF)} = 0 \ .
\ee

In order to describe the linear
multiplet we need to impose covariant constraints on the field strength $E_5$
\bsubeq\label{Linear-5-form}
\begin{align}
E_{abc}{}_\a^i{}_\b^j
= 4 \ri (\g_{abc})_{\a\b} L^{ij}
\ , \quad L^{ij} = L^{ji} \ , \quad S^\g_k L^{ij} = 0 \ , \quad \mathbb D L^{ij} = 4 L^{ij} \ ,
\label{O(2)constrants}
\end{align}
and require all lower mass-dimension components of $E_{ABCDE}$ to vanish.
The remaining components of $E_5$ are constrained by the Bianchi identities \eqref{BI_H5}
to be defined in terms of the superfield $L^{ij}$, and its descendants as
\bea
E_{abcd}{}_\a^i
&=&
\frac{1}{3} \eps_{abcdef} (\g^{ef})_\a{}^\b \nabla_{\b j} L^{ij}
\ , \\
E_{abcde} &=&\frac{\ri}{12} \eps_{abcdef} ({\tilde{\g}}^f)^{\a\b} \nabla_\a^k \nabla_\b^l L_{kl}
\equiv \eps_{abcdef} {\rm E}^f \ ,
\eea
\esubeq
where $L^{ij}$ is constrained 
to satisfy the defining constraint for the
linear multiplet\footnote{The linear multiplet in six dimensions
was introduced in \cite{HST,GrundbergLindstrom_6D} for the Minkowski case
extending the 4D results of \cite{LR,Siegel:1978yi,N=2tensor,SSW}.
See \cite{LT-M12,Kuzenko:2017jdy,Kuzenko:2017zsw,BSVanP}
for other references about the coupling of the linear multiplet to 6D supergravity.}
\be \nabla_{\a}^{(i} L^{jk)} = 0 \ .
\label{linear-constr}
\ee
The highest dimension component of the superform satisfies the Bianchi identity ${{\nabla}}_a {\rm E}^a = 0$
with ${\rm E}^a=\frac{1}{5!}\ve^{abcdef}E_{bcdef}$.\footnote{The reader should not confuse
${\rm E}^a$ with the vector supervielbein.}

In order to elaborate on the component structure of the superfield $L^{ij}$, the following
identities prove useful:
\bsubeq\label{SUSYTransinsuperspaceLinear}
\bea
\nabla_\a^i L^{jk} &=& - 2 \eps^{i(j} \varphi_\a^{k)} \ ,
 \\
\nabla_\a^i \varphi_\b^j &=& - \frac{\ri}{2} \eps^{ij} {\rm E}_{\a\b} - \ri {\nabla}_{\a\b} L^{ij} \ ,
 \\
\nabla_\g^k {\rm E}_{\a\b}
&=& - 8 {\nabla}_{\g[\a} \varphi_{\b]}^k
- 2 {\nabla}_{\a\b} \varphi_\g^k
+ 2 \eps_{\a\b\g\d} W^{\d \r} \varphi_\r^k
\ ,
\eea
\esubeq
where we have defined the descendant superfields
\bea
\varphi_\a^{i}
:=
-\frac{1}{3}\nabla_{\a j} L^{ij}
\ ,
\qquad
{\rm E}_a =-\frac{\ri}{4}(\tilde{\g}_a)^{\a\b}\nabla_{\a}^k \varphi_{\b k}
~,~~
{\rm E}_{\a\b}=(\g^a)_{\a\b}{\rm E}_a
\ .
\eea
These prove to be annihilated by $K^a$ and to satisfy
\bea\label{SSUSYTransinsuperspaceLinear}
S^\b_j \varphi_\a^i = 8 \d_\a^\b L^i{}_j  \ ,
\qquad
S^\g_k {\rm E}_{\a\b} = - 40 \ri \d^\g_{[\a} \varphi_{\b] k}
\ .
\eea
These turn out to be equivalent to the  condition that $E_5$ is a primary superform $K^C E_5=0$.

\subsubsection{The linear multiplet in components}

The covariant component fields of the linear multiplet can be identified
by the component projections of $L^{ij}$ and $\varphi_\a^i$. The component
gauge 4-form and its 5-form field strength can be identified by the component projections
$b_{mnpq}:=B_{mnpq}|$ and $h_{mnpqr}=5\pa_{[m}b_{npqr]}$, respectively.
In the paper we will also use the Hodge dual of  the 4-form 
$\tilde{b}^{mn}=\frac{1}{4!}\ve^{mnpqrs}b_{pqrs}$.\footnote{It is worth reminding that in our notation,
see appendix A of \cite{BKNT16}, 
the Levi-Civita tensors with curved indices include $e$ or $e^{-1}$ factors 
 since we use the following definitions
$\ve^{mnpqrs}:=\ve^{abcdef}e_{a}{}^me_{b}{}^ne_{c}{}^pe_{d}{}^qe_{e}{}^re_{f}{}^s$,
and $\ve_{mnpqrs}:=e_{m}{}^ae_{n}{}^be_{p}{}^ce_{q}{}^de_{r}{}^ee_{s}{}^f\ve_{abcdef}$,
with $\ve_{abcdef}$ such that $\ve_{012345}=-\ve^{012345}=1$.}
The supercovariant field strength
is just the component projection of $E_{abcde}$,
 or equivalently ${\rm E}^a$, 
 the top component of the superform $E_5$.
At the component level it holds
\bea
{\rm E}^a = h^a
- \psi_b{}_i \g^{ab}\varphi^i
- \frac{\ri}{2} \psi_b{}_i\g^{abc} \psi_c{}_j  L^{ij}
\ ,~~~~~~
h^a =\frac{1}{5!} \eps^{a mnpqr}h_{mnpqr}
~.
\eea
In the previous equation and in what follows we drop the component projection of the descendant fields
when it is clear from context what we mean.
The supersymmetry transformations of the covariant fields can be
read off from \eqref{SUSYTransinsuperspaceLinear}
and
\eqref{SSUSYTransinsuperspaceLinear}
which give
\bsubeq\label{linear-transf}
\bea
\d L^{ij} &=& 2\xi^{(i} \varphi^{j)} \ , \\
\d \varphi^i &=&
\frac{\ri}{2} \xi^{i} \g^a {\rm E}_a
-\ri \xi_j {{{\slashed\nabla}}}_a L^{ji}
-8 \eta_{j} L^{ij} \ , \\
\d {\rm E}_a &=&
2\xi_i \g_{ab}{{\nabla}}^b \varphi^i
+ \frac{1}{12} \xi^{i} \g_a\tilde{\g}^{bcd}T^-_{bcd} \varphi_{i}
- 10\ri \eta^i \tilde{\g}_a \varphi_{i} \ ,
\eea
\esubeq
where
\bsubeq
\bea
{\nabla}_a L^{ij} &=& {\cD}_aL^{ij} - \psi_a{}^{(i} \varphi^{j)}
\label{deL}
\ , \\
{\nabla}_a \varphi^i
&=&
{\cD}_a \varphi^i
- \frac{\ri}{4}\psi_a{}^{i}\g_b{\rm E}^b
-\frac{\ri}{2}\psi_a{}_j{\slashed{\nabla}}L^{ij}
+ 4 {\phi}_{a j} L^{ij}
\ .
\eea
\esubeq
The transformation of the component gauge field $b_{mnpq}$ can be obtained
by projecting to components the supergravity gauge transformation of $B_4$,
$\delta_\cG B_4 = \frac{1}{4!}E^{A_4}\wedge\cdots\wedge E^{A_1} \xi^B E_{BA_1\cdots A_4}$.
This leads to
\bea
\d b_{mnpq}
&=&
e_{[q}{}^{d}e_{p}{}^{c} e_{n}{}^{b}\big(
e_{m]}{}^{a}\x^\b_j E_{\b}^j{}_{abcd}|
+2\psi_{m]}{}^\a_i\x^\b_j E_{\b}^j{}_\a^i{}_{abc}|
\big)
\non\\
&=&
-\eps_{mnpqef}\, \x_j\g^{ef}\varphi^j
+ 8 \ri \psi_{[m}{}_i\g_{npq]}\x_j L^{ij}
\ .
\label{b4susy}
\eea
The transformations of the components of the linear multiplet reproduce, up to the change of notation
 described in appendix \ref{notation}, the results of \cite{BSVanP}.


\subsection{The gauge 3-form multiplet}
\label{gauge-3-form-multiplet-section}

In this section we introduce another off-shell multiplet that will play a central role in the construction
of an invariant action principle in our paper.
This was described in \cite{ALR14} and also in \cite{BKNT16,BNT-M17}.
It contains a closed 4-form field strength amongst its component fields, which can be solved in terms of
a gauge 3-form. We will refer to it as the gauge 3-form multiplet.

\subsubsection{The gauge 3-form multiplet in superspace}

The gauge 3-form multiplet can naturally be described in superspace
using a super 3-form $B_3 = \frac{1}{3!} E^C \wedge E^B \wedge E^A B_{ABC}$
possessing the gauge transformation
\be
\d_{\r_2} B_3 = \rd \r_2 \ ,
\ee
where $\r_2$ is a two-form gauge parameter. The corresponding field strength is
\bsubeq
\bea
H_4 &=& \rd B_3 = \frac{1}{4!} E^D \wedge E^C \wedge E^B \wedge E^A H_{ABCD} \ ,
\\
H_{ABCD} &=& 4 {\nabla}_{[A} B_{BCD )} + 6 {T}_{[AB}{}^F B_{|F|C)}
\ .
\eea
\esubeq
The 4-form
field strength satisfies the Bianchi identity
\bsubeq\label{four-form-000}
\be \rd H_4 = 0  \implies {\nabla}_{[A} H_{BCDE)} + 2 {T}_{[AB}{}^{F} H_{|F|CDE)} = 0 \ . \label{BI4formmult}
\ee

In order to describe an irreducible 3-form multiplet
it is necessary to impose the following covariant constraints \cite{ALR14} on the field strength $H_4$:
\be H_{ab}{}_\g^k{}_\d^l =~\ri (\g_{abc})_{\g\d}B^{c}{}^{kl} \ ,
\quad S^\g_k B_a{}^{ij} = 0 \ , \quad \mathbb D B_a{}^{ij} = 3 B_a{}^{ij}
~,
\ee
with all lower dimension components of the superform vanishing. The Bianchi identity \eqref{BI4formmult}
fixes the remaining components as
\begin{align}
H_{abc}{}_\d^l =&~ -\frac{1}{12}\ve_{abcdef}(\g^{de})_\d{}^\r \nabla_{\r p} B^{f}{}^{lp}
\ , \\
H_{abcd} =&~
\frac{\ri}{48}\ve_{abcdef}(\tilde{\g}^{e})^{\a\b}\de_{\a k}\nabla_{\b l} B^{f}{}^{kl}
~, \label{eq:Habcd}
\end{align}
\esubeq
and requires $B_a{}^{ij}$ to satisfy the constraints\footnote{It was shown in \cite{BKNT16,BNT-M17} that a reducible
multiplet described by a primary superfield $B_a{}^{ij}$ satisfying only \eqref{4FormConst-1} but not
\eqref{4FormConst-2} can be used to construct a non-primary super 6-form.
This plays an important role in, e.g., providing a density formula for the full superspace integral and to describe
one of the two $\cN=(1,0)$ conformal supergravity actions.}
\bsubeq \label{4FormConst}
\bea
\nabla_\a^{(i} B^{\b\g jk)} &=& - \frac{2}{3} \d_\a^{[\b} \nabla_\d^{(i} B^{\g]\d jk)} \ ,
\label{4FormConst-1}\\
{[}\nabla_\a^{(i},  \nabla_{\b k}{]}B^{\alpha \beta j) k}
&=& - 8 \ri \,\nabla_{\a\b} B^{\a\b i j} \ .
\label{4FormConst-2}
\eea
\esubeq

The superfield $B_a{}^{ij}$ has a large number of descendants, only some of which will appear in
the field strength $H_4$ or in the action principle we will construct based upon it.
The relevant ones are
\bsubeq\label{descendants-4-form}
\bea
\L^{\a i j k} &:=& \frac{\ri}{3} \nabla_{\b}^{(i} B^{\b\a j k)}
~,
\\
\L_{\a a}{}^i &:=& \frac{2\ri}{3} \nabla_{\a j} B_a{}^{ij}
~,
\label{descendants-4-form-2}
\\
C_{a b} &:=& \frac{1}{8} (\tilde\gamma_{a})^{\alpha \beta} \nabla_{\a k} \L_{\b b}{}^k~,
\label{descendants-4-form-3}
\eea
\esubeq
where the reader is cautioned that $C_{ab}$ is a generic rank-two tensor containing both a symmetric and
antisymmetric part.
These descendants satisfy the following useful relations, which are consequences 
of the constraints \eqref{4FormConst}\bsubeq\label{deB}
\bea
\nabla_\a^i B_a{}^{jk} &=& - \frac{\ri}{2} (\g_a)_{\a\b} \L^{\b ijk} - \ri \,\eps^{i(j} \L_\a{}_a{}^{k)}
\ , \\
\nabla_\a^i \L_{\b b}{}^j &=&
	- \frac{1}{2} (\gamma_{b})_{\alpha \gamma} C_{\beta}{}^{\gamma i j}
	+ \frac{1}{6} (\gamma_{b})_{\beta \gamma} C_{\alpha}{}^{\gamma i j}
	+ \veps^{i j} \,(\gamma^{a})_{\alpha \beta} C_{a b}
	\nonumber \\ &&
	- 2 (\gamma^{a})_{\alpha \beta} \nabla_{a}{B_{b}{}^{i j}}
	+ 4 \,(\gamma^{c})_{\alpha \beta} \, W_{c b a} B^{a}{}^{i j} ~, \\
\nabla_\a^{i} C_{ab} &=&
	- \frac{\ri}{8} \eta_{a b} \rho_{\alpha \beta}{}^{\beta i}
	- \frac{\ri}{8} (\gamma_{a b})_{\beta}{}^{\gamma} \rho_{\alpha \gamma}{}^{\beta i}
	+ \frac{\ri}{8} (\gamma_{b})_{\alpha \beta} (\tgamma_{a})^{\gamma \delta} \rho_{\gamma \delta}{}^{\beta i}
	+ \ri (\gamma_{a c}){}_{\alpha}{}^{\beta} \, \nabla^{c} \Lambda_{\beta b}{}^{i}
	\nonumber \\ &&
	+ \big( 2 \ri \,\eta_{a b} (\gamma_{c})_{\alpha \beta}
	- 2 \ri \,\eta_{a c} (\gamma_{b})_{\alpha \beta}
	- \ri (\gamma_{a b c})_{\alpha \beta} \big) X^{\beta}_j B^{c}{}^{i j}
	- 4 \ri \,(\gamma_{a})_{\alpha \delta} B^{c}{}^{i j}
	(\gamma_{b c})_{\gamma}{}^{\beta} X_{\beta j}{}^{\gamma \delta}
	\nonumber \\ &&
	- \frac{\ri}{2} W_{a c d} (\gamma^{c d})_\alpha{}^{\beta} \,\Lambda_{\beta b}{}^{i}
	+ 2 \ri \, W_{b}{}^{c d} \,(\gamma_{a d})_{\alpha}{}^{\beta} \,\Lambda_{\beta c}{}^{i}
\ ,
\eea
\esubeq
and
\bsubeq\label{SB}
\bea
S^\g_k \L_{\a a}{}^{i}
&=&
-\frac{44\ri}{3}\d_\a^{\g}B_a{}_{k}{}^{i}
-\frac{4\ri}{3}(\g_{ab})_\a{}^{\g}B^b{}_k{}^{i}
\ , \\
S^\g_{k} C_{ab}
&=&
-\frac{9}{2} (\tilde\g_{a})^{\g\d}\L_{\d}{}{}_b{}_{k}
-\hf(\tilde\g_{b})^{\g\d}\L_\d{}_{a}{}_{k}
+\hf\eta_{ab}(\tilde\g_{c})^{\g\d}\L_\d{}^{c}{}_{ k}
+\frac{1}{2} (\tilde\g_{abc})^{\g\d}\L_\d{}^{c}{}_{k}
\ ,~~~~~~
\eea
\esubeq
while each is annihilated by $K^a$.
These conditions are equivalent to the condition that $H_4$ is primary, $K^C H_4=0$.
Note in the supersymmetry transformations \eqref{deB}, additional component fields
appear. These are defined by the descendants
\bea
C_{\alpha}{}^{\beta i j} :=
	\frac{3}{4} \nabla_{\alpha k} \Lambda^{\beta i j k}~,  \qquad
\rho_{\alpha \beta}{}^{\gamma i} :=
	-\frac{2\ri}{3} \nabla_{[\alpha j} C_{\beta]}{}^{\gamma i j}
	~,
\eea
but we do not give their transformations here as they are unnecessary for what follows.
A more complete discussion of this multiplet can be found in \cite{BKNT16,BNT-M17}.

\subsubsection{The gauge 3-form multiplet in components}

The component structure of the 3-form multiplet
follows from the superspace constraints \eqref{4FormConst}.
Besides the lowest components of the descendant superfields $\L_{\a a}{}^i$ and $C_{ab}$,
there are other component fields of $B_a{}^{ij}$ that we do not summarise here, see \cite{BNT-M17}
for more details.
In later sections,
we will only make use of the component projections of $B_a{}^{ij}$,
$\L_{\a a}{}^i$, and
$C_{ab}$ whose $\d=\d_Q+\d_S+\d_K$ transformation,
that follow from \eqref{deB} and \eqref{SB}, prove to be
\bsubeq\label{deB-2}
\bea
\delta B_a{}^{jk} &=& - \frac{\ri}{2} \xi_i \g_a \L^{\b ijk} + \ri \,\xi^{(j} \L{}_a{}^{k)}
\ , \\
\delta \L_{\b b}{}^j &=&
	\frac{1}{2} C_{\beta}{}^{\gamma i j} (\gamma_{b} \xi_i)_\gamma
	+ \frac{1}{6} \xi^\alpha_i C_{\alpha}{}^{\gamma i j} (\gamma_{b})_{\beta \gamma}
	+ (\gamma^{a} \xi^j)_\beta C_{a b}
	+ 2  (\gamma^{a} \xi_i)_{\beta} \nabla_{a}{B_{b}{}^{i j}}
	\nonumber \\ &&
	+ 2 \,(\gamma^{c} \xi_i)_{\beta} \, T^-_{c b a} B^{a}{}^{i j}
	+ \frac{44\ri}{3} \eta_{\alpha j} B_a{}^{i j}
	+ \frac{4\ri}{3} (\g_{ab} \eta_j)_\alpha \,B^b{}^{i j}~, \\
\delta C_{ab} &=&
	- \frac{\ri}{8} \eta_{a b} \, \xi_i \gamma^c \rho_c{}^i
	+ \frac{\ri}{8} \xi_i \gamma^c \tgamma_{a b} \rho_c{}^{i}
	+ \frac{\ri}{2} \xi_i \gamma_{b} \rho_{a}{}^{i}
	+ \ri \,\xi_i \gamma_{a c} \nabla_{c} \Lambda_{ b}{}^{i}
	\nonumber \\ &&
	+ \Big( \frac{4}{15} \ri \,\eta_{a b} \,\xi_i \gamma_{c} \chi_j
	- \frac{4\ri }{15}\,\eta_{a c} \,\xi_i \gamma_{b} \chi_j
	- \frac{2\ri }{15} \xi_i \gamma_{a b c} \chi_j
	- 4 \ri \, \xi_i \gamma_{a} R(Q)_{b c}{}_j \Big) B^{c}{}^{i j}
	\nonumber \\ &&
	+ \frac{\ri}{4} T^-_{a c d} \,\xi_i \gamma^{c d} \Lambda_{b}{}^{i}
	- \ri \, T^-_{b}{}^{c d} \,\xi_i \gamma_{a d} \Lambda_{c}{}^{i}
	-\frac{9}{2} \eta^k \tilde\g_{a} \L_b{}_k
	-\hf \eta^k \tilde\g_{b} \L_{a}{}_{k}
	\nonumber \\ &&
	+\hf\eta_{ab} \, \eta^k \tilde\g_{c} \L^{c}{}_{ k}
	+\frac{1}{2} \, \eta^k \tilde\g_{abc} \L^{c}{}_{k}~.
\eea
\esubeq


\subsection{The tensor multiplet and the dilaton-Weyl multiplet}
\label{tensor-multiplet-section}

So far we have described various off-shell multiplets coupled to the standard-Weyl  (or type I) multiplet
of $\cN=(1,0)$ conformal supergravity \cite{BSVanP}.
In this subsection we introduce a variant off-shell formulation
in which the matter fields $(T^-_{abc},\chi^i,D)$ of the standard-Weyl multiplet are replaced by a scalar $\sigma$,
a gauge two-form tensor $b_{mn}$, and a chiral $\psi_i$ matter fields.
These fields belong to a 6D $(1,0)$ tensor multiplet \cite{HST,Koller:1982cs} which
once coupled to the standard-Weyl multiplet, can be used to define a new multiplet of conformal supergravity:
the dilaton-Weyl (or type II) multiplet \cite{BSVanP}.
This plays an important role in six-dimensional supergravity since, to date, it has proven to be the simplest 
formulation
that can be consistently used to construct actions for general  off-shell supergravity-matter systems.
In flat superspace, the tensor multiplet has been constructed as a super gauge two-form
\cite{BergshoeffQM}.
Extending the curved superspace analysis of \cite{LT-M12},
we will first introduce the dilaton-Weyl multiplet by consistently describing the tensor multiplet gauge two-form
in conformal superspace, see also \cite{Kuzenko:2017zsw}.
Successively, we will reproduce the description in components of \cite{BSVanP}.

\subsubsection{The tensor multiplet and the dilaton-Weyl multiplet in superspace}

Consider a super two-form gauge
potential $B_2 = \frac{1}{2} E^B \wedge E^A B_{AB}$ possessing the gauge transformation
\be \d_{\r_1} B_2 = \rd \r_1 \ ,
\label{dB2}
\ee
where $\r_1$ is a one-form gauge parameter. The corresponding field strength is
\be H_3 = \rd B_2 = \frac{1}{3!} E^C \wedge E^B \wedge E^A H_{A B C} \ ,
\label{rdB2=H3}
\ee
where
\be
H_{A B C} = 3 {\nabla}_{[A} B_{BC)} + 3 {T}_{[A B}{}^{F} B_{|F|C)} \ .
\ee
The field strength must satisfy the Bianchi identity
\be \rd H_3 = 0 \ \implies \ {\nabla}_{[A} H_{BCD )} + \frac{3}{2} {T}_{[AB}{}^{E} H_{|E|CD)} = 0 \ .
\label{BI_H3}
\ee

In order to describe the tensor multiplet we need to impose some covariant constraints on
the field strength $H_3$
 \cite{BergshoeffQM,LT-M12}
\begin{align}
H_\a^i{}_\b^j{}_\g^k &= 0 \ , \quad
H_{a}{}_\a^i{}_\b^j = 2 \ri \eps^{ij} (\g_a)_{\a\b} \Phi \ , \quad S^\a_i \Phi = 0 \ , \quad \mathbb D \Phi = 2 \Phi
\ .
\label{tensor-ansatz}
\end{align}
By using the Bianchi identities \eqref{BI_H3},
the remaining components of the superform $H_3$ are uniquely fixed in terms of $\Phi$ and its descendants.
The solution implies that $\Phi$ satisfies the differential constraint corresponding to the tensor multiplet
\be \nabla_{\a}^{(i} \nabla_\b^{j)} \Phi = 0 \ ,
\label{constr-tensor}
\ee
and also determines the higher dimension super-form components to be
\bsubeq\label{tensor-2}
\bea
H_{ab}{}_\a^i &=& (\g_{ab})_\a{}^\b \nabla_\b^i \Phi \ , \\
H_{abc} &=& - \frac{\ri}{8} (\tilde{\g}_{abc})^{\g\d} \nabla_\g^k \nabla_{\d k} \Phi
- 4 W_{abc} \Phi \ .
\eea
\esubeq
It is worth noting that the constraint
\eqref{constr-tensor}
can be solved in terms of a constrained prepotential $V^{\a i}$ as follows \cite{SokatchevAA,BergshoeffQM,LT-M12}
\be \Phi = \nabla_\g^k V^\g_k \ , \quad \nabla_\a^{(i} V^{\b j)} - \frac{1}{4} \d_\a^\b \nabla_\g^{(i} V^{\g j)} = 0 \ .
\label{tensor-prepotential}
\ee
It is in fact simple to prove that the following two-form
\bea
\cB_2
=
-8\ri E^\b_j\wedge E^a (\g_a)_{\b\g}V^{\g j}
-E^b\wedge E^a(\g_{ab})_{\b}{}^{\a}\de_{\a}^i V^{\b}_i
\eea
obeys $\rd\cB_2=H_3$.
The prepotential $V^{\a i}$ is defined up to a shift by a superfield describing an abelian vector multiplet,
$V^{\a i}\to V^{\a i}+\L^{\a i}$ which leaves $H_3$ of \eqref{tensor-ansatz} and \eqref{tensor-2} invariant.

We introduce the following descendants of $\Phi$
\bea
\psi_\a^i := \nabla_\a^i \Phi
~,\qquad
H_{\a\b}
=
\frac{1}{6}(\g^{abc})_{\a\b}H^+_{abc}
=
-\ri\nabla_{(\a}^k \psi_{\b)k}
~,
\eea
where we used the decomposition of $H_{abc}$ into self-dual and anti-self-dual parts
$H_{abc} = H^+_{abc} + H^-_{abc}$.
The superfields $\psi_\a^i$ and $H_{\a\b}$
satisfy
\bsubeq\label{detensor}
\bea
\nabla_\a^i \psi_\b^j
&=&
- \frac{\ri}{2} \eps^{ij} H_{\a\b} - \ri \eps^{ij} {\nabla}_{\a\b} \Phi
\ ,
 \\
\nabla_\g^k H_{\a\b}
&=&
- 4 {\nabla}_{\g(\a} \psi_{\b)}^k
~,
\eea
\esubeq
and
\bea
S^\b_j \psi_\a^i
=
4 \d_\a^\b \d_j^i \Phi
~ ,~~~~~~
S^\g_k H_{\a\b}
= - 24 \ri \d^\g_{(\a} \psi_{\b) k}
\ .
\label{Stensor}
\eea
Both $ \psi_{\a}^{i} $ and $H_{\a\b}$ are annihilated by $K^a$.
Note also that \eqref{Stensor}
turn out to be equivalent to  $H_3$ being a primary superform $K^C H_3=0$.

Assuming the tensor multiplet superfield $\Phi$ is everywhere nonvanishing,
$\Phi\ne 0$, which is a standard requirement for a conformal compensator,
it is straightforward to check that the constraint \eqref{constr-tensor} implies
the following relations
\bsubeq \label{WeylStandard->Dilaton-Weyl}
\bea
W_{abc} &=& - \frac{1}{4 \Phi} H^-_{abc} \ , \\
X^{\a i} &=& - \frac{\ri}{4 \Phi} \Big( {{\nabla}}^{\a\b} \psi_\b^i - W^{\a\b} \psi_\b^i \Big) \ , \\
Y &=& \frac{1}{2 \Phi} \Big(
{\nabla}^a {\nabla}_a \Phi
- 4 X^{\a i} \psi_{\a i} - \frac{2}{3} W^{abc} H_{abc}
\Big) \ .
\eea
\esubeq
This shows that the covariant superfield of the standard-Weyl multiplet, $W_{abc}$ and its descendants,
are now composed of the tensor multiplet covariant superfields $\F$, $H_{abc}$ and their descendants.
The result is a superspace description of the dilaton-Weyl (or type II) multiplet
of conformal supergravity \cite{BSVanP,LT-M12}.

\subsubsection{The tensor multiplet and the dilaton-Weyl multiplet in components}

The covariant component fields of the tensor multiplet can be identified
by the component projections $\s := \Phi|$ and $\psi_\a^i|$.
The supercovariant 3-form field strength
is just the component projection of the superfield $H_{abc}$.
By taking the double bar projection of \eqref{rdB2=H3} one can derive
\bea
H_{abc}=
 h_{abc}
+\frac{3}{2}\psi_{[a}{}^i \g_{bc]}\psi_i
+\frac{3\ri}{2}\psi_{[a}{}^i\g_{b}\psi_{c]}{}_{i} \s
~.
\eea

Using \eqref{detensor} and \eqref{Stensor},
and suppressing the component projection on the superfields $\psi_\a^i$ and $H^+_{abc}$,
 one can obtain the
$\d=\d_Q+\d_S+\d_K$
transformations of the components:\bsubeq
\bea
\d \s &=& \xi_i \psi^i \ , 
\label{dsigma}\\
\d \psi^i &=& \frac{\ri}{12}\xi^{i} \g_{abc} H^{abc}
+\ri \xi^{i} {\slashed{\nabla}} \s + 4 \eta^i \s \ , 
\label{dpsi}\\
\d H^+_{abc} &=&
- \hf  \xi_k \g^d\tilde{\g}_{abc}{{\nabla}}_{d} \psi^k
- 3 \ri  \eta^k \tilde{\g}_{abc}\psi_{k}
\ ,
\eea
\esubeq
where
\bsubeq
\bea
{\nabla}_a \s &=& {\cD}_a \s - \hf \psi_a{}_i \psi^i
\label{nabla-sigma}
\ , \\
{\nabla}_a \psi^i &=& {\cD}_a \psi^i
- \frac{\ri}{24} \g^{bcd} \psi_a{}^{i} H_{bcd}
- \frac{\ri}{2} \psi_a{}^{i} {\slashed{\nabla}} \s
- 2 \phi_a{}^i \s
\ .
\label{nabla-psi}
\eea
\esubeq
The transformation of the component gauge two-form $b_{mn} = B_{mn}|$
can be obtained by using the fact that $B_2$ is a primary superform and
by projecting to components the supergravity gauge transformation of $B_2$,
$\delta_\cG B_2 =\hf E^{B}\wedge E^{A} \xi^C H_{CAB}$,
leading to
\bea
\d b_{mn}&=&
e_{[n}{}^{b}\big(e_{m]}{}^{a}\x^\g_k H_{\g}^k{}_{ab}|
+\psi_{m]}{}^\a_i\x^\g_k H_{\g}^k{}_\a^i{}_{b}|\big)
=
\x_i\g_{mn}\psi^i
+2 \ri \x_i \g_{[m}\psi_{n]}{}^{i}\s
 \ .
 \label{db2}
\eea
The relations \eqref{WeylStandard->Dilaton-Weyl} 
that define the dilaton-Weyl multiplet in superspace
translate to the following at the component level
\bsubeq\label{dilaton-Weyl}
\bea
T^-_{abc} &=& \frac{1}{2 \s} H^-_{abc} \ , \\
\chi^{i} &=&
- \frac{15\ri}{8 \s}{\slashed{\nabla}} \psi^i
- \frac{5\ri}{32\s} T^-_{abc}\tilde{\g}^{abc} \psi^i
\ , \\
D &=&
\frac{15}{4 \s} {\nabla}^a {\nabla}_a \s
- 2\chi^{i} \psi_{i}
+\frac{15}{12 \s}   T^-_{abc} H^{abc}
\label{DdilatonWeyl}
\ ,
\eea
\esubeq
with
\bea
{\de}^a{\de}_a\s
&=&
\Big({\cD}_a{\de}^a
-4\frak{{f}}_{a}{}^{a}
+\frac{\ri}{2}\psi_a{}_i\g^a\chi^{i}\Big)\s
-\hf\psi_a{}_i\Big(
{\de}^a\psi^i
+\frac{1}{4}T^-_{bcd}\g^a\tilde{\g}^{bcd}\psi^i
\Big)
\non\\
&&
+\frac{\ri}{2}{\phi}_a{}^i\tilde{\g}^a\psi_{i}
~,
\eea
where ${\de}_a\s$ and ${\de}_a\psi^i$ are respectively given in \eqref{nabla-sigma} and \eqref{nabla-psi},
while
\bea
{\mathfrak{f}}_{a}{}^{a}
&=&
-\frac{1}{20}\cR
-\frac{\ri}{8}\psi^{a}{}_k\g^{b}{R}(Q)_{ab}{}^k
+\frac{\ri}{60} \psi_{a}{}_k\g^a\chi^k
+\frac{1}{20}\psi_{a}{}_k \g^{ab}{\f}_{b}{}^k
+\frac{\ri}{40}\psi_{a}{}^{ k}\g_b\psi_{c}{}_k\,T^-{}^{abc}
~.~~~~~~~~~
\label{faaconf}
\eea
In all the previous equations the reader should keep in mind that on the right hand side
$T^-_{abc}$, $\chi^i$ and $D$ are 
built from the tensor multiplet covariant fields $\s$, $\psi^i$
and $H_{abc}$.
Equations \eqref{dilaton-Weyl} in fact show
how in the dilaton-Weyl multiplet
the covariant fields of the standard-Weyl multiplet are replaced with the fields of the tensor multiplet.
The dilaton-Weyl multiplet will play a central role
in the construction of curvature squared terms in later sections.

The supersymmetry transformations of the independent component connections of the dilaton-Weyl multiplet
$(e_m{}^a,\psi_m{}_i,\cV_m{}^{kl},b_m)$
can be read straightforwardly  from the transformation rules of the same fields in the standard-Weyl multiplet
\eqref{eq:WeylSUSY2-a}--\eqref{eq:WeylSUSY2-d}
where now \eqref{dilaton-Weyl} should be used everywhere.
The transformation rules of the independent matter fields $(\s,\psi^i,b_{mn})$ 
of the dilaton-Weyl multiplet are respectively given by 
\eqref{dsigma}, \eqref{dpsi} and \eqref{db2}.

In this section we have described in detail various off-shell matter multiplets coupled to the
standard-Weyl multiplet. For each of these multiplets,
it is clear that the same analysis holds straightforwardly in the case of
the coupling to the dilaton-Weyl multiplet. In fact, to switch from one off-shell
 conformal supergravity multiplet to the other,
it is only necessary to  use equations \eqref{dilaton-Weyl} everywhere or their superfield equivalents.


\section{Locally superconformal action principles} \label{ActFormulas}
\label{section-3}

In this paper we will make use of three
action principles
to construct various locally superconformal invariants. One of these involves the product
of a linear multiplet with an Abelian vector multiplet and describes the supersymmetric extension
of a $B_{4}\wedge \cF_2$ term. This action formula was the main building block for the
supergravity invariants in \cite{BSVanP}. In superspace it may be described by a
full superspace integral
\be
\int \rd^6x\, \rd^8\q\, E\,U_{ij} \, {L}^{ij} \ ,
\label{BF0}
\ee
where $U_{ij}$ is the Mezincescu prepotential for the vector multiplet \cite{Mezincescu:1979af,Howe:1981qj},
$L^{ij}$ is a linear multiplet
and
$E={\rm Ber}(E_M{}^A)$ is the Berezinian (or superdeterminant) of the supervielbein.

A second locally superconformal invariant
is  the so-called $A$ action principle of \cite{BKNT16,BNT-M17}. This is based on a
primary dimension-9/2 superfield $A_\a{}^{ijk}$ whose differential constraint will be reviewed later in this section.
This describes the bottom component of a covariant closed super 6-form that was originally constructed in
\cite{ALR14}.

Another action formula involves the product of a tensor multiplet with
a gauge 3-form multiplet
and it describes the supersymmetric extension of
a $B_2\wedge H_4$ term.
Its existence was noted in \cite{BKNT16} where
it was described in terms of the $A$ action principle
with a superfield Lagrangian $A_\a{}^{ijk}$ chosen as
\be A_\a{}^{ijk} = \eps_{\a\b\g\d} V^{\b (i} B^{\g\d jk)} \ ,
\label{BH0}
\ee
where $V^{\a i}$ is a prepotential for the tensor multiplet \eqref{tensor-prepotential}
while $B^{\a\b}{}^{ij}=(\tilde{\g}^a)^{\a\b}B_{a}{}^{ij}$ is the primary superfield
describing a gauge 3-form supermultiplet as in section \ref{gauge-3-form-multiplet-section}.
The bosonic component structure of this new locally superconformal invariant
was given in \cite{NOPT-M17} while its complete structure is given for the first time
here in our paper.

The $A$ action principle was already studied in detail in \cite{BKNT16,BNT-M17}.
In this section we are going to present a construction of both the $B_{4}\wedge \cF_2$
and $B_{2}\wedge H_4$ action principle by using the superform approach
to construct supersymmetric invariants \cite{Hasler, Ectoplasm, GGKS, Castellani}.
The advantage compared to using \eqref{BF0} and \eqref{BH0} is two-fold:
firstly, we will see that no prepotential superfields, either for the vector or tensor multiplets, appear explicitly;
secondly, it is straightforward to reduce the results to component fields making direct contact with the superconformal
tensor calculus.
Our approach follows the one of, e.g., \cite{BKN12, KT-M12, BKNT-M13, KN14}.


\subsection{Superform construction of locally superconformal invariants}

For a 6D $\cN = (1,0)$ superspace, we introduce a closed 6-form $J$
\begin{align}
J = \frac{1}{6!}
\rd z^{M_6} \wedge \cdots \wedge \rd z^{M_1}  \,J_{M_1 \cdots M_6}(z)
\ ,\qquad
 \rd J = 0 ~.
\end{align}
Such a closed superform leads to the supersymmetric action principle
\cite{Hasler, Ectoplasm, GGKS, Castellani}
\bsubeq
\label{ectoS}
\bea
S &=& \int_{\cM^6} i^* J = \int \rd^6 x \,e \,{}^* J |_{\q=0}\ , \qquad
e:=\det{e_m{}^a}
~,
\\
&&{}^*J := \frac{1}{6!} \eps^{m_1m_2m_3m_4m_5m_6} J_{m_1m_2m_3m_4m_5m_6} \ ,
\eea
\esubeq
where $i:\cM^6 \rightarrow \cM^{6|8}$ is the inclusion map and
$i^*$ is its pullback
which effectively acts as the double bar projection,
$\q^\mu_i =\rd \q^\mu_i=0$.
Due to the transformation rule of a closed 6-form
\be
\d_{\xi} {J} = \cL_\xi {J} \equiv i_{\xi} \rd {J} + \rd i_{\xi} {J}
= \rd i_{\xi} {J} \ ,
\ee
up to boundary terms that we are going to neglect,
the closure of $J$ guarantees that the action is
invariant under general coordinate transformations of superspace
generated by a vector field  $\x = \x^A E_A = \x^M \pa_M$.

The action principle \eqref{ectoS} is manifestly invariant under  superdiffeomorphisms.
In addition, the action must be invariant under all other gauge transformations.
Among other possible gauge transformations of matter multiplets in a specific model,
for conformal supergravity we need to include the other superconformal transformations
describing the structure group of conformal superspace, which form the subgroup $\cH$.
To ensure the invariance of \eqref{ectoS} then $J$ is demanded to transform.

A special case is when the closed 6-form is itself invariant,
$\d_{\cH} J =  0$.
This implies that if one instead decomposes $J$ in
the tangent frame,
\begin{align}
J &= \frac{1}{6 !} E^{A_6} \wedge
\cdots
\wedge E^{A_1}
J_{A_1
\cdots
A_6}~,
\end{align}
the components $J_{A_1 \cdots A_6}$ transform covariantly and obey the
covariant constraints
\be \label{eq:CovClosure}
\nabla_{[A_1} J_{A_2 \cdots A_7 ) } + 3 T_{[A_1 A_2}{}^{B} J_{|B|A_3 \cdots A_7 )} =  0 \ .
\ee
In particular, their $S$ and $K$ transformations are given by
\be
S^\b_j J_{a_1 \cdots a_{n}}{}_{\a_1}^{i_1}{\cdots}{}_{\a_{6 - n}}^{i_{6-n}} =
- \ri \,n (\tilde{\g}_{[a_{1}})^{\b \g} J_{\g j}{}_{a_2 \cdots a_{n}]}{}_{\a_1}^{i_1}{\cdots}{}_{\a_{6- n}}^{i_{6-n}}{}  \ , \qquad
K^b J_{A_1 \cdots A_6} = 0~.\label{recurs}
\ee
These are equivalent to demanding $J$ to be a primary six-form $K^AJ=0$.

Once constructed an appropriate invariant closed super 6-form $J$, it is straightforward to describe
a gauge invariant action principle in components.
In fact,  by expressing the action  \eqref{ectoS} in terms of tangent frame indices one obtains
\bea
S &=&
\frac{1}{6!}
 \int \rd^6 x
\,e\,\eps^{m_1 \cdots m_6}
E_{m_6}{}^{A_6} \cdots
E_{m_1}{}^{A_1}
{J}_{ A_1 \cdots A_6}
|_{\q=0}
\ , \qquad
\non\\
&=&
\frac{1}{6!}\int \rd^6 x
\, e\,\eps^{a_1 \cdots a_6} \Big{[}\,
J_{a_1\cdots a_6}
-3\psi_{a_1}{}^\a_i
J_{a_2\cdots a_6}{}_\a^i
-\frac{15}{4}\psi_{a_1}{}^{\a_2}_{i_2}\psi_{a_2}{}^{\a_1}_{i_1}
J_{a_3\cdots a_6}{}_{\a_1}^{i_1}{}_{\a_2}^{i_2}
\non\\
&&~~~~~~
+\frac{5}{2}\psi_{a_1}{}^{\a_3}_{i_3}\psi_{a_2}{}^{\a_2}_{i_2}\psi_{a_3}{}^{\a_1}_{i_1}
J_{a_4a_5a_6}{}_{\a_1}^{i_1}{}_{\a_2}^{i_2}{}_{\a_3}^{i_3}
+\frac{15}{16}\psi_{a_1}{}^{\a_1}_{i_1}
\cdots
\psi_{a_4}{}^{\a_4}_{i_4}
J_{a_5a_6}{}_{\a_1}^{i_1}
{}_{\cdots\,}^{\cdots\,}
{}_{\a_4}^{i_4}
\non\\
&&~~~~~~
-\frac{3}{16}
\psi_{a_1}{}^{\a_1}_{i_1}
\cdots
\psi_{a_5}{}^{\a_5}_{i_5}
J_{a_6}{}_{\a_1}^{i_1}
{}_{\cdots\,}^{\cdots\,}
{}_{\a_5}^{i_5}
-\frac{1}{64}
\psi_{a_1}{}^{\a_1}_{i_1}
\cdots
\psi_{a_6}{}^{\a_6}_{i_6}
J_{\a_1}^{i_1}
{}_{\cdots\,}^{\cdots\,}
{}_{\a_6}^{i_6}
 \Big{]} |_{\q=0} \ ,
\label{superform-action}
\eea
which is the standard expansion of supergravity actions as a supercovariant power series in the gravitini.
Let us now turn to the examples relevant for our paper.


\subsection{The $B_4 \wedge \cF_2$ action principle}

Consider the gauge 4-form multiplet associated to a linear multiplet, as in section \ref{linear-multiplet-section},
together with an Abelian vector multiplet, see section \ref{YMmultiplet}.
By considering the gauge 4-form $B_4$ and the vector multiplet two-form field strength $\cF_2$
 we can construct the primary super 6-form $B_4 \wedge \cF_2$.
This is manifestly invariant under vector multiplet transformations, that leave invariant $\cF_2$,
while it transforms as an exact form under the gauge transformation of the 4-form
\eqref{dB4}, $\d_{\r_3}(B_4 \wedge \cF_2)=\rd(\r_3\wedge \cF_2)$,
so that its exterior derivative is gauge invariant
\bea
\rd (B_4 \wedge \cF_2)
=E_5 \wedge \cF_2
~.
\eea
It turns out that the covariant constraints on $\cF_2$ and $E_5$ mean that
$E_5\wedge \cF_2 $ is Weil trivial \cite{BPT}. In other words,
it turns out that there is a second super 6-form $\S_{E_5 \wedge \cF_2}$
obeying $\rd \S_{E_5 \wedge \cF_2}={E_5 \wedge \cF_2}$
such that $\S_{E_5 \wedge \cF_2}$ is constructed entirely in terms of the covariant superfields in 
$\cF_2$ and $E_5$.
This implies that $\S_{E_5 \wedge \cF_2}$ is primary and gauge invariant.
We call this super 6-form the curvature induced form.
This means that by construction the primary super 6-form
\be
J_{B_4 \wedge \cF_2} = B_4 \wedge \cF_2 - \S_{{E_5 \wedge \cF_2}}
\label{JB4F2}
\ee
is closed, $\rd J_{B_4 \wedge \cF_2} = 0$, such that $\d_{\r_3}J_{B_4 \wedge \cF_2}=\rd(\r_3\wedge \cF_2)$
and then, according to the discussion in the previous section,
it leads to a locally superconformal and gauge invariant action.

It is not difficult to solve explicitly for $\S_{E_5 \wedge \cF_2}$.
It obeys
\bea \rd \S_{E_5 \wedge \cF_2} = {E_5 \wedge \cF_2} \ \Longleftrightarrow \
{\nabla}_{[A_1} \S_{A_2\cdots A_7)}
+ 3 {T}_{[A_1A_2}{}^{B} \S_{|B|A_3\cdots A_7)}
= 3 \cF_{[A_1A_2} E_{A_3\cdots A_7 )} \ ,
~~~~~~
\eea
with $\cF_{AB}$  the components of the abelian vector multiplet super two-form $\cF_2$, eqs.\,\eqref{Vector-2-form},
and $E_{A_1\cdots A_5}$ the components of  the linear multiplet super $5$-form $E_5$, eq.\,\eqref{Linear-5-form}.
A covariant superform solution $\S_{E_5 \wedge \cF_2}$ is given by
\bsubeq
\bea
\S_{abcde}{}_\a^i
&=& 2 \eps_{abcdef} (\g^{f})_{\a\b} \L^{\b}_j L^{ij}
\ ,
\\
\S_{abcdef} &=& - \eps_{abcdef} \Big( X^{ij} L_{ij} - 2 \ri \L^{\a i} \varphi_{\a i} \Big) \ ,
\eea
\esubeq
with all components at lower dimension  vanishing.

If we now plug the resulting closed 6-form
\eqref{JB4F2} into \eqref{superform-action} we obtain the component action principle
\bea
S_{B_4\wedge \cF_2}
=
 \int \rd^6 x \,e \, \Big(
-\hf f_{mn} \tilde{b}^{mn}
+X^{ij} L_{ij}
-2 \ri \L^{ i} \varphi_{i}
- \psi_a{}_i \g^a \L_j L^{ij}
\Big)
~,
\label{compB4F2action}
\eea
where we have suppressed spinor indices and we remind that we defined
\bea
\tilde{b}^{mn}:=\frac{1}{4!}\ve^{mnpqrs}b_{pqrs}
~.
\eea
The action \eqref{compB4F2action}, which we will refer to as $B_4\wedge \cF_2$ action principle,
was obtained for the first time in \cite{BSVanP}.


\subsection{The $B_2 \wedge H_4$ action principle}
\label{sectionB2H4action}

By starting from
the gauge two-form multiplet associated to a tensor multiplet, as in section \ref{tensor-multiplet-section},
together with a gauge 3-form multiplet of  section \ref{gauge-3-form-multiplet-section}, we can construct another
gauge invariant action principle using the same logic as in the previous subsection.

Considering the gauge two-form $B_2$ and the closed
4-form field strength $H_4=\rd B_3$
we construct the primary super 6-form $B_2 \wedge H_4$.
This is invariant under the 3-form gauge transformation, $\d_{\r_2}B_3=\rd \r_2$,
it transforms as an exact form under the two-form gauge  transformation
\eqref{dB2}, $\d_{\r_1}(B_2 \wedge H_4)=\rd(\r_1\wedge H_4)$, so that
\bea
\rd (B_2 \wedge H_4)
=H_3 \wedge H_4
~,
\eea
is gauge invariant.
The covariant constraints on $H_3$, eqs.\,\eqref{tensor-ansatz} and \eqref{tensor-2},
 and on $H_4$, eq.\,\eqref{four-form-000}, are such that the super 7-form
$H_3 \wedge H_4$ is again Weil trivial  and then it is possible to construct
a curvature induced super 6-form $\S_{H_3\wedge H_4}$, so that
\be
J_{B_2 \wedge H_4} =\S_{H_3\wedge H_4}- B_2 \wedge H_4
\label{JB2H4}
\ee
is closed, $\rd J_{B_2 \wedge H_4} = 0$, and such that $\d_{\r_1}J_{B_2 \wedge H_4}=\rd(\r_1\wedge H_4)$
leading to a locally superconformal and gauge invariant action.

The superform equation defining $\S_{H_3\wedge H_4} $ is
\be \rd \S_{H_3\wedge H_4} = H_3 \wedge H_4\ \Longleftrightarrow \ {\nabla}_{[A} \S_{BCDEFG)}
+ 3 T_{[AB}{}^{H} \S_{|H|CDEFG)}
= 5 H_{[ABC} H_{DEFG )} \ ,
\label{eq-H3H4}
\ee
with $H_3$ corresponding to a tensor multiplet, see section \ref{tensor-multiplet-section},
 and $H_4$ corresponding to the gauge 3-form multiplet, see section \ref{linear-multiplet-section}.
A covariant solution of equation \eqref{eq-H3H4} is given by
a super $6$-form $\S_{H_3\wedge H_4} $ possessing the following nontrivial components
\bsubeq
\bea
\S_{abcd}{}_{\r}^p{}_{\t}^q&=&
\frac{\ri}{2}\ve_{abcdef}(\g^{efg})_{\r\t}\F B_{g}{}^{pq}
\ ,\\
\S_{abcde}{}_{\d}^l
&=&
-\frac{1}{4}  \ve_{abcdef}\Big{[}\,
2\psi_{\d k}B^{f}{}^{kl}
+\ri\Phi\L_{\d}{}^{f}{}^l
+(\g^{fg})_{\d}{}^{\r}\big(2\psi_{\r k}B_{g}{}^{kl}-\ri\Phi\L_{\r g}{}^l \big)\Big{]}
 \ ,~~~~~~\\
\S_{abcdef}
&=&~
\frac{1}{4}\ve_{abcdef}(\tilde{\g}^{g})^{\a\b}\psi_{\a k}  \L_{\b g}{}^k
+\frac{1}{4}\ve_{abcdef}\Phi C
\ ,
\eea
\esubeq
with all lower dimension components vanishing and the descendant superfield $\L_{\a a}{}^i$ defined in
\eqref{descendants-4-form-2}
and $C$ given by the trace of $C_{ab}$, defined in eq.\,\eqref{descendants-4-form-3},
\bea
C:=
\eta^{ab}C_{ab}
=\frac{\ri}{12}(\tilde{\g}^a)^{\a\b}\de_{\a k}\de_{\b l}B_a{}^{kl}
=\frac{\ri}{12}\de_{\a k}\de_{\b l}B^{\a\b}{}^{kl}
= \frac{1}{8} (\tilde\gamma^{a})^{\alpha \beta} \nabla_{\a k} \L_{\b a}{}^k
~.
\eea
The antisymmetric part of $C_{ab}$ is related to the supercovariant four-form
field strength $H_{abcd}$ via \eqref{eq:Habcd}, leading to
\begin{align}
C_{[ab]} &:=
    - \frac{1}{12} \veps_{a b c d e f} h^{c d e f}
    - \frac{3i}{4} \psi^c{}_{j} \gamma_{[a b} \Lambda_{c]}{}^{j}
    - \frac{i}{8} \veps_{a b c d e f} \psi^c{}_{j} \gamma^{d e g} \psi^{f}{}_{k} \, B_g{}^{j k}~,
\end{align}
where $h_{mnpq} = 4 \pa_{[m} b_{npq]}$.
If we now plug the resulting closed 6-form
\eqref{JB2H4} into \eqref{superform-action} we obtain the component action principle
\bea
S_{B_2\wedge H_4} &=&
\int \rd^6 x \, e \,
\Big\{
-\frac{1}{4}\s C
-\frac{1}{4}\psi_{j}\tilde{\g}^{g} \L_{g}{}^j
+\frac{1}{4}b_{ab}C^{ab}
\non\\
&&~
+\frac{1}{8}   \psi_{a}{}_i
\Big{[}\,
2\psi_{j}B^{a}{}^{ij}
+\ri\s \L^{a}{}^i
+\g^{ab}
\big(
2\psi_{ j}B_{b}{}^{ij}
-\ri\s\L_{b}{}^i
\big)
+\frac{3\ri}{2}b_{bc} \g^{[ab}\L^{c]}{}^i
\Big{]}
\non\\
&&~
+\frac{\ri}{4} \psi_a{}_i
\Big{[}\,
\frac{1}{2}\g^{abc}\s B_{c}{}^{ij}
-b_{cd}\g^{[abc}B^{d]}{}^{ij}
\Big{]}
\psi_b{}_j
\Big\}
~,
\label{BH-action}
\eea
where spinor indices have been suppressed. Note that the couplings involving $b_{ab}$ can be
rewritten as $\veps^{abcdef} b_{ab} h_{cdef}$.
We will refer to the invariant \eqref{BH-action} as the $B_2\wedge H_4$ action principle.
The bosonic part of the action \eqref{BH-action} appeared for the first time in \cite{NOPT-M17}.
It is important to mention that although one can formally work in the standard-Weyl multiplet
in the construction of the above invariant, the tensor multiplet is on-shell. To have an
off-shell description, one must identify the tensor multiplet as the one of the dilaton-Weyl
multiplet.
Note that by turning off the supergravity multiplet in \eqref{BH-action} 
one obtains the action principle that was first used
in \cite{HST} to describe the action for the rigid supersymmetric Yang-Mills multiplet.
Let us see how \eqref{BH-action} allows us to extend that result to the curved case.

\subsubsection{Non-Abelian vector multiplet action}

It is worth presenting a first nontrivial application of the $B_2\wedge H_4$ action
principle \eqref{BH-action}: a direct construction of the non-Abelian vector multiplet action
in a general dilaton-Weyl multiplet background. Previously, this action had been constructed
in \cite{BSVanP} undertaking the following steps:
(i) restrict to an Abelian vector multiplet;
(ii) in a dilaton-Weyl background,
construct a composite linear multiplet in terms of an Abelian vector multiplet;\footnote{In superspace,
with $\L^{\a i}$ the superfield strength of an Abelian vector multiplet,
the composite linear multiplet is described by the superfield
$\cF^{ij}
:=
\big(\ri (\de_\a^{(i}\F)\L^{\a j)}
+\frac{\ri}{4} \F\de_\a^{(i}\L^{\a j)}\big)
=
\big(\ri \psi_\a^{(i}\L^{\a j)}
+\F X^{ij}\big)
$
which can be easily proven to be a dimension-4 primary such that $\de_\a^{(i}\cF^{jk)}=0$ \cite{LT-M12}.}
(iii) plug the composite linear multiplet in the $B_4\wedge \cF_2$ action principle;
(iv) realise that the resulting action is invariant also in the case of a non-Abelian vector multiplet.

In the case of a general non-Abelian vector multiplet, described by the superfield strength ${\bm \L}^{\a i}$,
as observed in \cite{BKNT16}
it is straightforward to construct a composite gauge 3-form multiplet
in terms of the primary dimension three superfield
\bsubeq
\bea
{B}^{\a\b ij} &=&-4 \ri \Tr \Big( {\bm \L}^{\a (i} {\bm \L}^{\b j)} \Big) \ ,
\eea
which, due to \eqref{vector-Bianchies},
satisfies the constraints \eqref{4FormConst}.
The descendants of the composite ${ B}^{\a\b ij} $ are then
\bea
\L_{\a a}{}^i &=&  \frac{2}{3} (\g_a)_{\b\g} \Tr \Big{[}
\ri \d_\a^\b {\bm X}^{ij} {\bm \L}^\g_j
- 6 \ri {\bm \cF}_\a{}^\b {\bm \L}^{\g i}
\Big{]}
\ , \\
C &=&2 \Tr \Big{[} {\bm \cF}^{ab} {\bm \cF}_{ab}
-{\bm X}^{ij} {\bm X}_{ij}
-2\ri {\bm \L^{\a k}} {\bm \nabla}_{\a\b} \bm \L^\b_k \Big{]} \ , \\
C_{[ab]}
&=&
\hf\Tr \Big{[} \eps_{abcdef} {\bm \cF}^{cd} {\bm \cF}^{ef}
-\ri {{\bm \nabla}}^c \big( (\g_{abc})_{\a\b} {\bm \L}^{\a k} {\bm \L}^{\b}_k \big)
\Big{]}
\ .
\eea
\esubeq

If we plug these results into the $B_2 \wedge H_4$ action principle \eqref{BH-action} we find
the action for a general non-Abelian vector multiplet
\bea
S_{\s F^2}
&=&
\int \rd^6 x \,e \,\Tr
\Big\{
-\hf\s  {\bm f}^{ab} {\bm f}_{ab}
+\hf\s {\bm X}^{ij} {\bm X}_{ij}
+\frac{1}{8}\eps^{abcdef}b_{ab} {\bm f}_{cd} {\bm f}_{ef}
\Big\}
\non\\
&&
+{\rm fermions}
~,
\label{SYMaction}
\eea
where we neglected the fermionic terms.
The action, up to a change of notations, coincides with the result of \cite{BSVanP}.

\subsection{The $A$ action principle}
\label{sectionAaction}

Let us conclude this section by reviewing the salient results concerning the $A$ action principle.
The reader should refer to \cite{BKNT16,BNT-M17} for a complete analysis.

The $A$ action principle is based on
a primary dimension $9/2$ superfield $A_\a{}^{ijk}=A_\a{}^{(ijk)}$ obeying
the reality condition $\overline{A_\a{}^{ijk}} = A_{\a\, ijk}$ and
satisfying the differential constraint
\be
\nabla_{(\a}^{(i} A_{\b)}{}^{jkl)} = 0 \ .
\label{Aconst}
\ee
As already mentioned, this superfield arises as the bottom component of a covariant closed super 6-form
\cite{ALR14, BKNT16}.
By using the superform approach for the construction of supersymmetric invariants
it is possible to obtain the  invariant \cite{BNT-M17}
\bea \label{AactionPrinc}
S_{A} = \int \rd^6 x \, e & \Big\{&
	F
	- \frac{\ri}{4} \psi_{a i} \Omega'{}^{a i}
	- \frac{\ri}{144} \psi_{d i} \gamma^{de} \gamma^{abc} \psi_{e j} S_{abc}^+{}^{ij}
	- \frac{\ri}{12} \psi_{a i} \gamma^{abc} \psi_{b j} E_c{}^{i j}
	\non\\
	&&
	+ \frac{1}{16} (\psi_{a i} \gamma^{abc} \psi_{b j})\, (\psi_{ck} A^{ijk})
	\Big\}
	~,
\eea
where the following descendant component fields of $A_\a{}^{ijk}$ have been introduced:
\bsubeq \label{AsuperfieldDescend}
\begin{align}
S_{abc}^+{}^{ij} &:= \frac{3}{32} (\tgamma_{abc})^{\a\b} \nabla_{\a k} A_{\b}{}^{ijk}\loco \ , \qquad
E_{a}{}^{ij} := \frac{3}{16} (\tgamma_a)^{\a\b} \nabla_{\a k} A_{\b}{}^{ijk}\loco \ ,
\\
\Omega'_{a\a}{}^{i} &:=
	\frac{\ri}{32} (\tgamma_a)^{\b\g} (
	\nabla_{\b j} \nabla_{\g k} A_\a{}^{ijk}\loco
	+ \nabla_{\a j} \nabla_{\b k} A_\g{}^{ijk}\loco
	)~, \\
F &:=
\frac{\ri}{2^4 4!}\eps^{\a\b\g\d} \nabla_{\a i} \nabla_{\b j} \nabla_{\g k} A_\d{}^{ijk}\loco \ .
\label{4.15e}
\end{align}
\esubeq
The superconformal transformations of the previous multiplet are rather involved,
see \cite{BNT-M17} for details. The main point is that \eqref{AactionPrinc} proves to be manifestly 
locally superconformal
invariant. For the scope of this paper we will only need to know that the purely bosonic part of this action can be
extracted from the $F$ component \eqref{4.15e} of the $A_\a{}^{ijk}$ multiplet.


\section{Einstein-Hilbert Poincar\'e supergravity}
\label{EH}

In this section we describe how to construct in conformal
superspace a supersymmetric extension of the Einstein-Hilbert term
and then describe the off-shell and on-shell two-derivative Poincar\'e supergravity theory
reproducing the results of \cite{BSVanP}.

\subsection{Linear multiplet action}

An action for the linear multiplet can be constructed using the $B_4 \wedge \cF_2$ action principle
\eqref{compB4F2action}
with the vector multiplet built out of the linear multiplet \cite{BSVanP}.
The appropriate composite vector multiplet
is described by the primary dimension-3/2 superfield strength
\bea
\L^{\a i}
&=& - \frac{1}{2L} {\nabla}^{\a\b} \varphi_\b^i
- \frac{1}{2L} \big( W^{\a\b} \varphi_\b^i - 2 \ri X^\a_j L^{ji} \big)
+ \frac{1}{4L^3} L_{jk}  ({\nabla}^{\a\b} L^{ij}) \varphi_\b^k
\non\\
&&
- \frac{1}{8 L^3} L^{ij} {\rm E}^{\a\b} \varphi_{\b j}
- \frac{\ri}{8 L^5} \eps^{\a\b\g\d} \varphi_{\b j} \varphi_{\g k} \varphi_{\d l} L^{ij} L^{kl}
\ ,
\label{composite-vector-linear}
\eea
where $L^2 := \hf L^{ij} L_{ij}$  is assumed to be nonvanishing, $L\ne 0$.
The component fields 
of the  composite vector multiplet 
can be   computed straightforwardly.
They include the $\q=0$ projection of $\L^{\a i}$
together with the descendant components of the composite vector multiplet,
$X^{ij}=\frac{\ri}{4}\de_\g^{(i}\L^{\g j)}|$ and 
$\cF_{ab}=-\frac{\ri}{8}(\g_{ab})_\b{}^\a\de_{\a}^k\L^{\b}_k|$,\footnote{Here and in what follows, 
whenever we only explicitly give the bosonic sectors, 
 any supercovariant field (which involves fermionic terms) can be replaced with its purely bosonic analogue, 
e.g. ${\rm E}^a$ with $h^a$, $H_{abc}$ with $h_{abc}$ and so on,
 since it will only change the suppressed fermionic terms.}
\bsubeq\label{composite-vector-linear-2}
\bea
X^{ij} &=& - \frac{1}{2}L^{-1} {{\nabla}}^a {{\nabla}}_a L^{ij}
- \frac{1}{15}L^{-1}  L^{ij} D
+ \frac{1}{16 }L^{-3} {\rm E}^a {\rm E}_a L^{ij}
- \frac{1}{4}L^{-3} {\rm E}^a L^{k(i} {\cD}_a L^{j)}{}_k
\non\\
&&
+ \frac{1}{4}L^{-3} L_{kl} ({{\cD}}^a L^{k(i}) {{\cD}}_a L^{j) l}
+ {\rm fermions}
\ ,
\\
\cF_{ab} &= &
\hf {{\cD}}_{[a} \big( L^{-1}{\rm E}_{b]} \big)
+ \frac{1}{2}L^{-1} \cR_{ab}{}^{ij} L_{ij}
- \frac{1}{4}L^{-3} L_{ij} ({\cD}_{[a} L^{ki}) {\cD}_{b]} L^{j}{}_k
+ {\rm fermions}
\ ,~~~~~~
\eea
\esubeq
where
\be
{\nabla^a} {\nabla}_a L^{ij} =
{{\cD}}^a {\nabla}_a L^{ij}
- 8 \frak{f}_a{}^a L^{ij}
+ {\rm fermions}
\ ,
\ee
the expression for ${\de}_aL^{ij}$ is given in \eqref{deL}, and
 $\frak{f}_a{}^a$ is given in \eqref{faaconf}.
The $f_{mn}$ field strength \eqref{VM-field-strengths} in the case of the composite vector multiplet proves to be
\bsubeq
\bea
f_{mn}
&=&
2 \partial_{[m} \G_{n]}
- \frac{1}{4}L^{-3} L_{ij} (\partial_{[m} L^{ki}) \partial_{n]} L^{j}{}_k
\ ,
\\
 \G_m &:=& \frac{1}{2}L^{-1} \Big(\cV_m{}^{ij} L_{ij}+\hf h_m\Big)
+ {\rm fermions}
~.
\eea
\esubeq
The previous composite vector multiplet coincides, up to change of notation, to the one originally constructed in
\cite{BSVanP} where the reader can also find  the completion with the fermionic terms.

Using \eqref{compB4F2action} and disregarding a total derivative,
the action for the linear multiplet takes the form
\bea
S_{\rm EH}
&=&
\int \rd^6 x \,e \, \Big\{
- \frac{2}{5} \cR L
- \frac{2}{15} D L
- \frac{1}{8 L} {\rm E}^a {\rm E}_a
- \frac{1}{2L} {\rm E}^a \cV_a{}^{ij} L_{ij}
-  \cD^a {\cD}_a L
\non\\
&&
+ \frac{1}{4L} ({\cD}^a L^{ij}) {\cD}_a L_{ij}
+ \frac{1}{8 L^3} \tilde{b}^{mn} L_{ij} (\partial_m L^{k i}) \partial_n L^{j}{}_k
\Big\}
+ {\rm fermions}
\ .
\label{linear-action}
\eea

The previous action is by construction locally superconformal invariant
independently of the conformal supergravity background chosen, either the standard- or dilaton-Weyl multiplets
which we have not specified so far in this section.
In the standard-Weyl multiplet background the $D L $ coupling in \eqref{linear-action} 
implies $L=0$ on-shell. This is inconsistent
with the requirement of $L$ being a conformal compensator, $L\ne0$,
and makes the action \eqref{linear-action} alone inconsistent in a standard-Weyl
multiplet background.
A standard resolution of this issue is to consider \eqref{linear-action} in a dilaton-Weyl multiplet background
where, thanks to \eqref{DdilatonWeyl}
\bea
D &=&
\frac{15}{4 \s}{\cD}^a{\cD}_a \s
+\frac{3}{4} \cR
+\frac{5}{4 \s}   T^-_{abc} H^+{}^{abc}
+{\rm fermions}
\label{D-sigma}
\eea
and the action \eqref{linear-action} becomes
\bea
&&S_{\rm EH}
=
\int \rd^6 x \,e \, \Big\{
-\hf L\cR
+ \frac{1}{4L}( \partial^m L_{ij}) \partial_m L^{ij}
+\frac{1}{L} \cV^m{}_{i}{}^kL_{jk}( \partial_m L^{ij})
+\frac{1}{L}\cV_m{}^{(i}{}_kL^{j)k}\cV^m{}_{il}L^l{}_{j}
\non\\
&&~~~
- \frac{L}{2\s}{\cD}^a {\cD}_a \s
- \frac{L}{6\s}T^-_{abc} H^+{}^{abc}
-\frac{1}{8 L} {\rm E}^a{\rm E}_a
- \frac{1}{2L} {\rm E}^a \cV_a{}^{ij} L_{ij}
- \frac{1}{8 L^3} \tilde{b}^{mn} L_{i}{}^{j} (\partial_m L^{k i}) \partial_n L_{jk}
\Big\}
\non\\
&&~~~
+ {\rm fermions}
~.
\label{linear-dil-background}
\eea

Another option to solve the inconsistent dynamics of \eqref{linear-action} in a standard-Weyl multiplet
background would be to add to \eqref{linear-action} other invariants 
such that the resulting dynamical system
possesses equations of motions for $D$ that are self consistent.
We will show an example of this option later in section \ref{new-R2}
when we will introduce a new curvature squared invariant in a standard-Weyl multiplet background
that will contain a $D^2$ term in its action.

\subsection{Off-shell Einstein-Hilbert Poincar\'e supergravity}

The six dimensional $\cN = (1,0)$ supersymmetric extension of the Einstein-Hilbert term and consequently
the off-shell Poincar\'e supergravity action
can be given by coupling a linear multiplet compensator to a dilaton-Weyl multiplet, eq.\,\eqref{linear-dil-background},
followed by the gauge fixing of the redundant superconformal symmetries.
The gauge fixing conditions we impose in superspace are\bsubeq\label{GaugeFixing-0}
\bea
&B_M=0~,~~~
\F=1
~,
\label{GaugeFixing-0-conf}\\
&L_{ij} = \d_{ij} L
~.
\label{GaugeFixing-0-su2}
\eea
\esubeq
At the component level these imply
\bsubeq\label{GaugeFixing}
\bea
&b_m = 0
~,~~~
\s = 1
~,~~~
\psi^i = 0
~ ,
\label{GaugeFixing-conf}
\\
&L_{ij} = \d_{ij} L \ .
\label{GaugeFixing-su2}
\eea
\esubeq
The gauge conditions \eqref{GaugeFixing-0-conf} and \eqref{GaugeFixing-conf} fix dilatations,
conformal boosts and  $S$-supersymmetry transformations
while \eqref{GaugeFixing-0-su2} and \eqref{GaugeFixing-su2}
breaks the ${\rm SU}(2)_R$ down to a residual ${\rm U}(1)_R$ gauge symmetry.
After fixing the gauge, the remaining physical fields are
\bea
\{ e_m{}^a \,,  \psi_{m}{}^i\,, b_{mn}\,, V_{m} \,, V_m^{\prime}{}^{ ij}\,, L \,, \vf^i \,, b_{mnrs} \} \,,
\label{FieldContent}
\eea
which form the minimal off-shell Poincar\'e supergravity multiplet.
Note that we have decomposed the gauge field $\cV_m{}^{ij}$ of the ${\rm SU}(2)_R$
symmetry into its trace and traceless parts as
\bea
\cV_m{}^{ij} = V_m^{\prime}{}^{ij} + \frac1{2} \d^{ij} V_m \,, \qquad V_m^{\prime}{}^{ ij} \d_{ij} = 0 \,.
\eea
To preserve the gauge fixing conditions (\ref{GaugeFixing}) under the residual 
Poincar\'e supergravity transformations 
 the following decomposition rules for the dependent compensating
gauge parameters have to be used \cite{Bergshoeff:2012ax}
\bsubeq
\bea
\eta^i&=& -\frac{\ri}{96} \g^{abc}  \xi^i H_{abc}
\,,\\
\l_{m}
&=&
 - \frac12\xi^i {\phi}_m{}_{i}
- \frac{\ri}{30} \xi^i \g_{m} \chi_i
+ \frac12\eta^i  \psi_{m}{}_i
\,,
\\
\l^{\prime ij}
&=&
\frac{1}{L}
S^{\prime}{}_{k}{}^{(i} \d^{j)k}
\,, \qquad
S^{\prime}{}^{ij} = -\xi^{(i} \vf^{j)} + \frac12 \d^{ij} \d_{kl} \x^k \vf^l \,.
\eea
\esubeq
Note that, besides the $Q$-supersymmetry transformations parametrized by local $\x^\a_i|$ parameter,
the trace $\l$ of the SU(2)$_R$ parameter,
$\l^{ij}:=\L^{ij}|=\l^{\prime ij}+\hf\d^{ij}\l$,
generates a residual arbitrary ${\rm U}(1)_R$ gauge transformation.
The other residual gauge transformations are arbitrary
covariant general coordinate  and Lorentz transformations parametrized by $\x^a|$ and $\l^{ab}:=\L^{ab}|$,
respectively.
As a consequence, an off-shell $\cN = (1,0)$ Poincar\'e supergravity background
 with the field content given in (\ref{FieldContent}) has,  up to cubic fermions,
  the following supersymmetry transformation rules \cite{Bergshoeff:2012ax}
\bsubeq\label{susy2-full}
\bea
\delta e_{m}{}^a
&=&-\ri \x_i\g^a\psi_{m}{}^i
\ ,\\
\delta \psi_{m}{}_i&=&
2\Big(\pa_{m} +\frac14\o_{m}{}^{ab}\tilde{\g}_{ab}\Big)\x_i
+2\cV_m{}_i{}^j\x_j
-\frac18 H_{mbc}\tilde{\g}^{bc}\x_i
\ ,
\\
\delta b_{mn}
&=&
2\ri\x_i \g_{[m}\psi_{n]}{}^i
\ ,
\\
\delta \varphi^i
&=&
\ri\d^{ij}\g^m\x_j\pa_m L
-\frac\ri2\g^a \x^i{\rm E}_a
-2\ri V^\prime_m{}^{(i}{}_k \d^{j)k}\g^m \x_j L
+ \frac{\ri}{6}L\d^{ij}\g^{abc}\x_j H_{abc}
\ ,
\\
\delta L &=& \x^i\varphi^j\delta_{ij}
\ ,
\non\\
\d b_{mnpq}
&=&
-\eps_{mnpqef}\, \x_j\g^{ef}\varphi^j
+ 8 \ri \psi_{[m}^i\g_{npq]}\x^j\d_{ij} L
\ ,
\label{b4-1}
\\
\delta {V}_{m}^\prime{}^{ij}
&=&
-\ri\x^{(i}\g^{n }\widehat{R}_{mn}{}^{j)}(Q)
+\frac{\ri}{2}\x^{k}\g^{n}\widehat{R}_{mn}{}^\ell (Q)\delta_{k\ell }\delta^{ij}
+\frac{\ri}{6}\x^{(i}\g^{abc}\psi_{m}^{j)}H^{abc}
\non\\
&&
-\frac{\ri}{12}\x^{k}\g^{abc}\psi_{m}^\ell \delta_{k\ell }\delta^{ij}H_{abc}
\ ,
\\
\delta {V}_{m}
&=&
-\ri\x^{i}\g^{n}\widehat{R}_{mn}{}^{j}(Q)\delta_{ij}
+\frac{\ri}{6}\x^{i}\g^{abc}\psi_{m}^{j}\delta_{ij}H_{abc}
\ ,
\label{susy2}
\eea
\esubeq
where
\bsubeq
\bea
\widehat{R}_{mn}{}^i(Q)
&=&
2D_{[m}\psi _{n]}{}^i
+2V^\prime_{[m}{}^{ij}\psi _{n]}{}_{j}
+\frac{1}{4}\g^{ab}\psi_{[m}H_{n]ab}
\ ,
\label{RQ}
\\
D_{m}\psi _{n}^i&=&
\Big(\pa_m
+\frac{1}{4}\o_m{}^{cd}\tilde{\g}_{cd}\Big)
\psi_n{}^i
+\hf V_m\d^{ij}\psi_n{}_j
~.
\eea
\esubeq

By imposing the gauge fixing \eqref{GaugeFixing} in the action \eqref{linear-dil-background}
it follows that the off-shell two-derivative Poincar\'e supergravity Lagrangian takes the form
\cite{Bergshoeff:2012ax}
\bea
e^{-1} \cL_{\rm EH}
&=&
-\hf L \cR
+\hf L^{-1} \partial_{m} L \partial^m L
- \frac{1}{24} L H^{abc} H_{abc}
+L V_m^{\prime}{}^{ ij} V^{\prime}{}^{m}{}_{ij}
- \frac{1}{8L} {\rm E}^a {\rm E}_a
- \hf {\rm E}^m V_m
\non\\
&&
+{\rm fermions}
\ .
\label{Poincare}
\eea
The off-shell Poincar\'e supergravity action is invariant under the transformation rules (\ref{susy2-full}).

\subsection{On-shell Poincar\'e supergravity}

When considering the Einstein-Hilbert supergravity (\ref{Poincare}), the equations of motion for the
$V_m^{\prime}{}^{ij}, V_m$ and $b_{mnpq}$ fields simply imply \cite{Bergshoeff:2012ax}
\bea
V_m = V_m^{\prime}{}^{ ij} =b_{mnpq} = 0\,,
\label{EOM2Derivative}
\eea
leaving the following on-shell Einstein-Hilbert supergravity Lagrangian
\bea
e^{-1} \cL &=& -\hf e^{-2 v} \Big( \cR - 4 \partial^m v\,\partial_m v + \frac1{12} H_{mnp} H^{mnp} \Big)
+{\rm fermions}
\,
\label{Poincare-on-shell}
\eea
where we have set $L = e^{-2v}$ with $v$ being the dilaton field.
The massless supermultiplet of the two-derivative theory is therefore given by the following component fields
\bea
\{e_m{}^a \,,\psi_{m}{}_i \,, b_{mn} \,,v \,, \vf^i \}
\,,
\eea
whose supersymmetry transformation rules, that follow from \eqref{susy2-full},
up to cubic fermion terms are
\bsubeq\label{susy3}
\bea
\delta e_{m}{}^a
&=&\ri\x^i\gamma^a\psi_{m}{}_i
\ ,
\\
\delta \psi_{m}{}^i
&=&
2\Big(\partial_{m} \x^i
+\frac14\omega_{m}{}^{ ab}\tilde{\g}_{ab}\Big) \x^i
-\frac14 H_{mab}\tilde{\g}^{ab}\x^i
\ ,
\\
\delta b_{mn}
&=&
2\ri\x_i\g_{[m}\psi_{n]}{}^i
\ ,
\\
\delta \varphi^i &=&
-2\ri e^{-2v}\delta^{ij} \g^m\x_j  \pa_m v
+ \frac{\ri}{6}e^{-2v}\d^{ij} \g^{abc}\x_j  H_{abc}
 \ ,
\\
\delta v
&=&
-\hf e^{2v} \x^i\varphi^j\delta_{ij}
 \ .
\eea
\esubeq
Note that, since the fields $V_m^{\prime}{}^{ij}, V_m$ and $b_{mnpq}$ have been integrated out, the previous 
supersymmetry transformations close only on the mass-shell.


\section{Curvature squared  invariants in a dilaton-Weyl background}
\label{main-curvature-squared}

Here we turn to constructing curvature squared supergravity invariants.
All the locally superconformal invariants described in this section
are based on the $B_2 \wedge H_4$ action principle;
hence they will all implicitly be in a dilaton-Weyl multiplet  background.
Within our framework, we will describe 
the off-shell locally $\cN\!=\!(1,0)$ supersymmetric extensions of all the
possible purely gravitational curvature squared terms that can be described as linear combinations of 
Riemann squared, 
Ricci tensor squared, and scalar curvature squared terms.


\subsection{The Riemann curvature squared invariant}

A supersymmetric extension of the Riemann curvature squared term
was constructed in \cite{BSS1,BSS2,Nishino:1986da,BR, Bergshoeff:2012ax}. It has been coupled to
the gauged chiral supergravity in six dimensions
 extending the Salam-Sezgin model with curvature squared corrections. The exact spectrum of the Riemann squared
extended Salam-Sezgin model around the half-BPS Minkowski$_4\times {\rm S}^2$ background was analyzed 
for the first time in \cite{Pang:2012xs}. 
The construction of the  Riemann curvature squared invariant of 
\cite{BSS1,BSS2,Nishino:1986da,BR, Bergshoeff:2012ax}
 was based on the action for a Yang-Mills multiplet coupled to conformal supergravity \cite{BSVanP}
and  the feature that in the gauge \eqref{GaugeFixing-conf}
the dilaton-Weyl multiplet can be mapped to
a Yang-Mills vector multiplet taking values in the 6D Lorentz algebra \cite{BSS1}.
This observation is sometimes referred to as the ``Yang-Mills trick''.
The superspace analogue of the ``Yang-Mills trick'' was elaborated in Sec.~6 of \cite{BNT-M17}.
Let us here further elaborate on this analysis.

In  a dilaton-Weyl multiplet background, with tensor superfield $\Phi\ne0$,
following \cite{BKNT16} one can introduce the spinor covariant derivative,
\bea
\mathscr D_\a^i &=&
\F^{- \frac{1}{4}} \Big( \nabla_\a^i
+ (\nabla_\b^i \ln \F) M_\a{}^\b
- 2 (\nabla_\a^j \ln \F) J_j{}^i
- \hf (\nabla_\a^i \ln \F) \mathbb D \Big)
\ .
\label{Dtensor}
\eea
This is chosen so that given a primary tensor superfield $U$ of dimension $\Delta$,
the superfield $\mathscr D_\a^iU$
is also a primary superfield of dimension $\D$. Moreover, $\F$ is annihilated
by $\mathscr D_\a^i$, $\mathscr D_\a^i \F = 0$.
When acting on a primary superfield, the algebra of
covariant derivatives becomes\footnote{The
resulting superspace geometry 
for the covariant derivatives ${\mathscr D}_A=({\mathscr D}_\a^{i},{\mathscr D}_a)$ is equivalent to the SU(2)
superspace formulation of conformal supergravity of \cite{LT-M12} with the torsion superfield 
component $ C_a{}^{ij}$ set to zero, $C_{a}{}^{ij}=0$. This is also equivalent to the
superspace geometry of \cite{BR}.
Using in \eqref{Dtensor} instead
a generic real unconstrained dimension-2 primary superfield $X$ instead of $\Phi$, as described in
 \cite{BKNT16}, one would obtain the general SU(2) superspace geometry of \cite{LT-M12}
 with $C_a{}^{ij}\ne 0$.}
\bea
\{ \mathscr D_\a^i , \mathscr D_\b^j \}
&=&
- 2  \ri \eps^{ij} {\mathscr D}_{\a\b}
- 4 \ri\eps^{ij} {\mathscr W}^{abc} (\g_a)_{\a\b} M_{bc}
- 4 \ri\eps^{ij} {\mathscr N}^{abc} (\g_a)_{\a\b} M_{bc}
- 16 \ri {\mathscr N}_{\a\b} J^{ij}
\ ,~~~~~~
\eea
where
\bea
\mathscr D_{\a\b} &=&
- \frac{\ri}{4} \{ \mathscr D_\a^k , \mathscr D_{\b k} \}
- 2 \mathscr N^{bcd} (\g_b)_{\a\b} M_{cd}
- 2 \mathscr W^{bcd} (\g_b)_{\a\b} M_{cd}
\eea
and the primary dimensionless  superfields $\mathscr N_{\a\b}$ and $\mathscr W^{\a\b}$
are defined as
\bea
\mathscr N_{\a\b} :=
\frac{1}{3!}(\g^{abc})_{\a\b}\mathscr N_{abc}
= - \frac{\ri}{16} \F^{\frac{3}{2}} \nabla_{(\a}^{k} \nabla_{\b) k} \F^{-2}
~,~~~~~~
\mathscr W^{\a\b} := \F^{-\hf} W^{\a\b}
\ .
\eea

By using the previous composite superfields,
one can construct a primary dimension-3/2
superfield valued in the Lorentz algebra
$\bm \L^{\a i}:=\bm \L^{\a i}{}_\b{}^\g M_\g{}^\b$ \cite{BNT-M17}
\be \label{YMtrick-0}
\bm \L^{\a i}
=\Phi^{3/4} \Big( {\mathscr D}_\b^i \mathscr W^{\a\g}
- \frac{2}{3} \eps^{\a\g\d\r} \mathscr D_\d^i \mathscr N_{\r \b}
- \frac{1}{3} \d^\a_\b {\mathscr D}_\d^i \mathscr W^{\g\d} \Big) M_\g{}^\b
\ .
\ee
This satisfies constraints that are formally the same  as those of a non-Abelian vector multiplet:
\bea
 \bm \L^{\a i}{}_{\b}{}^\b=0~,~~~
{\mathscr D}_\a^{(i} {\bm \L}^{\b j)}{}_\g{}^\d - \frac{1}{4} \d_\a^\b {\mathscr D}_\r^{(i} {\bm \L}^{\r j)}{}_\g{}^\d = 0
~,~~~
{\mathscr D}_{\a i} \bm \L^{\a i}{}_{\b}{}^\g = 0 \ .
\eea
The primary $\bm \L^{\a i}{}_\b{}^\g$ can be equivalently rewritten as
\bea
{\bm \L}^{\a i}{}_{\b}{}^\g
&=&
4\ri X_\b^i{}^{\a\g}
- \frac{4\ri}{3} \d^\a_\b X^{\g i}
+ \frac{\ri}{3}\d_\b^\g  X^{\a i}
-\F^{-1}\Big(W^{\a\g}\psi_\b^i
-\frac{1}{3} \d^\a_\b W^{\g\d}\psi_\d^i
+ \frac{1}{3}\d_\b^\g W^{\a\d}\psi_\d^i\Big)
\non\\
&&
+\eps^{\a\g\d\r}\F^{-2}\Big{[}\frac{1}{2} ({\de}_{\d(\r}\F)\psi_{\b)}^{i}
- \frac{1}{3}\F{\de}_{\d(\r} \psi_{\b)}^{i}
-\frac{1}{8} H_{\r\b}\psi_{\d}^{i}
- \frac{\ri}{4}\F^{-1}\psi_\d^i\psi_{(\r}^{k}\psi_{\b) k}
\Big{]}
~,~~~~~~~~~
\label{Rieman2composite}
\eea
where we have used the tensor multiplet relations
\bea
\psi_\a^i
=
\de_\a^i\F
~,~~~~~~
H_{\a\b}
:=
-\ri \nabla_{(\a}^k \psi_{\b) k}
~,~~~~~~
\nabla_\a^{i} \nabla_\b^{j} \Phi
=
-\frac{\ri}{2}\ve^{ij}H_{\a\b}
-\ri\ve^{ij}{\de}_{\a\b}\Phi
~,
\eea
and it should be kept in mind that in a dilaton-Weyl background $W^{\a\b} =-\frac{1}{4}\F^{-1}H^{\a\b}$.
A remarkable property of the composite superfield \eqref{Rieman2composite}
 is that, in the gauge where
$B_M=0$ and $\F=1$, equivalent to $b_m = 0$, $\s = 1$ and $\psi^i = 0$, 
the following relations for the descendants of $\bm \L^{\a i}{}_\b{}^\g$ hold:
\bsubeq\label{Riemann-multiplet}
\bea
{\bm\cF}_{ab}{}^{cd}&:=&
-\frac{\ri}{16}(\g_{ab})_\b{}^\a(\g^{cd})_\d{}^\g\de_{\a}^{ k} {\bm \L}_k^{\b}{}_\g{}^\d|
=
\cR_{ab}{}^{cd}({\o}_+)
+ {\rm fermions }
~,
\\
{\bm X}^{ij}{}^{ab}
&:=&
\frac{\ri}{8}(\g^{ab})_\g{}^\b\de_\a^{(i}{\bm \L}^{\a j)}{}_\b{}^\g|
=-2\cR^{ab}{}^{ij}
+{\rm fermions}
~.
\eea
\esubeq
Here  $\cR_{ab}{}^{cd}({\o}_+)$ is the torsionful Riemann tensor which is defined
through \eqref{Romega}
in terms of the shifted Lorentz connection
\bea
\o_\pm{}_m{}^{cd}
:=
{\o}_m{}^{cd}
\pm\frac{1}{2}e_m{}^aH_a{}^{cd}
~,
\eea
such that
\bea
\cR_{ab}{}^{cd}({\o}_\pm)
=
\cR_{ab}{}^{cd}({\o})
\pm
{\cD}_{[a}H_{b]}{}^{cd}
+\frac{1}{2}H_{e[a}{}^{[c}H_{b]}{}^{d]e}
~.
\label{TorsionfulRiemann}
\eea

Considering the analogy of $\bm \L^{\a i}{}_\b{}^\g$ with a Yang-Mills multiplet and
the known construction of the Riemann squared action from a vector multiplet one,
it is natural to argue that
a gauge 3-form multiplet is described by
the following composite superfield
\bea
B^{\a\b ij} &=&
-16\ri \Tr \big{[} {\bm \L}^{\a (i} {\bm \L}^{\b j)} \big{]}
=-16\ri {\bm \L}^{\a (i}{}_\g{}^\d {\bm \L}^{\b j)}{}_\d{}^\g
=8\ri {\bm \L}^{\a (i}{}_{cd} {\bm \L}^{\b j)}{}^{cd}
\ .
\label{BRieman2}
\eea
Thanks to the fact that
$\de_\g^k\Tr \big{[} {\bm \L}^{\a (i} {\bm \L}^{\b j)} \big{]}
=\F^{\frac{1}{4}}\mathscr D_\g^k\Tr \big{[} {\bm \L}^{\a (i} {\bm \L}^{\b j)} \big{]}$,
one can prove that the composite  \eqref{BRieman2}
satisfies the constraints \eqref{4FormConst}.
Then, by plugging \eqref{BRieman2} in the $B_2\wedge H_4$ action principle,
or equivalently by plugging \eqref{Riemann-multiplet} in \eqref{SYMaction},
after imposing the gauge  condition \eqref{GaugeFixing-conf}, one obtains the action
\bea
S_{{\rm{Riem}}^2}
=
\int \rd^6 x \,e &
\Big\{&
\cR^{ab}{}^{cd}({\o}_+)\,\cR_{ab}{}_{cd}({\o}_+)
-\frac{1}{4}\eps^{abcdef}b_{ab}\,\cR_{cd}{}^{gh}({\o}_+)\,\cR_{ef}{}_{gh}({\o}_+)
\non\\
&&
-4\cR^{ab}{}^{ij}\cR_{ab}{}_{ij}
\Big\}
+{\rm fermions}
~.
\label{Riemann2}
\eea
Up to a change of notation, the previous result coincides with the known Riemann squared invariant
\cite{BSS1,BSS2,Nishino:1986da,BR, Bergshoeff:2012ax}.


\subsection{The scalar curvature squared invariant}
\label{scalar-squared}

In section \ref{EH} we have shown how to construct the locally supersymmetric extension of the Einstein-Hilbert
term by using the composite Abelian vector multiplet based on the superfield
$\L^{\a i}$ defined by eq.\,\eqref{composite-vector-linear}.
It is clear that the composite
\bea
B^{\a\b ij}:=-4\ri \L^{(\a i}\L^{\b) j}
~,
\label{Bscalar2}
\eea
satisfies the constraints \eqref{4FormConst}.
Moreover, it is straightforward  to check that by plugging
 \eqref{Bscalar2} in the $B_2\wedge H_4$ action principle, in the gauge \eqref{GaugeFixing-0},
 one can find the supersymmetric extension of
a scalar curvature squared that was first constructed in \cite{OzkanThesis}.
Let us here review the salient results for this invariant in components.

 In the gauge (\ref{GaugeFixing}),
by using  the components of the composite vector multiplet \eqref{composite-vector-linear-2},
it is straightforward to obtain the following useful expressions
\bsubeq
\bea
f_{mn} &=& \partial_{[m} \Big( V_{n]} + \frac1{2} L^{-1} {\rm E}_{n]}\Big)
+{\rm fermions}
~,
\\
X_{ij} X^{ij}
&=&
\hf\Upsilon^2
-\hf |\Xi|^2
+{\rm fermions}
~,
\eea
where
\bea
\Upsilon &=&
-\hf\cR
- \frac1{24} H_{mnp} H^{mnp}
-L^{-1} \pa^m\pa_m  L
+\hf L^{-2}( \partial^m L) \partial_m L
\non\\
&&
- 2 Z^m \bar{Z}_m
+ \frac{1}{8} L^{-2} {\rm E}^m {\rm E}_m
 ~,
\\
\Xi &=&
-2 L^{-1} \cD^m (L Z_m)
+ \ri L^{-1} {\rm E}^m Z_m
~,
\eea
with the complex vector field $Z_m$  defined as
\bea
Z_m = V_m^{\prime}{}^{12}  + {\ri} V_m^{\prime}{}^{11} \,.
\eea
\esubeq
Using the action \eqref{SYMaction},
we obtain (up to a numerical factor)
the following Lagrangian for the Ricci scalar curvature squared
invariant \cite{OzkanThesis}
\bea
e^{-1} \cL_{R^2} &=&\frac{1}{16}
 \Big{[} \cR
+ \frac1{12} H_{mnr} H^{mnr}
+2 L^{-1}\pa^m\pa_m L
- L^{-2} (\partial_m L) \partial^m L
\non\\
&&~~~~~
+ 4 Z^m \bar{Z}_m
- \frac{1}{4} L^{-2} {\rm E}^m {\rm E}_m   \Big{]}^2
 \non\\
&&
- \frac{1}{4} \Big{[}2 L^{-1} \cD^m (L Z_m)
- \ri L^{-1} {\rm E}^m Z_m \Big{]}
\Big{[}2 L^{-1} \cD^n (L \bar Z_n)
+ \ri L^{-1} {\rm E}^n \bar{Z}_n \Big{]} \non\\
&&
+\frac{1}{8} \ve_{mnpqrs} b^{mn} \partial^{p} \Big( V^q + \frac1{2} L^{-1} {\rm E}^q \Big)
\partial^{r} \Big( V^s + \frac1{2} L^{-1} {\rm E}^s \Big)
 \non\\
&&
- \frac{1}{2} \partial_{[m} \Big( V_{n]} + \frac1{2} L^{-1} {\rm E}_{n]} \Big)
\partial^{m} \Big( V^{n} + \frac1{2} L^{-1} {\rm E}^{n} \Big)
+{\rm fermions}
\ .
\label{scalar-squared-Lagrangian}
\eea
The resulting Ricci scalar curvature squared 
action is invariant under the off-shell supersymmetry transformation rules (\ref{susy2-full}).


\subsection{The new curvature squared invariant}
\label{new-invariant-subsection}

Recently, a particular composite gauge 3-form multiplet defined solely using the the standard-Weyl multiplet
has been constructed in \cite{BKNT16}. This was used to describe one of the
$\cN = (1,0)$ conformal supergravity actions \cite{BKNT16, BNT-M17}.
Such a composite gauge 3-form multiplet is defined in terms of the following primary
dimension 3 superfield \cite{BKNT16}
\bea
B^{\a\b\, i j} =
- 4 W^{\gamma [\alpha} Y_{\gamma}{}^{\beta] i j}
- 32 \ri\, X_{\gamma}{}^{\alpha \delta (i} X_{\delta}{}^{\beta \gamma j)}
+ 10 \ri\, X^{\alpha (i} X^{\beta j)}
~,
\label{C2gauge3form}
\eea
satisfying the constraints \eqref{4FormConst}.
Its lowest component is
\begin{align}\label{eq:BforCBoxC}
B_a{}^{ij} &= T^-_{a b c}\, R(J)^{b c\, i j}
	+ \ri\, R(Q)_{b c}{}^i \gamma_a R(Q)^{b c}{}^j
	+ \frac{2\ri}{45} \chi^i \gamma_a \chi^j~.
\end{align}
The structure of its descendants, including the
$\L_{\a a}{}^i$ and $C_{ab}$ components appearing in the $B_2\wedge H_4$ action principle \eqref{BH-action},
were worked out in \cite{BNT-M17} and are collected for convenience of the reader
in appendix \ref{descendants-C2gauge3form}.
As we first described in \cite{NOPT-M17},
by plugging these results into \eqref{BH-action}
and performing some integrations by parts,
one obtains the new curvature squared invariant
\bea
S_{{\rm new}} &=&
\frac{1}{32} \int \rd^6 x \,e \,
\Big\{
\s C_{ab}{}^{cd}  C_{cd}{}^{ab}
- 3\s  \cR_{ab}{}^{ij} \cR^{ab}{}_{ij}
+ \frac{4}{15}\s  D^2
+ 4\s   (\cD_c T^{- abc}) \cD^d T^-_{abd}
\non\\
&&
 - 8\s  T^{- dab}\Big(
 \cD_d \cD^c T^-_{ab c}
- \frac12 \cR_d{}^c T^-_{abc}
\Big)
+ 4\s  T^{- abc} T^-_{ab}{}^d T^{- ef}{}_c T^-_{efd}
-2 H_{abc} C^{ab}{}_{de} T^{- cde}
\non\\
&&
- \frac{8}{45} H_{abc} T^{- abc} D
+ 4  H_{abc} T^-_d{}^{ab} \cD_e T^{- cde}
- \frac{4}{3} H_{abc} T^{- dea} T^{- bcf} T^-_{def}
\non\\
&&
-\frac{1}{4} \ve^{abcdef} b_{ab} \Big(
C_{cd}{}^{gh} C_{efgh}
- \cR_{cd}{}^{ij}  \cR_{ef}{}_{ij} \Big)
\Big\}
+{\rm fermions}
~,
\label{newInvariant}
\eea
where the reader should keep in mind that, in a dilaton-Weyl background,
$T_{abc}^- = \frac12 H_{abc}^-$
and $D$ satisfies \eqref{D-sigma}.
Note that in \eqref{newInvariant} we focused only on the bosonic part of the invariant.
In principle,
by plugging into  \eqref{BH-action}
the expressions for
$B_a{}^{ij}$, eq.\,\eqref{C2gauge3form},
together with its descendants
$\L_{\a a}{}^i$ and $C_{ab}$
given in eq.\,\eqref{descendants-C2gauge3form-equations},
the reader can obtain the full result including all the fermionic terms.

In the gauge \eqref{GaugeFixing-conf} the new invariant
contains the following linear combination of Riemann squared, Ricci squared and scalar curvature squared terms
\bea
e^{-1}\cL_{{\rm new}}&=&
\frac{1}{32} \Big(
\cR_{abcd} \cR^{abcd}
-\cR_{ab}\cR^{ab}
+ \frac{1}{4} \cR^2
\Big)
+\cdots
~.
\eea

This completes the construction of the three independent 
supersymmetric curvature squared invariants. 
Let us now turn to describing some physical applications.


\section{Einstein-Gauss-Bonnet supergravity}
\label{Einstein-Gauss-Bonnet}

In  section \ref{EH} we reviewed the off-shell supersymmetrization of the Einstein-Hilbert action
while in the previous section we have described three linearly independent curvature squared invariants
of the six dimensional $\cN = (1,0)$ supergravity.
Due to the off-shell nature of these  models,
we can combine them to form a general curvature squared extended Poincar\'e supergravity Lagrangian
\bea
\cL &=&  \a\cL_{\rm Riem^2} +\b \cL_{R^2}+ \g \cL_{\rm new}   
\,,
\eea
without having to modify the supersymmetry transformations.
As described for the first time in \cite{NOPT-M17},
the  off-shell
Gauss-Bonnet combination corresponds to $\a=-3$, $\b = 0$ and $\g=128$. For this particular
choice of parameters, in the gauge \eqref{GaugeFixing-conf},
if one uses the following identities
\bsubeq
\bea
0 &=& H_{abc} H^{ade} \cD^b H^c{}_{de}
\,,\\
0 &=&
\ve_{abcdef}\big(
 \cR_{gh}{}^{ef} H^{gha} H^{bcd}
+ \cR_g{}^f H^{gab} H^{cde} \big)
\,,\\
0 &=&  \ve^{abcdef}\big(
 H_{abc} H_{deg} \cD^h H^g{}_{fh}
- 2 H_{abc} H_{dgh} \cD_e H^{gh}{}_f \big)
\,,
\eea
\esubeq
and introduces the notation,
\bsubeq
\bea
H^2_{ab}:=H_{a}{}^{cd}H_{bcd}
~,~~~~~~
H^2:=H_{abc}H^{abc}
~,~~~~~~
H^4:=H_{abe}H_{cd}{}^{e}H^{acf}H^{bd}{}_{f}
~,
\eea
\esubeq
 up to fermionic terms the Lagrangian for the supersymmetric Gauss-Bonnet invariant \eqref{newInvariant}
takes the following very compact form
\bea
e^{-1} \cL_{\rm GB} &=&
\cR_{ab}{}_{cd} \cR^{ab}{}^{cd}
- 4 \cR_{ab} \cR^{ab}
+ \cR^2
\non\\
&&
- \frac12 \cR_{ab}{}_{cd} H^{abe} H^{cd}{}_{e}
+ \cR^{ab} H_{ab}^2
- \frac16 \cR H^2
 + \frac{5}{24} H^4
 + \frac{1}{144} (H^2)^2
 - \frac18 (H_{ab}^2)^2
 \non\\
&&
+ \ve^{abcdef} b_{ab} \cR_{cd}{}^{ij} \cR_{ef ij}
 - \frac14 \ve^{abcdef} b_{ab} \cR_{cd}{}^{gh} (\o_-) \cR_{ef}{}_{gh}(\o_-)
 \non\\
 &&
 +{\rm fermions}
 \ .
\label{GBSUGRA}
\eea
This can equivalently be written
\bea
e^{-1} \cL_{\rm GB} &=&
6\cR_{[ab}{}^{ab}(\o_-) \cR_{cd]}{}^{cd}(\o_-)
-4\cR_{ab}{}_{cd}(\o_-) H^{abe} H^{cd}{}_{e}
+4 \cR^{ab}(\o_-) H_{ab}^2
- \frac23 \cR(\o_-) H^2
\non\\
&&
 - \frac{2}{3} H^4
 + \frac{1}{9} (H^2)^2
+ \ve^{abcdef} b_{ab} \Big{[}
\cR_{cd}{}^{ij} \cR_{ef ij}
 - \frac14 \cR_{cd}{}^{gh} (\o_-) \cR_{ef}{}_{gh}(\o_-)
 \Big{]}
  \non\\
 &&
 +{\rm fermions}
 \ .
\label{GBSUGRA-2}
\eea
In this form it becomes evident that the $b_2$-field dependence of the
supersymmetric Gauss-Bonnet invariant cannot be captured solely in terms of a
Riemann tensor with a torsionful connection.
This explains the unsuccessful
attempts to construct the Gauss-Bonnet invariant in previous works
where this was assumed as a working condition \cite{BSS1,BSS2}.

By using the action \eqref{GBSUGRA},
we can now easily construct the 6D $\cN=(1,0)$
off-shell Einstein-Gauss-Bonnet (EGB) supergravity.
In the gauge \eqref{GaugeFixing},
this is given by the following combination of the supersymmetric completion of the Einstein-Hilbert term
(\ref{Poincare}) and the Gauss-Bonnet term \eqref{GBSUGRA}
\bea
2\k^2\cL_{\rm EGB} &=& \cL_{\rm EH} + \frac{1}{16} \a^\prime \cL_{\rm GB} \,.
\label{EGBSugra}
\eea
We have introduced the Newton constant $\k^2$.
The Einstein-Gauss-Bonnet supergravity (\ref{EGBSugra})
is invariant under the off-shell supersymmetry transformation rules (\ref{susy2-full}).

\subsection{On-shell Einstein-Gauss-Bonnet supergravity}

In  section \ref{EH}
we have shown
how the on-shell Einstein-Hilbert supergravity (\ref{Poincare-on-shell})
arises from (\ref{Poincare}) by using
the equations of motion $V_m^{\prime}{}^{ij}= V_m=b_{mnpq}=0$.
A remarkable feature of the supersymmetric Einstein-Gauss-Bonnet model is that the massless supermultiplet of
the two-derivative theory does not acquire mass due to the inclusion of higher-derivative terms.
Furthermore,
the kinetic term $ \cR_{ab}{}^{ij} \cR^{ab}{}_{ij}$ for the ${\rm SU(2)}_R$ gauge field $\cV_m{}^{ij}$
 that exists in both the  Riemann-squared and the new invariants,
 cancels out in the Gauss-Bonnet combination.
 Therefore, there are no new dynamical degrees of freedom induced by the higher-derivative terms and
 the solution
(\ref{EOM2Derivative}) remains consistent in the Einstein-Gauss-Bonnet supergravity,
leading to the on-shell theory associated to the following Lagrangian
\bea
2\k^2e^{-1} \cL &=&
-e^{-2 v} \Big(\cR - 4 \partial^m v\, \partial_m v + \frac1{12} H_{mnp} H^{mnp} \Big)
\non\\
&&
 + \frac{1}{16} \a^\prime \Big{[}\,
6 \cR_{[mn}{}^{mn} \cR_{rs]}{}^{rs}
- \frac12 \cR_{mnrs} H^{mnt} H^{rs}{}_{t}
 + \cR^{mn} H_{mn}^2
  - \frac16 \cR H^2
+ \frac{5}{24} H^4
 \non\\
&&
\qquad \quad\,
+ \frac{1}{144} (H^2)^2
- \frac18 (H_{mn}^2)^2
- \frac14 \ve^{mnpqrs} b_{mn} \cR_{pq}{}^{ab} (\o_-) \cR_{rs}{}_{ab}(\o_-)
\Big{]}
 \non\\
 &&
 +{\rm fermions}
~.
\label{EGBSugra-onshell}
\eea
As a consequance, up to cubic fermions,
the on-shell $\cN = (1,0)$ Einstein-Gauss-Bonnet supergravity
has the same supersymmetry transformations as the on-shell Einstein-Hilbert supergravity given by
\eqref{susy3}.

Note that our on-shell Einstein-Gauss-Bonnet supergravity action
precisely matches the result derived in \cite{Liu:2013dna}.
In fact, we have actually fixed the proportionality constant of the $\alpha^\prime$ term in \eqref{EGBSugra} 
by comparing the on-shell Lagrangian \eqref{EGBSugra-onshell} with \cite{Liu:2013dna}.

To conclude this section, let us comment further on the relevance of our result in comparison with the analysis of
Liu and Minasian in \cite{Liu:2013dna}.
There, it was conjectured that for the type II-string, the NSNS $b_2$-field dependence in 
the $R^4$ corrections is
nearly completely captured in terms of the torsionful Riemann tensor (\ref{TorsionfulRiemann})
(except for the CP-odd sector).
The claim was further investigated in \cite{Liu:2013dna}
by fixing the one-loop four-derivative corrections by means of a K3 reduction to six dimensions of
type IIA and requiring that the dyonic string remains as a solution
and that the model remain dual to the heterotic
string compactified on $T^4$. In our work, we provided an alternative derivation of the four-derivative corrections by
exact supersymmetrization
of the curvature squared invariants.
The fact that our results for the Einstein-Gauss-Bonnet supergravity match, and extend off-shell,
the results of \cite{Liu:2013dna} thereby provides strong evidence for 
their conjecture.


\section{New curvature squared invariants with a linear multiplet compensator}
\label{new-R2}

In a conformal supergravity background described by the dilaton-Weyl multiplet,
we constructed three independent curvature squared supergravity invariants in 
section \ref{main-curvature-squared}. These were all based on the $B_2 \wedge H_4$ action 
principle of subsection \ref{sectionB2H4action}. One of these invariants was a 
locally superconformal extension of a four-derivative invariant for a linear multiplet. A natural 
question one could ask is whether this is the only four-derivative superconformal invariant 
for a linear multiplet. It turns out that this is not the case as one can simply write down another 
local four-derivative superconformal invariant using a full superspace integral as follows
\be
\int \rd^6x \rd^8\q\,E\,L^{1/2}
~,
~~~~~~
E:={\rm Ber}[E_M{}^A]~,
\label{newintL}
\ee
where $L^{1/2}=\sqrt{\ft{1}{2} L^{ij}L_{ij}}$ and $E$ is the Berezinian (or superdeterminant) 
of the supervielbein.\footnote{The reader can look at \cite{LT-M12,BKNT16} 
for discussions about general properties 
of the full superspace integral
under local superconformal transformations.}
Moreover, \eqref{newintL} does not coincide with the four-derivative linear multiplet invariant of the 
previous sections since it is well-defined in the standard-Weyl multiplet (as it does not involve the 
tensor multiplet).

As of yet, we do not have a manifestly superconformal component (or superform) representation 
of the full superspace integral
based on a primary density formula (or superform).\footnote{See \cite{BKNT16} for a non-primary
superform description of the full superspace integral based on the so-called $B$ action principle.}
For this reason, we will endeavour to make use of the manifestly 
superconformal $A$ action principle for the construction of a new local four-derivative invariant based on the 
linear multiplet.\footnote{This invariant presumably coincides with the one defined by the full superspace 
integral above.} In this section, we will then show that this new invariant leads to curvature squared terms such that:
(i) once added  to the Einstein-Hilbert term in a standard-Weyl multiplet background,
the new invariant leads to a dynamical system with consistent equations of motion for the $D$ field
which dynamically generates a cosmological constant term that leads to (non-supersymmetric) 
${\rm (A)dS}_6$ solutions;
(ii) in a dilaton-Weyl multiplet background the new invariant 
proves to have a nontrivial coupling 
to the dilaton and is distinct from the curvature squared 
invariants constructed in section \ref{main-curvature-squared}.

\subsection{A new locally superconformal higher-derivative action for the linear multiplet}

In \cite{Kuzenko:2017jdy} it was recently proven that, given
a so-called real $\cO^*(4)$ multiplet, which is
a real dimension $-4$ primary superfield $T$ satisfying
\bea
\overline{T}=T
~,~~~~~~
\de_{(\a}^k\de_{\b)k}T=0
~~~\Longrightarrow~~~
\de_{(\a}^i\de_{\b)}^jT=0
~,
\eea
and a real $\cO(4)$ multiplet which is a primary superfield $T^{ijkl}$ of dimension 8 
satisfying
\bea
T^{ijkl}=T^{(ijkl)}
~,~~~~~~
\overline{(T^{ijkl})}=T_{ijkl}
~,~~~~~~
\de_\a^{(p}T^{ijkl)}=0
~,
\eea
it is possible to construct a manifestly locally superconformal action with the A action principle.
In fact, for any pair of $\cO(4)$ and $\cO(4)^*$ multiplets,
the following  superfield
\bea
A_\alpha{}^{ijk} &:=
-\ri T \nabla_{\alpha l} T^{i j k l}
- 5\ri T^{i j k l} \nabla_{\alpha l} T
~,
\eea
proves to be a primary of dimension $9/2$ satisfying the symmetry and reality conditions 
$\overline{A_\a{}^{ijk}} = A_{\a\, ijk}$ and $A_\a{}^{ijk}=A_\a{}^{(ijk)}$
together with  the differential constraint $\nabla_{(\a}^{(i} A_{\b)}{}^{jkl)} = 0$, \eqref{Aconst}.
Then, given any pair of real $\cO(4)$ and $\cO^*(4)$
multiplets one can construct a manifestly locally superconformal
action.

Using a linear (real $\cO(2)$) multiplet as in subsection \ref{linear-multiplet-section}
that is nowhere vanishing, $L^2=\hf L^{ij}L_{ij}\ne 0$,
it is possible to construct a pair of composite real $\cO(4)$ and $\cO^*(4)$ multiplets.
The real $\cO(4)$ multiplet superfield is
\bea
L^{i j k l} &:=& -\frac{1}{96} \veps^{\alpha\beta\gamma\delta}
\nabla_{\alpha}^{(i} \nabla_\beta^j \nabla_\gamma^k \nabla_\delta^{l)} L^{3/2}
~,
\label{compositeO4}
\eea
and the real $\cO^*(4)$ multiplet is $L^{-1}$, which due to \eqref{linear-constr}
obeys $\de_{(\a}^k\de_{\b)k}L^{-1}=0$.
This implies that we can straightforwardly
construct a new higher-derivative action for the linear multiplet 
using the $A$ action principle with
\bea
A_\alpha{}^{ijk} &:=
-\ri L^{-1} \nabla_{\alpha l} L^{i j k l}
- 5\ri L^{i j k l} \nabla_{\alpha l} L^{-1}
~.
\label{AcompositeL}
\eea
To show that, it suffices to focus on the bosonic part of this invariant.
With the assistance of
the computer algebra program \emph{Cadabra} \cite{Cadabra-1,Cadabra-2},
we computed the bosonic part of the action.
The full result of this calculation is presented appendix \ref{full-nasty-new-invariant}.
For the scope of this section, we can simplify the result \eqref{new_Ricci+scalar_invariant-full}
by going to a conformal and ${\rm SU}(2)_R\to{\rm U}(1)_R$ gauge where
\bea
b_m\equiv 0~,~~~~~~
L^{ij}=\d^{ij}L
~,
\eea
and neglect, besides the fermion fields, also
the SU(2)$_R$ connection and curvature together with the ${\rm E}_a$ and $T^{-}_{abc}$ fields.
This leads to
 \bea
e^{-1}\cL_{\rm new-linear}&=&
 -\frac{1}{12}\, L^{\frac{1}{2}}\Box D
  - \frac{1}{20}L^{\frac{1}{2}}D^2
 -  \frac{1}{12}D L^{-\frac{1}{2}}  {\Box}L
  +\frac{1}{24} L^{-\frac{3}{2}} D(\cD^{a}L)  \cD_{a}L
 \non\\
&&
 - \frac{5}{8} L^{-\frac{1}{2}}  {\Box}^2L
 + \frac{15}{8}L^{-\frac{3}{2}} (\cD^{a}L)\nabla_{a} {\Box}L
 + \frac{15}{16}L^{-\frac{3}{2}}( \Box L) \Box L
\non\\
&&
 + \frac{5}{32}L^{-\frac{3}{2}}( \nabla^{a}\nabla^{b}L)  \nabla_{a}\nabla_{b}L
- \frac{295}{128} L^{-\frac{5}{2}}(\cD^{a}L)(\cD_{a}L)  {\Box}L
 \non\\
 &&
- \frac{85}{64}L^{-\frac{5}{2}}(\cD^{a}L)(\cD^{b}L)\nabla_{a}\nabla_{b}L
+\frac{1355}{1024}L^{-\frac{7}{2}}(\cD^{a}L)(\cD_{a}L)(\cD^{b}L) \cD_{b}L
\non\\
&&
+\cdots
~.
\label{new_Ricci+scalar_invariant}
\eea
Here we note that given a field $\phi$  constrained to be a Lorentz scalar
 conformal primary ($K^a\phi=0$) of dimension $\D$ ($\mathbb{D}\phi=\D\phi$),
 as for instance the fields $L$ and $D$,
one finds\bsubeq\label{usefulRR}
\bea
 {\de}_a {\de}_b\phi
&=&
 {\cD}_a {\cD}_b\phi
-2\D {\frak{f}}_{ab}\phi
~,
\\
 {\Box}\phi
&=&
 {\cD}^2\phi
-2\D {\frak{f}}_{a}{}^{a}\phi
~,
\\
 {\de}_a {\Box}\phi
&=&
 {\cD}_a {\cD}^2\phi
-2\D {\frak{f}}_{b}{}^{b} {\cD}_a\phi
+4(2-\D) {\frak{f}}_{a}{}^{b} {\cD}_b\phi
-2\D( {\cD}_a {\frak{f}}_{b}{}^{b})\phi
~,
\\
 {\Box}^2\phi
&=&
 {\cD}^4\phi
+8(2-\D) {\frak{f}}^{ab} {\cD}_a {\cD}_b\phi
-4(\D+1) {\frak{f}}_{a}{}^{a} {\cD}^2\phi
\non\\
&&
+4\Big((2-\D)( {\cD}^a {\frak{f}}_{a}{}^{b})-\D( {\cD}^b {\frak{f}}_{a}{}^{a})\Big) {\cD}_b\phi
\non\\
&&
-2\D\Big(( {\cD}^2 {\frak{f}}_{a}{}^{a})
-2(\D+2)( {\frak{f}}_{a}{}^{a})^2
-4(\D-2) {\frak{f}}^{ab} {\frak{f}}_{ab}
\Big)\phi
~,
\eea
\esubeq
with
\bea
 \Box:=\de^a\de_a
 ~,~~~~~~
  {\cD}^2:= {\cD}^a {\cD}_a
~,~~~~~~
 {\cD}^4:=( {\cD}^2)^2
~,
\eea
and
\bea
 {\frak{f}}_{ab}=-\frac{1}{8}\cR_{ab}+\frac{1}{80}\eta_{ab}\cR
~,~~~~~~
 {\frak{f}}_a{}^a=-\frac{1}{20}\cR
~.
\eea

The action $S_{\rm new-linear}$ based on \eqref{new_Ricci+scalar_invariant-full}
is locally superconformal invariant both in the standard and dilaton-Weyl
backgrounds for conformal supergravity.
In the second case, the reader should keep in mind that the $D$ and $T^-_{abc}$ fields
are composite fields that should be expressed  in terms of $\s$ and $H_{abc}$ by using \eqref{dilaton-Weyl}.
By looking at the relations \eqref{usefulRR} it is clear that the new invariant $S_{\rm new-linear}$
includes $\cR^{ab}\cR_{ab}$ and $\cR^2$ curvature squared terms.
The exact contributions depend on the conformal supergravity background  used.
We leave a detailed analysis and application of the new invariant for future work but
we now describe some simple properties of $S_{\rm new-linear}$
in a standard-Weyl and dilaton-Weyl background, respectively.

\subsection{The invariant in a standard-Weyl multiplet background, dynamically generated cosmological constant,
 ${\rm dS}_6$ and  ${\rm AdS}_6$  solutions}

Given two real constant parameters $A$ and $B$, let us consider a linear combination
  \bea
 S=A S_{\rm EH}+B S_{\rm new-linear}
 ~,
 \label{peculiar-sugra}
 \eea
of the Einstein-Hilbert term described by the action $S_{\rm EH}$ \eqref{linear-action} for a linear multiplet,
 and the new invariant $S_{\rm new-linear}$ constructed in the previous subsection.
In this subsection we are going to analyse properties of the action \eqref{peculiar-sugra} in a
standard-Weyl multiplet background of conformal supergravity coupled only to a linear multiplet.
We fix dilatations, special conformal transformations and ${\rm SU}(2)_R\to {\rm U}(1)_R$
by imposing the following gauge conditions
\bsubeq \label{gauge-section-7.2}
 \bea
B_M=0~,~~~~~~
 L^{ij}=\d^{ij}
 ~,~~~~~~
 L=1
 ~,
 \eea
which in components imply
 \bea
b_m=0~,~~~~~~
 L^{ij}=\d^{ij}
 ~,~~~~~~
 L=1
 ~,~~~~~~
 \varphi^i=0
 ~.
  \eea
\esubeq
In this section, besides the fermions,
we neglect the SU(2)$_R$ fields together with the ${\rm E}_a$ and $T^{-}_{abc}$ fields.
Then, once imposed these restrictions and the gauge fixing \eqref{gauge-section-7.2}
in \eqref{linear-action} and \eqref{new_Ricci+scalar_invariant},
up to total derivatives the action \eqref{peculiar-sugra} becomes
 \bea
S
&=&
-\int \rd^6 x \,e \, \Big\{
 \frac{2A}{5}\Big( \cR
+ \frac{1}{3} D \Big)
+\frac{B}{20}\big(\cR+D\big)D
+\frac{15B}{32}\Big(\cR^{ab}\cR_{ab}
-\frac{21}{150} \cR^2 \Big)
  \Big\}
+\cdots
\ .~~~~~~~~~
\label{peculiar-sugra-2}
\eea
It is then clear that, thanks to the $D^2$ term in the new invariant \eqref{new_Ricci+scalar_invariant},
 the previous action possesses a remarkable feature compared to
the two-derivative linear multiplet action \eqref{linear-action}:
provided the constant $B\ne0$, the dynamics for the $D$ field is completely consistent without the need of a second
compensating multiplet.
Note that this feature was noticed also in \cite{Kuzenko:2015jxa} for models  of 4D $\cN=2$ supergravity
 including an off-shell scalar curvature squared invariant constructed from a linear multiplet.
What is also interesting about \eqref{peculiar-sugra-2} is that a cosmological constant is generated dynamically
by integrating out the $D$ field.
This is especially remarkable since, to the best of our knowledge, no supersymmetric $\cN=(1,0)$ 
cosmological constant is known.

Let us integrate out the auxiliary field $D$ from \eqref{peculiar-sugra-2}.
Its equation of motion implies
\bea
D
=
 -\frac{4A}{3B}
-\frac{1}{2}\cR
+\cdots
~,
\label{D-section-7.2}
\eea
where the ellipses
refer to terms including SU(2)$_R$ gauge fields, 
${\rm E}_a$, and $T^{-}_{abc}$ fields, together with fermions, that we have neglected
in this section.
Inserting this result into \eqref{peculiar-sugra-2} we obtain
 \bea
S
&=&
\frac{2A}{3}\int \rd^6 x \,e \, \Big\{
 \frac{2A}{15B}
- \frac{1}{2} \cR
   -\frac{45B}{64A}\cR^{ab}\cR_{ab}
+\frac{15B}{128A}\cR^2
  \Big\}
+\cdots
~.
\label{peculiar-sugra-3}
\eea
After defining the cosmological constant as
\bea
\L
:=
-\frac{2A}{15B}
~,
\eea
and choosing for convenience $A=3/2$,
 \bea
S
&=&
-\int \rd^6 x \,e \, \Big\{\,
\L
+\frac{1}{2} \cR
-\frac{3}{32\L}\Big(
\cR^{ab}\cR_{ab}
-\frac{1}{6}\cR^2
\Big)
  \Big\}
+\cdots
~.
\label{dynamical-cosm-const-action}
\eea
Note that the dynamically generated cosmological constant $\L$ can take both positive and negative values.
It is straightforward to prove that the previous action possesses both ${\rm dS}_6$ and ${\rm AdS}_6$ solutions.
In fact, the relevant part of the metric equations of motion of \eqref{dynamical-cosm-const-action} reads
\bea
 \cG_{mn}
 -g_{mn}\L
- \frac{3}{16\L} \left(
2 \cR_{m}{}^r \cR_{nr}
- \frac12 g_{mn} \cR^{rs} \cR_{rs}
-\frac13 \cR \cR_{mn}
+ \frac1{12} g_{mn} \cR^2
\right ) = 0 \,,
~~~~~~
\label{dynamical-cosm-const-gEOM}
\eea
with $\cG_{mn}$ the Einstein tensor.
The  trace of \eqref{dynamical-cosm-const-gEOM} reads
\bea
\cR + 3 \L  - \frac{3}{32\L} \left(  \cR^{mn} \cR_{mn} - \frac16 \cR^2 \right) = 0  \,.
\label{dynamical-cosm-const-g-trace-EOM}
\eea
Let us consider an ansatz  for an ${\rm (A)dS}_6$ metric
\bea
\cR_{mn}{}_{rs} =\ k \left( g_{mr} g_{ns} - g_{ms} g_{nr} \right) ~~~\Longrightarrow~~~
\cR= 30\, k
\, ,
\eea
for a given real constant parameter $k$.
Then we find that the trace equation \eqref{dynamical-cosm-const-g-trace-EOM} implies
\bea
k = - \frac{1}{10} \L \,,
\label{AdS-solution}
\eea
which solves \eqref{dynamical-cosm-const-gEOM}.
Given that $\L \neq 0$, we can have both ${\rm dS}_6$ or ${\rm AdS}_6$
solutions depending on the sign of $\L$.

It is also straightforward to show that these solutions cannot be supersymmetric.
Note that the gauge fixing conditions \eqref{gauge-section-7.2} leave the $\chi^i$ chiral matter
fermion of the standard-Weyl multiplet in the spectrum.
Neglecting contributions from the ${\rm SU}(2)_R$ and $T^-_{abc}$ fields,
the residual gauge preserving supersymmetry transformation of $\chi^i$
follow from \eqref{eq:WeylSUSY2-f}
\be
\d \chi^{ i}
=
\hf  D\x^{ i}
+\cdots
~.
\ee
A necessary condition for the ${\rm dS}_6$ or ${\rm AdS}_6$ backgrounds to possess unbroken supersymmetry
is then $0\equiv\d \chi^{ i}=D$.
The $D$ equation of motion \eqref{D-section-7.2} for these solutions reads
\bea
D = - \frac{4A}{3B} - \frac12 \cR = 10 \L - 15\,k \,,
\eea
which implies that in order for us to be able to set $D = 0$ we must have
\bea
k = \frac23 \L \,.
\eea
This result clearly conflicts with \eqref{AdS-solution}
 given that $\L \neq 0$.
Thus the ${\rm (A)dS}_6$ solution \eqref{AdS-solution} cannot preserve supersymmetry.
This is expected since it is known that the supersymmetric extensions
of the ${\rm dS}_6$ and ${\rm AdS}_6$ symmetry groups
${\rm SO}(1,6)$ and ${\rm SO}(2,5)$
describe two real forms of the exceptional supergroup ${\rm F}(4)$,
respectively ${\rm F}^1(4)$ and ${\rm F}^2(4)$,
that possess 16 fermionic generators,
see \cite{Nahm:1977tg,Lukierski:1984it,Pilch:1984aw,DAuria:2002xnk,Fre:2002pd}.
Hence it is necessary to consider extended 6D supergravities if one is interested in supersymmetric
${\rm (A)dS}_6$ backgrounds.
Nevertheless, we find intriguing the existence of the non-supersymmetric ${\rm (A)dS}_6$ solutions
triggered by the non-trivial higher-derivative term  described in this section.

\subsection{The invariant in a dilaton-Weyl background}

A natural question to be asked is whether the new invariant constructed in this section
is independent of the ones described in section  \ref{main-curvature-squared}.
It is simple to see that the answer is yes.
To prove this, it is first necessary to remember that all the invariants in section \ref{main-curvature-squared}
were defined in a dilaton-Weyl background for conformal supergravity.
Hence, to compare the results, in the new invariant described by  \eqref{new_Ricci+scalar_invariant}
we need to express the $D$ and $T^-_{abc}$ fields in terms of the fields $\s$ and $H_{abc}$
of the tensor multiplet by using \eqref{dilaton-Weyl}.
It then suffices to compare the results in
the conformal and ${\rm SU}(2)_R$ gauge
given by \eqref{GaugeFixing}.
Besides the gauge conditions mentioned before, for simplicity 
we neglect all fields in the supergravity multiplet
except the vielbein and the dilaton $L=\re^{-2v}$. 
In the equations in this subsection the ellipses will indicate all the terms neglected as  described above.
It turns out that by using the following relations that are consequences of \eqref{usefulRR}
\bsubeq
\bea
D|_{\s=1}&=&
\frac{3}{4}\cR
+\cdots
~,
\\
\Box D|_{\s=1}
&=&
\frac{3}{4}{\cD}^2\cR
+\frac{9}{20}\cR^2
+\cdots
~,
\\
 {\de}_a {\de}_bL
&=&
 {\cD}_a {\cD}_bL
+\cR_{ab}L
-\frac{1}{10}\eta_{ab}\cR L
+\cdots
~,
\\
 {\Box}L
&=&
 {\cD}^2L
+\frac{2}{5}\cR L
+\cdots
~,
\\
 {\de}_a {\Box}L
&=&
 {\cD}_a {\cD}^2L
+\cR_{ab}{\cD}^bL
+\frac{3}{10}\cR {\cD}_aL
+\frac{2}{5}L {\cD}_a \cR
+\cdots
~,
\\
 {\Box}^2L
&=&
{\cD}^4L
+2\cR^{ab} {\cD}_a {\cD}_bL
+\frac{4}{5}\cR {\cD}^2L
+\frac{6}{5}( {\cD}^a\cR) {\cD}_aL
\non\\
&&
+L\Big(
\frac{2}{5} {\cD}^2\cR
-\cR^{ab}\cR_{ab}
+\frac{19}{50}\cR^2
\Big)
+\cdots
~,
\eea
\esubeq
one can obtain from \eqref{new_Ricci+scalar_invariant} the following Lagrangian for the new invariant in a 
dilaton-Weyl background
 \bea
e^{-1}\cL_{\rm new-linear}&=&
\re^{-v}\Big{[}\,
\frac{25}{32}\cR^{ab}\cR_{ab}
 -\frac{1}{5} \cR^2
 -\frac{5}{16}{\cD}^2\cR
+\frac{15}{8} \cR^{ab}{\cD}_a{\cD}_bv
-\frac{5}{16}\cR {\cD}^2v
  \non\\
&&~~~~~~
- \frac{25}{16} \cR^{ab}({\cD}_av) {\cD}_bv
  -\frac{5}{32}\cR(\cD^{a}v)\cD_{a}v
+ \frac{5}{4}{\cD}^4v
+ \frac{5}{2}(\cD^av)\cD_a{\cD}^2v
\non\\
&&~~~~~~
- 5 ({\cD}^av)\cD^2{\cD}_av
 - \frac{35}{8}({\cD}^a{\cD}^bv)\cD_a\cD_bv
  + \frac{5}{4}({\cD}^2v)^2
\non\\
&&~~~~~~
 -\frac{15}{8}(\cD^av)(\cD^bv){\cD}_a{\cD}_bv
-\frac{25}{16}(\cD^{a}v)(\cD_{a}v){\cD}^2v
\non\\
&&~~~~~~
+\frac{35}{64}(\cD^{a}v)(\cD_{a}v)(\cD^{b}v)\cD_{b}v
\Big{]}
  \non\\
&&
+\cdots
\label{inv-7_dilatonWeyl}
\eea
It is evident that it not possible to express the previous invariant as a linear combination
of the three invariants constructed in section \ref{main-curvature-squared}.
The new invariant appears multiplied by a different factor of the dilaton and it includes higher-derivative terms for
$v$.
We leave  the analysis of possible dynamical features of the new invariant studied in this section
for future work and now 
come back to analysing in detail physical properties of the Einstein-Gauss-Bonnet supergravity
of section \ref{Einstein-Gauss-Bonnet}.


\section{Spectrum of Einstein-Gauss-Bonnet supergravity about ${\rm AdS}_3\times {\rm S}^3$}
\label{spectrum}

In this section, we elaborate  on
the derivation of the spectrum of the Einstein-Gauss-Bonnet supergravity about the supersymmetric
${\rm AdS}_3\times {\rm S}^3$ vacuum whose main results were originally presented in \cite{NOPT-M17}.
Although the spectrum of the short multiplets can be inferred from \cite{Deger:1998nm,deBoer:1998kjm}, the 
spectrum of the long multiplets computed here comprises new results. 
As the AdS energies of the long-multiplet states are not protected by shortening conditions and depend on
the detail of the theory, they cannot be obtained by the group theory based approach of \cite{deBoer:1998kjm}. 
We therefore adopted the same method as \cite{Deger:1998nm} by solving the linearized equations for supegravity 
fields about the supersymmetric AdS vacuum.

Before turning to the main technical discussion, it is important to underline that,
in order to precisely adhere to the results presented in \cite{NOPT-M17}, 
in this section we decided to use the same 
conventions as \cite{NOPT-M17}. These differ compared to the conventions we use elsewhere
in this paper. To avoid possible confusion, the reader should consider results presented in
this section as self-contained.

In particular, as in \cite{NOPT-M17}, in this section  we adopt the conventions of
\cite{CVanP,Bergshoeff:2012ax}
with the exception of a sign difference in the parity transformation and, consequently, the Levi-Civita tensor 
(which agrees with the one in our paper).\footnote{See appendix 
\ref{notation} to compare the conventions of \cite{CVanP} with the ones of the rest of our paper.}
Other differences of conventions used in this section are:
the six-dimensional vector indices here are labeled by ${\mu},{\nu},\cdots$;
the Lorentz and ${\rm SU}(2)_R$ connections and curvatures, which here will be denoted as 
$(\o_\mu{}^{ab},R_{\mu\nu}{}^{ab})$ instead of 
$(\o_m{}^{ab},\cR_{mn}{}^{ab})$
 and 
 $(V_\mu{}^{ij},F_{\mu\nu}{}^{ij})$ instead of 
$(\cV_m{}^{ij},\cR_{mn}{}^{ij})$,
all have a sign flip compared to the same tensors in the rest of the paper;
the vielbein $e_\mu{}^a$ 
(up to the nomenclature for the curved 6D indices)
and the field $\s$ are identical 
to the ones in the rest of the paper;
the gauge 2-form and its field strength 3-form here will be denoted by $B_{\mu\nu}$ and $H_{\mu\nu\rho}$
but they are otherwise identical to $b_{mn}$ and $h_{mnp}$ of the rest of the paper;
the scalar field $L^{ij}$ of the linear multiplet is identical to the one of the rest of the paper with the difference that 
in this section we use the definition $L^2:= L^{ij}L_{ij}$; 
the derivatives $\de_\mu$ used in this section (which should not be confused with the locally superconformally 
covariant derivatives of the rest of the paper)
 are solely defined in terms of the metric and are the standard 
general coordinate covariant derivatives possessing no other connections other than the Levi-Civita connection.

Before continuing our analysis, it is also worth giving the Einstein-Hilbert and Gauss-Bonnet supergravity invariants
in these conventions.
After assuming the gauge fixing conditions
\be
\s=1\,,\quad L_{ij}=\frac1{\sqrt2}\delta_{ij}L\,,\quad \psi^i=0\,,\quad b_{ {\m}}=0\, ,
\ee
and applying the decomposition of the original ${\rm SU}(2)_R$ 
gauge field $V_{ {\mu}}^{ij}$ under the residual ${\rm U}(1)_R$ symmetry
\be
V_{ {\m}}^{ij}=V'_{ {\m}}{}^{ij}+\frac12\delta^{ij}V_{ {\m}}\, ,
\ee
up to fermionic terms,
the Lagrangian of Einstein-Hilbert supergravity takes the form
\bea
e^{-1}{\cal L}_{\rm EH}
&=&
LR
+L^{-1}\partial_{\mu}L\partial^{\mu}L
-\frac1{12}LH_{\mu\nu\r}H^{\mu\nu\r}
-4LZ_{\mu}\bar{Z}^{\mu}
-\frac12L^{-1}{\rm E}^{\mu}{\rm E}_{\mu}
+\sqrt{2}{\rm E}^{\mu}V_{\mu}
 \non\\
 &&
 +{\rm fermions}
\, ,
\label{eha}
\eea
where we have defined
\be
Z_{\m}=V'^{11}_{\m}+{\rm i}V'^{12}_{\m}\,.
\ee
The off-shell Gauss-Bonnet invariant \eqref{GBSUGRA}
and in the conformal and 
${\rm SU}(2)_R\to {\rm U}(1)_R$ gauge takes the form
\bea
e^{-1}{\cal L}_{\rm GB}&=&
R_{\mu\nu\r\s} R^{\mu\nu\r\s} 
- 4 R_{\mu\nu} R^{\mu\nu} + R^2+\ft12 R_{\mu\nu\r\s} H^{\mu\nu\l} H^{\r\s}{}_{\l} 
- R^{\mu\nu} H_{\mu\nu}^2+ \ft{1}{6} R H^2
\non\\
&& + \ft{1}{144} (H^2)^2 -\ft18 (H^2_{\mu\nu})^2 + \ft{5}{24} H^4
+\ft14\e^{\mu\nu\r\s\l\tau} B_{\mu\nu} R_{\r\s}{}^{{\a}}{}_{{\b}}(\omega_+) R_{\l{\tau}}{}^{{\b}}{}_{{\a}}(\omega_+)
\non\\
&&-2\e^{\mu\nu\r\s\l{\tau}} B_{\mu\nu} F_{\r\s}(Z) \bar{F}_{\l{\tau}}(Z)
+\ft12\e^{\mu\nu\r\s\l{\tau}} B_{\mu\nu} F_{\r\s}(V) F_{\l{\tau}}(V)
 \non\\
 &&
 +{\rm fermions}
\, ,
\label{GBs8}
\eea
where
\be
F_{\mu\nu}(Z)=2\partial_{[\mu}Z_{\nu]}-2{\rm i}V_{[\mu}Z_{\nu]}
\,,\quad 
F_{\mu\nu}(V)=2\partial_{[\mu}V_{\nu]}+4{\rm i}Z_{[\mu}\bar{Z}_{\nu]}\,.
\ee
Note that $R_{\mu\nu}{}^{{\a}}{}_{{\b}}(\omega_+)$, 
which later will also be simply denoted by $R^+{}_{\mu\nu}{}^{{\a}}{}_{{\b}}$, 
 is the Riemann tensor defined with a torsionfull connection
$\o_\pm{}_\mu{}^{cd}:={\o}_\mu{}^{cd}\pm\ft{1}{2}e_\mu{}^aH_a{}^{cd}$.\footnote{The fact that in \eqref{GBs8}
 $R_{ab}{}^{cd}(\omega_+)$ appears
instead of  $R_{ab}{}^{cd}(\omega_-)$, as in \eqref{GBSUGRA}, is simply due to the sign difference in the definition
of the Lorentz connection and curvature in this section.}

The total Lagrangian of the Einstein-Gauss-Bonnet supergravity is parametrized as
(for convenience we set $\k^2=1$)
\be
{\cal L}_{\rm tot}={\cal L}_{\rm EH}+\ft{\a'}{16}{\cal L}_{\rm GB}\,.
\ee
Accordingly, the equation of motion for the dilaton field $L$ is given by
\be
R+L^{-2}\partial_{\mu}L\partial^{\mu}L
-2L^{-1}\Box L-\ft1{12}H^2-4Z_{\mu}\bar{Z}^{\mu}
+\ft12L^{-2}{\rm E}^{\mu}{\rm E}_{\mu}=0\,,
\ee
where we underline that the GB invariant is independent of the dilaton field $L$.
Using the equation above, the Einstein equation with Gauss-Bonnet correction can be written as
\bea
LR_{\mu\nu}&=&\nabla_{\mu}\nabla_{\nu }L-L^{-1}\partial_{\mu} L\partial_{\nu} L+\ft14LH_{\mu\r\l}H_{\nu}{}^{\r\l}
+4LZ_{(\mu}\bar{Z}_{\nu)}
+\ft12L^{-1}{\rm E}_{\mu} {\rm E}_{\nu}
\non\\
&&
-\sqrt2{\rm E}_{(\mu} V_{\nu)}
+\ft12g_{\mu\nu}(\sqrt2 {\rm E}_{\r}V^{\r}-L^{-1}{\rm E}_{\r}{\rm E}^{\r})-\ft{\a'}{16}{\cal E}_{\mu\nu}\,,
\eea
where ${\cal E}_{\mu\nu}$ is the contribution from supersymmetric Gauss-Bonnet action and takes the form
\bea
{\cal E}_{\m\n}&=&2RR_{\mu\nu}-4R_{\mu\l}R^{\l}{}_{\nu}-4R_{\mu\r\nu\l}R^{\r\l}+2R_{\mu}{}^{\r\l\s}R_{\nu\r\l\s}\non\\
&&-\nabla_{\s}\nabla_{\r}\big(H_{(\mu}{}^{\s\l}H^{\r}{}_{\nu)\l}\big)
+\ft12R^{{\a}}{}_{{\b}\r\s}H_{{\a}}{}^{{\b}}{}_{(\mu}H^{\r\s}{}_{\nu)}
+\ft32R^{{\a}}{}_{{\b}\s(\mu}H_{\nu)\l}{}^{\s}H_{{\a}}{}^{{\b}\l}\non\\
&&-2R_{\r(\mu}H^2_{\nu)}{}^{\r}-2R^{\r\s}H_{\mu\r}{}^{\l}H_{\nu\s\l}-\ft12\Box H^2_{\mu\nu}
-\ft12g_{\mu\nu}\nabla_{\r}\nabla_{\l}H^{2\r\l}+\nabla_{\r}\nabla_{(\mu}H^2_{\nu)}{}^{\r}\non\\
&&+\ft16R_{\mu\nu}H^2+\ft12RH^2_{\mu\nu}+\ft16g_{\mu\nu}\Box H^2-\ft16\nabla_{\mu}\nabla_{\nu}H^2
+\ft1{24}H^2H^2_{\mu\nu}\non\\
&&-\ft14H^2_{\mu\l}H^{2\l}{}_{\nu}-\ft12H_{\mu\l}{}^{{\a}}H_{\nu\s{\a}}H^{2\l\s}
+\ft5{12}H_{\mu}{}^{{\a}\l}H_{\nu}{}^{{\b}\s}H^2_{{\a}{\b},\,\l\s}
+\ft5{6}H^2_{\mu\l,\,{\b}\s}H^2_{\nu}{}^{{\b},\,\l\s}\non\\
&&-2\nabla_{{\b}}\big(R^{+\l{\t}{\b}}{}_{(\mu}*H_{\nu)\l{\t}}\big)
-R^+{}_{\l{\t}}{}^{{\b}}{}_{(\mu}H_{\nu){\b}\s}*H^{\s\l{\t}}
-\ft12g_{\mu\nu}{\cal L}_{\rm GB\,P-even~term}
\,.~~~~~~
\eea
In the expression above, ${\cal L}_{\rm GB\,P-even~term}$ indicates the parity-even part of the GB Lagrangian
\eqref{GBs8} (meaning every term in \eqref{GBs8} except the last three) and
we have defined
\be
H^{2\,\mu\l,\nu\s}:= H^{\mu\l\r}H_{\r}{}^{\nu\s},\quad H^{2\mu\nu}=g_{\l\s}H^{2\,\mu\l,\,\nu\s}\,.
\ee
The two-form equation of motion takes the form
\bea
0&=&\nabla_{\r}\big(LH^{\r\mu\nu}+\ft{\a'}{8}{S^{\r\mu\nu}}\big)
\non\\
&&
+\ft{\a'}{16}\e^{\mu\nu\r\s\l{\t}}\Big(F_{\r\s}(V) F_{\l{\t}}(V)
-4F_{\r\s}(Z) F_{\l{\t}}(Z)
+\frac14 R_{\r\s}{}^{\a}{}_{\b} (\o_+) R_{\l\t}{}^{\b}{}_{\a}(\o_+)
\Big)
\,,
~~~
\eea
where ${S^{\r\mu\nu}}$ is given by
\bea
{S^{\r\mu\nu}}
&=&
-3R^{\l\s[\r\mu}H^{\nu]}{}_{\l\s}
+6R^{\l[\r}H_{\l}{}^{\mu\nu]}
-RH^{\r\mu\nu}
-\ft1{12}H^2H^{\r\mu\nu}
+\ft32H^{2\,\l[\r}H_{\l}{}^{\mu\nu]}
\non\\
&&
-\ft5{2}H^{[\r}{}_{\l\s}H^{2\,\mu|\l|,\,\nu]\s}
+3R^+{}_{\l\s}{}^{[\mu\nu}*H^{\r]\l\s}\,.
\eea
The equations of motion for the auxiliary fields $V_\mu$, $Z_\mu$ and its complex conjugate, and 
the gauge 4-form $b_{\mu\nu\r\l}$, whose field strength is associated to ${\rm E}_\mu$,
 are given as follows:
\bsubeq
\be
LZ^{\mu}
-\ft{\a'}{48}\epsilon^{{\t}\r\s\mu\nu\l}H_{{\t}\r\s}F_{\nu\l}(Z)=0\,,
\label{eqZ}
\ee
\be
\sqrt{2}V_{\mu}-L^{-1}{\rm E}_{\mu}=0\,,
\label{eqVE}
\ee
\be
\sqrt2{\rm E}^{\mu}
-\ft{\a'}{24}\epsilon^{{\t}\r\s\mu\nu\l}H_{{\t}\r\s}F_{\nu\l}(V)=0\,.
\label{eqV}
\ee
\esubeq
It is well known that the off-shell 6D $\cN=(1,0)$ Einstein-Hilbert action admits 
a maximally supersymmetric ${\rm AdS}_3\times {\rm S}^3$ vacuum with a metric, 
3-form flux, and dilaton field of the form
\be
ds^2_6={ \varrho}^2\big(ds^2_{{\rm AdS}_3}+ds^2_{{\rm S}^3}\big)
\,,\quad H_{(3)}=2{\varrho}\big(\Omega_{{\rm AdS}_3}-\Omega_{{\rm S}^3}\big)
\,,\quad L=1\,,
\label{ads3s3}
\ee
where $ds^2_{{\rm AdS}_3}$ and $ds^2_{{\rm S}^3}$ are the metrics on the unit radius 
${\rm AdS}_3$ and ${\rm S}^3$, and $\Omega_{{\rm AdS}_3}$ and $\Omega_{{\rm S}^3}$ 
denote their volume forms.
This metric has vanishing Weyl tensor and Ricci scalar 
\be
C_{\mu\nu\r\l}=0\, ,\quad R=0\,,
\ee
similarly to the maximally supersymmetric
 ${\rm AdS}_5\times {\rm S}^5$ 
solution in 10D IIB superstring theory.
By substituting the ansatz \eqref{ads3s3} into the $\alpha^\prime$-corrected
 equations of motion given above, we have explicitly checked that \eqref{ads3s3} remains 
a solution. Moreover, the supersymmetry of the solution is also unaffected by the inclusion 
of the Gauss-Bonnet supergravity action, 
because the off-shell supersymmetry transformations are independent of the equations of motion.

In the following, we shall study the spectrum of fluctuations around the vacuum solution \eqref{ads3s3}. We split the 
six-dimensional indices into to 3 external indices on ${\rm AdS}_3$ 
labeled by $\hmu,\, \hnu,\, \hr,\cdots$ and 3 internal indices on ${\rm S}_3$ labeled by $\hm,\,\hn,\,\hp,\cdots$.
The metric fluctuations around the supersymmetric ${\rm AdS}_3\times {\rm S}^3$ vacuum are parametrized as 
(here we adopt the strategy used in \cite{Deger:1998nm})
\bsubeq
\bea
h_{\hmu\hnu}&=&H_{\hmu\hnu}+\bar{g}_{\hmu\hnu}M\,,\quad \bar{g}^{\hmu\hnu}H_{\hmu\hnu}=0\,,
\\
h_{\hmu \hm}&=&K_{\hmu \hm}\,,\\
h_{\hm\hn}&=&L_{\hm\hn}+\bar{g}_{\hm\hn}N\,,\quad \bar{g}^{\hm\hn}L_{\hm\hn}=0\,,
\eea
\esubeq
where $\bar{g}_{\hmu\hnu}$ and $\bar{g}_{\hm\hn}$ are the metric of unit radius
on ${\rm AdS}_3$  and ${\rm S}^3$, respectively.
The zero- and two-form fields fluctuations are parametrized as
\be
L=1+\phi\,,\quad b_{\hmu\hnu}=\epsilon_{\hmu\hnu\hr}X^{\hr}
\,,\quad 
b_{\hm\hn}=\e_{\hm\hn\hr}U^{\hr}
\,,\quad 
b_{\hmu \hm}=W_{\hmu \hm}
\,,
\ee
and for fluctuations associated with the one-forms $V_{\mu}$ and $Z_{\mu}$, 
we retain the same symbols. 
Since ${\rm E}_{\mu}$ is fully determined by $V_{\mu}$ via \eqref{eqVE}, 
which should be used in \eqref{eqV} to obtain a closed equation similar to \eqref{eqZ},
we do not give a separate discussion on ${\rm E}_{\mu}$. 
We impose the following gauge fixing conditions
\be
\nabla^\hm h_{\{ \hm\hn\}}=0\,,\quad \nabla^\hm h_{\hm\hmu}=0\,,\quad \nabla^\hm b_{\hm\mu}=0\,,\quad 
\nabla^\hm V_\hm=0\,,
\ee
where  $\{\,\cdots\}$ denotes the complete traceless symmetrization of a set of indices.
As a consequence of the previous gauge conditions, 
the fluctuations can be expanded in terms of harmonic functions on  ${\rm S}^3$
 as
\bsubeq
\bea
H_{\hmu\hnu}(x,y)&=&\sum H^{(\ell\,,0)}_{\hmu\hnu}(x)Y^{(\ell\,,0)}(y)\,,
\\
M(x,y)&=&\sum M^{(\ell\,,0)}(x)Y^{(\ell\,,0)}(y)\,,
\\
K_{\hmu \hm}(x,y)&=&\sum K_{\hmu}^{(\ell\,,\pm1)}(x)Y_\m^{(\ell\,,\pm1)}(y)\,,\\
L_{\hat{m} \hat{n}}(x,y)&=&\sum L^{(\ell\,,\pm2)}(x)Y_{\hat{m}\hat{n}}^{(\ell\,,\pm2)}(y)\,,\\
N(x,y)&=&\sum N^{(\ell\,,0)}(x)Y^{(\ell\,,0)}\,,\\
\phi(x,y)&=&\sum \phi^{(\ell\,,0)}(x)Y^{(\ell\,,0)}\,,\\
X_{\hmu}(x,y)&=&\sum X_{\hmu}^{(\ell\,,0)}(x)Y^{(\ell\,,0)}(y)\,,\\
U_{\hat{m}}(x,y)&=&\sum U^{(\ell\,,0)}(x)\partial_{\hat{m}} Y^{(\ell\,,0)}(y)\,,\\
W_{\hmu \hat{m}}&=&\sum  W_{\hmu}^{(\ell\,,\pm1)}(x)Y_{\hat{m}}^{(\ell\,,\pm1)}(y)\,,
\\
Z_{\hmu}(x,y)&=&\sum Z_{\hmu}^{(\ell\,,0)}(x)Y^{(\ell\,,0)}(y)\,,\\
Z_{\hat{m}}(x,y)&=&\sum \Big[Z^{(\ell\,,\pm1)}(x)Y_{\hat{m}}^{(\ell\,,\pm 1)}(y)
+Z^{(\ell\,,0)}(x)\partial_{\hat{m}} Y^{(\ell\,,0)}(y)\Big]\,,\\
V_{\hmu}(x,y)&=&\sum V_{\hmu}^{(\ell\,,0)}(x)Y^{(\ell\,,0)}(y)\,,\\
V_{\hat{m}}(x,y)&=&\sum V^{(\ell\,,\pm1)}(x)Y_{\hat{m}}^{(\ell\,,\pm1)}(y)\,,
\eea
\esubeq
where, in the parametrization we are using, 
we have denoted with $x$ the coordinates of ${\rm AdS}_3$
and with $y$ the coordinates of ${\rm S}^3$, respectively. 
Here we omitted several mathematical details with regard to harmonic
expansion of the fluctuations which can be found in \cite{Deger:1998nm} (for instance, all longitudinal gauge modes 
can be removed by a field dependent gauge transformation generated by the harmonic non-zero modes).
The harmonic functions $Y^{(\ell_1,\ell_2)}$ with various spins satisfy
\bsubeq
\bea
\nabla_y^2Y^{(\ell\,,0)}&=&-\ell(2+\ell)Y^{(\ell\,,0)}\,,
\\
\e_\hm{}^{\hn\hp}\partial_\hn Y^{(\ell\,,\pm1)}_\hp&=&\pm(\ell+1)Y^{(\ell\,,\pm1)}_\hm\,,
\\
\e_\hm{}^{\hp\hq}\nabla_\hp Y^{(\ell\,,\pm2)}_{\hq\hn}&=&\pm(\ell+1)Y^{(\ell\,,\pm2)}_{\hm\hn}\,.
\label{s3hf}
\eea
\esubeq
Since the isometry group of ${\rm S}^3$ is ${\rm SU}(2)\times {\rm SU}(2)$, the harmonic functions 
can also be labelled using the two ${\rm SU}(2)$ quantum numbers denoted by $(j\,,\bar{j})$. 
The relation between $(j\,,\bar{j})$ and $(\ell_1\,,\ell_2)$ is given by
\be
j=\ft12(\ell_1+\ell_2)\,,\quad \bar{j}=\ft12(\ell_1-\ell_2)\,.
\ee

The ${\rm AdS}_3$ part of the fluctuations can also be expanded in terms of harmonics on ${\rm AdS}_3$. 
We denote the ${\rm AdS}_3$ harmonics by $\Xi^{(E\,,s)}(x)$ where $(E\,,s)$ are related to the
${\rm SL}(2,{\mathbb R})\times {\rm SL}(2,{\mathbb R})$ quantum numbers $(h,\bar{h})$ by
\be
h=\ft12(E+s)\,,\quad \bar{h}=\ft12(E-s)\,.
\ee
Then we have
\bsubeq
\bea
H^{(\ell\,,0)}_{\hmu\hnu}(x)
&=&
\sum \Big[H^{(E\,,\pm2)\otimes(\ell\,,0)}\Xi^{(E\,,\pm2)}_{\hmu\hnu}(x)
+H^{(E\,,\pm1)\otimes(\ell\,,0)}\nabla_{\{\hmu}\Xi^{(E\,,\pm1)}_{\hnu\}}(x)
\non\\
&&\qquad +H^{(E\,,0)\otimes(\ell\,,0)}\nabla_{\{\hmu}\nabla_{\hnu\}}\Xi^{(E\,,0)}(x)\Big]\,,
\\
M^{(\ell\,,0)}(x)&=&\sum M^{(E\,,0)\otimes(\ell\,,0)}\Xi^{(E\,,0)}(x)\,,
\\
K_\hmu^{(\ell\,,\pm1)}
&=&
\sum \Big[K^{(E\,,\pm1)\otimes(\ell\,,\pm1)}\Xi_{\hmu}^{(E\,,\pm1)}
+K^{(E\,,0)\otimes(\ell\,,\pm1)}\partial_\hmu\Xi^{(E\,,0)}\Big]\,,
\\
L^{(\ell\,,\pm2)}(x)&=&\sum L^{(E\,,0)\otimes(\ell\,,\pm2)}\Xi^{(E\,,0)}(x)\,,
\\
N^{(\ell\,,0)}(x)&=&\sum N^{(E\,,0)\otimes(\ell\,,0)}\Xi^{(E\,,0)}(x)\,,
\\
\phi^{(\ell\,,0)}(x)&=&\sum \phi^{(E\,,0)\otimes(\ell\,,0)}\Xi^{(E\,,0)}(x)\,,
\\
X_\hmu^{(\ell\,,0)}&=&\sum \Big[X^{(E\,,\pm1)\otimes(\ell\,,0)}\Xi_{\hmu}^{(E\,,\pm1)}
+X^{(E\,,0)\otimes(\ell\,,0)}\partial_\hmu\Xi^{(E\,,0)}\Big]\,,
\\
U^{(\ell\,,0)}(x)&=&\sum U^{(E\,,0)\otimes(\ell\,,0)}\Xi^{(E\,,0)}(x)\,,
\\
W_\hmu^{(\ell\,,\pm1)}&=&\sum \Big[W^{(E\,,\pm1)\otimes(\ell\,,\pm1)}\Xi_{\hmu}^{(E\,,\pm1)}
+W^{(E\,,0)\otimes(\ell\,,\pm1)}\partial_\hmu\Xi^{(E\,,0)}\Big]\,,
\\
Z_\hmu^{(\ell\,,0)}&=&\sum \Big[Z^{(E\,,\pm1)\otimes(\ell\,,0)}\Xi_{\hmu}^{(E\,,\pm1)}
+Z^{(E\,,0)\otimes(\ell\,,0)}\partial_\hmu\Xi^{(E\,,0)}\Big]\,,
\\
Z^{(\ell\,,\pm1)}&=&\sum Z^{(E\,,0)\otimes(\ell\,,\pm1)}\Xi^{(E\,,0)}\,,
\\
Z^{(\ell\,,0)}&=&\sum Z^{(E\,,0)\otimes(\ell\,,0)}\Xi^{(E\,,0)}\,,
\\
V_\hmu^{(\ell\,,0)}
&=&
\sum \Big[V^{(E\,,\pm1)\otimes(\ell\,,0)}\Xi_{\hmu}^{(E\,,\pm1)}
+V^{(E\,,0)\otimes(\ell\,,0)}\partial_\hmu\Xi^{(E\,,0)}\Big]\,,\\
V^{(\ell\,,\pm1)}
&=&
\sum V^{(E\,,0)\otimes(\ell\,,\pm1)}\Xi^{(E\,,0)}\,,
\eea
\esubeq
where the harmonic functions $\Xi^{(E\,,s)}$ on ${\rm AdS}_3$ satisfy
\bsubeq
\bea
\nabla_x^2\Xi^{(E\,,0)}&=&-E(E-2)\Xi^{(E\,,0)}\,,\\
\e_\hmu{}^{\hnu\hr}\partial_\hnu\Xi^{(E\,,\pm1)}_\hr&=&\pm(E-1)\Xi^{(E\,,\pm1)}_\hmu\,,\\
\e_\hmu{}^{\hr\hs}\nabla_\hr\Xi^{(E\,,\pm2)}_{\hs\hnu}&=&\pm(E-1)\Xi^{(E\,,\pm2)}_{\hmu\hnu}\,.
\label{ads3hf}
\eea
\esubeq
The harmonic functions on ${\rm S}^3$ and ${\rm AdS}_3$ are known explicitly, and we have implemented them in 
a {\it Mathematica} code. After substituting the harmonic expansion of the fluctuations to the equations of motion 
and noticing that modes with different quantum numbers decouple from each other, we can derive a set of algebraic 
equations relating $E$ to $\ell$, from which we can solve $E$ in terms of $\ell$. 
The results are schematically listed 
below where we have also introduced the dimensionless parameter
\be
\tilde{\a}:=\a'/\varrho^2\,. 
\ee
Without any ambiguity and for succinctness, in the following we suppress the $(E\,,\ell)$ 
in the labelling of various expansion coefficients. 
\begin{itemize}
  \item{\bf $(|s|\,,\ell_2)=(2\,,0)$ sector}\vspace{0.3cm}\\
  We obtain
\be
(E-2 -\ell) \left(1 + \tilde{\a}-\ft12 E \tilde{\a} \right)H^{(2\,,0)}=0
\,,\quad 
(E-2 -\ell) \left(\ft12 E \tilde{\a}+1  \right)H^{(-2\,,0)}=0\,,
\ee
from which we find 4 towers of propagating degrees of freedom labelled by their quantum numbers
\be
\Big((E=2+\ell\,,2)\oplus(E=2+\ft{2}{\tilde{\a}}\,,2)\oplus(E=2+\ell\,,-2)\oplus(E=-\ft{2}{\tilde{\a}}\,,-2)\Big)\otimes(\ell\,,0)
\,.
\ee

\item{\bf $(|s|\,,|\ell_2|)=(1\,,1)$ sector}\vspace{0.3cm}\\
In this sector, from the $g_{\hat{\mu}\hat{\nu}}$ and $B_{\hat{\mu}\hat{\nu}}$ equations , 
 we obtain

\be
\left(
  \begin{array}{cc}
 A_{\pm\pm}& B_{\pm\pm} \\
 B_{\pm\pm}& A_{\pm\pm} \\
  \end{array}
\right)\left(
          \begin{array}{c}
           K^{(\pm1\,,\pm1)}\\
            W^{(\pm1\,,\pm1)}\\
          \end{array}
        \right)=0\,,
\ee
where the elements of the $2\times2$ mixing matrices are given by
\bsubeq
\bea
A_{++}&=&2 - E + \ell 
+ \big(1 - \ft12E+\ft32 \ell  + \ft14 (E-\ell)^2\big)\tilde{ \a}\,,
 \\
A_{+-}&=&(E-2-\ell)\Big(E + \ell + \big(1  - \ft12( E-\ell) -\ft14 (E+\ell)^2 \big)\tilde{\a}\Big)\,,
\\
A_{-+}&=&(E-2-\ell)\Big(E + \ell + \big(1  - \ft12( E-\ell) +\ft14 (E+\ell)^2 \big)\tilde{\a}\Big)\,,
\\
A_{--}&=&2 - E + \ell - \big(1 - \ft32E+\ft12 \ell  + \ft14 (E-\ell)^2\big)\tilde{ \a}\,,
\\
B_{++}&=&\ft14 \big(E^2-(2+\ell)^2\big)\tilde{\a}-2\,,
\\
B_{+-}&=&(E-2-\ell)\big(2  + \ft14(\ell^2-E^2) \tilde{\a}\big)\,,
\\
B_{-+}&=&-(E-2-\ell)\Big(2 +(E+\ell)\big(1-\ft14(E-\ell)\big)\tilde{\a}\Big)\,,
\\
B_{--}&=&2+\ft14 \left(\ell^2-(E-2)^2\right)\tilde{\a}
\,.
\eea
\esubeq
The existence of nontrivial solutions for $K^{(1\,,1)}$ and $W^{(1\,,1)}$ requires
\be
(E-4 -\ell) (E - \ell)  \left(-1 - \ft12 \tilde{\a} + \ft12 E \tilde{\a}\right)=0\,,
\ee
which leads to 3 infinite towers of propagating degrees of freedom
\be
\Big((E=4+\ell\,,1)\oplus (E=\ell\,,1)\oplus  (E=1+\ft{2}{\tilde{\a}}\,,1)\Big)\otimes(\ell\,,1)\,.
\ee
The existence of nontrivial solutions for $K^{(1\,,-1)}$ and $W^{(1\,,-1)}$ requires
\be
(E-2 -\ell)^2 \left(-1 - \ft12 \tilde{\a} + \ft12 E \tilde{\a}\right)=0\,,
\ee
which leads to 3 infinite towers of propagating degrees of freedom
\be
\Big(2\times (E=2+\ell\,,1)\oplus (E=1+\ft{2}{\tilde{\a}}\,,1)\Big)\otimes(\ell\,,-1)\,.
\ee
The existence of nontrivial solutions for $K^{(-1\,,1)}$ and $W^{(-1\,,1)}$ requires
\be
(E-2 -\ell)^2 \left(1 - \ft12 \tilde{\a} + \ft12 E \tilde{\a}\right)=0\,,
\ee
which leads to 3 infinite towers of propagating degrees of freedom
\be
\Big(2\times (E=2+\ell\,,-1)\oplus (E=1-\ft{2}{\tilde{\a}}\,,-1)\Big)\otimes(\ell\,,1)\,.
\ee
The existence of nontrivial solutions for $K^{(-1\,,-1)}$ and $W^{(-1\,,-1)}$ requires
\be
(E-4 -\ell) (E - \ell)  \left(1 - \ft12 \tilde{\a} + \ft12 E \tilde{\a}\right)=0\,,
\ee
which leads to 3 infinite towers of propagating degrees of freedom
\be
\Big((E=4+\ell\,,-1)\oplus (E=\ell\,,-1)\oplus  (E=1-\ft{2}{\tilde{\a}}\,,-1)\Big)\otimes(\ell\,,-1)\,.
\ee

 \item{\bf $(|s|\,,|\ell_2|)=(1\,,0)$ sector}\vspace{0.3cm}\\
In this sector we obtain the following conditions
\bsubeq
\bea
&\tilde{\a}X^{(1\,,0)}-\big(4-\tilde{\a}(E-3)\big)H^{(1\,,0)}=0
\,,\\
&\tilde{\a}(E-3)(E+1)H^{(1\,,0)}-\big(4-\tilde{\a}(E+1)\big)X^{(1\,,0)}=0
\,,
\\
&\big(4+\tilde{\a}(E+1)\big)H^{(-1\,,0)}-\tilde{\a}X^{(-1\,,0)}=0\,,
\\
&\tilde{\a}(E-3)(E+1)H^{(-1\,,0)}-\big(4+\tilde{\a}(E-3)\big)X^{(-1\,,0)}
=0
\,.
\eea
\esubeq
Nontrivial solutions for $H^{(\pm1\,,0)}$ and $X^{(\pm1\,,0)}$ exist if
\be
E=1\pm\ft2{\tilde{\a}}\,.
\ee
From the equations of motion of $Z_{\mu}$ and $V_{\mu}$, it can readily be deduced that nonvanishing 
$Z^{(\pm1\,,0)}$ and $ V^{(\pm1\,,0)}$ requires
\be
E=1\pm\ft2{\tilde{\a}}\,.
\ee
In total, this sector contains 8 infinite towers of propagating degrees of freedom
\be
\Big(4\times (E=1\pm\ft{2}{\tilde{\a}}\,,\pm1)\Big)\otimes(\ell\,,0)\,.
\ee
\item{\bf $(s\,,|\ell_2|)=(0\,,2)$ sector}\vspace{0.3cm}\\
We obtain the conditions
\be
(E-2 -\ell) (\ft12 \ell \tilde{\a}+1  )L^{(0\,,2)}=0\,,\quad (E-2 -\ell) (1 -\tilde{\a}-\ft12 \ell \tilde{\a} )L^{(0\,,-2)}=0\,.
\ee
For $L^{(0\,,\pm2)}$ to be nonvanishing, it is necessary to impose
\be
E=2+\ell\,.
\ee
Thus this sector contains 2 infinite towers of propagating degrees of freedom
\be
(E=2+\ell\,,0)\otimes(\ell\,,\pm2)\,.
\ee
 \item{\bf $(s\,,|\ell_2|)=(0\,,1)$ sector}\vspace{0.3cm}\\
 Explicit computation shows that this sector does not contain any dynamical degrees of freedom.

 \item{\bf $(s\,,\ell_2)=(0\,,0)$ sector}\vspace{0.3cm}\\
In this sector we obtain the following conditions
\bsubeq
 \bea
0&=& \big(1+\ft{\tilde{\a}}2\big)
\Big(E(E-2)+3\ell(\ell+2)\Big)H^{(0\,,0)}-3(1+2\tilde{\a})M^{(0\,,0)}-9N^{(0\,,0)}
\non\\
&&
-6\phi^{(0\,,0)}
+\ft32(E+\ell)(E-\ell-2)\tilde{\a}X^{(0\,,0)}\,,
\\
0&=&\Big(24+6E-3E^2+8\ell+4\ell^2+\big(24+6E-3E^2+14\ell+7\ell^2\big)\tilde{\a}\Big)N^{(0\,,0)}
\non\\
&&-\Big(12\ell(\ell+2)+\tilde{\a}\ell(\ell+2)\big(12+2E-E^2+2\ell+\ell^2\big)\Big)U^{(0\,,0)}
\non\\
&&+3(\ell+2)\ell M^{(0\,,0)}+2\ell(\ell+2)\phi^{(0\,,0)}
\,,
\\
0&=&3M^{(0\,,0)}+(1-2\tilde{\a})N^{(0\,,0)}+2\phi^{(0\,,0)}-\ft{\tilde{\a}}2(E+\ell)(E-\ell-2)U^{(0\,,0)}\,,
\\
0&=&(E-3)(E+1)H^{(0\,,0)}
-3\big(M^{(0\,,0)}+N^{(0\,,0)}+\phi^{(0\,,0)}-V^{(0\,,0)}-U^{(0\,,0)}\big)
\,,~~~~~~~~~
\\
0&=&
\ft{\tilde{\a}}2(E+\ell)(E-\ell-2)(E+1)(E-3)H^{(0\,,0)}
-9N^{(0\,,0)}-6\phi^{(0\,,0)}
\non\\
&&
-3\Big((E+\ell)(E-\ell-2)-\ft{\tilde{\a}}2(5E^2-10E-2\ell-\ell^2)\Big)V^{(0\,,0)}
\non\\
&&
+\Big(9-\ft{3\tilde{\a}}{2}\big(12-2E+E^2-2\ell-\ell^2\big)\Big)M^{(0\,,0)}\,,
\\
0&=&
-3M^{(0\,,0)}
-2\phi^{(0\,,0)}
+\Big(3+\ft{\tilde{\a}}2\big(12+2E-E^2+2\ell+\ell^2\big)\Big)N^{(0\,,0)}
\non\\
&&+\Big((E+\ell)(E-\ell-2)+\ft{\tilde{\a}}2\big(E^2-2E-5\ell^2-10\ell\big)\Big)U^{(0,0)}\,.
\eea
\esubeq
After diagnonalizing the equations above, we obtain the following condition
\be
(E-\ell-4)^2(E-\ell)^2(E+\ell-2)^2(E+\ell+2)^2\big(\ft{\tilde{\a}}2E-1\big)
\big(1-\tilde{\a}+\ft{\tilde{\a}}2E\big)\big(\ft{\tilde{\a}}2\ell-1\big)\big(1+\tilde{\a}+\ft{\tilde{\a}}2\ell\big)
=0
\,.
~~~~~~
\ee
Therefore, this sector contains 6  infinite towers of propagating degrees of freedom
\be
\Big(2\times \big(E=4+\ell\,,0\big)\oplus2\times \big(E=\ell\,,0\big)\oplus \big(E=\ft{2}{\tilde{\a}}\,,0\big)\oplus
 \big(E=2-\ft{2}{\tilde{\a}}\,,0\big)\Big)\otimes(\ell\,,0)\,.
\ee
\end{itemize}

We now proceed to arrange the states above into multiplets of ${\rm SU}(1,1|2)$.
The spectrum of the 2-derivative theory \eqref{eha} has been studied before in various works 
\cite{Deger:1998nm,deBoer:1998kjm}.
It contains only the short multiplets of ${\rm SU}(1,1|2)$ dressed by irreducible representations of the extra 
${\rm SL}(2,{\mathbb R})\times {\rm SU}(2)$.
A short multiplet of ${\rm SU}(1,1|2)$ has the structure
\be
(h,j)\,,\quad 2\times (h+\ft12,j-\ft12)\,,\quad (h+1,j-1)\,,\quad h=j\,,
\label{sm}
\ee
where $h$ and $j$ label the representations of the 
${\rm SL}(2,{\mathbb R})\times {\rm SU}(2)$ bosonic subgroup inside ${\rm SU}(1,1|2)$. 
Since the total isometry group associated with the maximally supersymmetric 
${\rm AdS}_3\times {\rm S}^3$ vacuum is ${\rm SU}(1,1|2)\times {\rm SL}(2,{\mathbb R})\times {\rm SU}(2)$, we also
introduce $(\bar{h},\bar{j})$ to label the irreducible representations 
of the extra ${\rm SL}(2,{\mathbb R})\times {\rm SU}(2)$ 
group. Eventually, we denote the short multiplet \eqref{sm} by ${\rm DS}^{(\bar{h},\bar{j})}(h,j)_{\rm S}$. 
The spectrum of the 2-derivative theory \eqref{eha} consists of the following multiplets labeled
by an integer $n\geq0$
\bea
{\rm DS}^{(\ft{n+4}2,\ft{n}2)}(\ft{n}2,\ft{n}2)_{\rm S}&\,,&
\quad {\rm DS}^{(\ft{n+1}2,\ft{n+1}2)}(\ft{n+3}2,\ft{n+3}2)_{\rm S}\,,\quad 
{\rm DS}^{(\ft{n+3}2,\ft{n+3}2)}(\ft{n+1}2,\ft{n+1}2)_{\rm S}\,,\non\\
{\rm DS}^{(\ft{n+4}2,\ft{n}2)}(\ft{n+2}2,\ft{n+2}2)_{\rm S}&\,,&
\quad{\rm DS}^{(\ft{n+2}2,\ft{n+2}2)}(\ft{n+2}2,\ft{n+2}2)_{\rm S}\,,\quad
{\rm DS}^{(\ft{n+4}2,\ft{n}2)}(\ft{n+4}2,\ft{n+4}2)_{\rm S}\, .
\eea
We see from the spectrum of the Einstein-Gauss-Bonnet supergravity that the $\tilde{\alpha}$-independent 
spectrum fits nicely into the multiplets structure above. 
In fact the short multiplets are still present in the EGB theory and are unaffected by $\alpha^\prime$-corrections,
which has to be the case since BPS conditions render these multiplets protected.
Note that we have not studied the fermionic spectrum here. 
However, there seems to be a unique way to arrange the bosonic spectrum into the supermultiplets
which strongly indicates that the fermions should just naturally arrange 
to complete the supermultiplets we have obtained.
We leave the explicit calculation of the fermionic spectrum to future work.

On the top of the infinite tower of short multiplets described above,
the inclusion of the supersymmetric Gauss-Bonnet invariant gives rise 
to 4 new long multiplets possessing the following structure
\bea
(h,j)&\,,&\quad2\times (h+\ft12,j-\ft12)\,,\quad (h+1,j-1)\,,\non\\
(h+\ft12,j+\ft12)&\,,&\quad2\times (h+1,j)\,,\quad (h+\ft32,j-\ft12)\, .
\eea
The 4 long multiplets whose ${\rm AdS}_3$ energies are independent of the KK level are denoted by
\bea
&&{\rm DS}^{(\ft{1}{{\tilde{\alpha}}},\ft{n}2)}(\ft{1}{{\tilde{\alpha}}}+\ft12,\ft{n+1}2)_{\rm L}
\,,\quad 
{\rm DS}^{(1-\ft{1}{{\tilde{\alpha}}},\ft{n}2)}(-\ft{1}{{\tilde{\alpha}}}-1,\ft{n}2)_{\rm L}\,\non\\
&&{\rm DS}^{(1-\ft{1}{{\tilde{\alpha}}},\ft{n}2)}(-\ft{1}{{\tilde{\alpha}}}-\ft12,\ft{n+1}2)_{\rm L}
\,,\quad 
{\rm DS}^{(\ft{1}{{\tilde{\alpha}}},\ft{n}2)}(\ft{1}{{\tilde{\alpha}}},\ft{n}2)_{\rm L}\,,\quad n\ge0
~.
\eea
For a ${\rm SU}(1,1|2)$ long multiplet, $h$ is not equal to $j$ and unitarity requires $h>j$. 

In summary, (8.50) and (8.52) comprise the full
spectrum of the Einstein-Gauss-Bonnet supergravity about the supersymmetric AdS$_3\times$S$_3$ vacuum, where it 
is understood that states with negative SU(2) label should be removed. The massless states are the usual graviton and a massless scalar
residing
in the two multiplets
\begin{equation}
{\rm DS}^{2,0}(0,0)_{\rm S}\,,\qquad {\rm DS}^{1,1}(1,1)_{\rm S}\,,
\end{equation}
consistent with the results of \cite{Deger:1998nm}. 
Although gauge fields in AdS$_{3}$ do not have local degrees of freedom, we prefer to keep the massless graviton in the spectrum as it carries
information about the non-trivial boundary dynamics. 
The fact that the $\alpha'$ dependent states can be nicely arranged into 
${\rm SU}(1,1|2)$ long multiplets provides a further check of the supersymmetry of the GB invariant.


\section{Conclusion}
\label{conclusion}

In this paper we have described the supersymmetric completion of several curvature-squared invariants for 
${\cal N}=(1,0)$ supergravity in six dimensions. 
By employing the dilaton-Weyl multiplet 
 of conformal supergravity
we have described  supersymmetric completions for the three possible purely 
gravitational curvature-squared terms, Riemann, Ricci, and scalar 
curvature-squared, where in the last case a coupling to a conformal compensator, 
which we have chosen to be a linear multiplet, was necessary.

We also constructed a novel locally superconformal invariant based on a higher-derivative action
for the linear multiplet which can be defined both in the standard-Weyl or dilaton-Weyl multiplet for 
conformal supergravity leading to new classes of curvature-squared terms  in both cases. 
In the case of the dilaton-Weyl multiplet, the new invariant leads to the Lagrangian \eqref{inv-7_dilatonWeyl},
which includes Ricci and scalar curvature-squared terms together with a nontrivial 
dependence on the dilaton field, as for instance an overall multiplicative factor of $\re^{-v}$,
which clearly distinguishes 
this invariant from the other three invariants described in section \ref{main-curvature-squared}.

To our knowledge, our analysis of curvature-squared invariants for 
$\cN=(1,0)$ supergravity in six dimensions is the most complete to date and it 
has already allowed us to study some
interesting applications of these results.
For instance, by extending the results presented in \cite{NOPT-M17}, 
we described  the Gauss-Bonnet invariant in detail and added to it 
the off-shell supersymmetric extension of the Einstein-Hilbert term 
to obtain the Einstein-Gauss-Bonnet supergravity, 
which plays a central role in the effective low-energy description of $\alpha^\prime$-corrected 
string theory compactified to six dimensions.
We gave the supersymmetry transformations for the on-shell 
 Einstein-Gauss-Bonnet supergravity  for the first time, 
 showing that up to cubic fermion terms there is no $\alpha^\prime$ correction.
 Moreover, we provided a detailed analysis of the spectrum about the 
${\rm AdS}_3\times {\rm S}^3$ solution which is relevant to holographic studies.

As an application of the new invariant described in section \ref{new-R2},
 we have shown how a linear combination of such an invariant and the supersymmetric Einstein-Hilbert term 
leads to a dynamically generated cosmological constant and non-supersymmetric (A)dS$_6$ solutions.
This result was based on the fact that in a standard-Weyl multiplet there exist $D^2$ terms in the action
for the auxiliary field $D$. 
Such a term is remarkable since, unlike the pure Einstein-Hilbert supergravity, the 
equation of motion for $D$ is consistent even in a standard-Weyl multiplet background and 
remains the case when coupled to the supersymmetric Einstein-Hilbert term.
Moreover, $D$ remains an auxiliary field even in the higher-derivative theory
and can be algebraically integrated out leading to a cosmological constant term on-shell.
To underline the importance of this mechanism in regards to the cosmological constant, 
we should stress that, to our knowledge, no supersymmetric pure cosmological constant term 
has ever been constructed in the literature for 6D $\cN=(1,0)$ minimal supergravity.

We believe that our studies open the avenue for various generalizations and, as already pointed out in the 
introduction, we expect the results in our paper might  find several applications. Let us now briefly comment
on some of these possibilities.

The fact that all the invariants presented in this paper possess manifest off-shell supersymmetry
makes it trivial to add them to other known off-shell models and matter couplings.
For instance, it is straightforward to consider vector multiplets coupled to the higher-derivative supergravity invariants
in our paper.
Since in our paper we have focused on Poincar\'e supergravity models 
based on the use of a compensating linear multiplet, which preserves a U(1)$_R$ symmetry off-shell, 
it is clear that in general the addition of a coupling between the vector multiplets and the linear multiplet 
\cite{BSVanP} on-shell 
will generate a potential for the dilaton field together with a gauged $R$-symmetry exactly as 
in \cite{Bergshoeff:2012ax}.
It was noted in \cite{Bergshoeff:2012ax} that the simple case of  the gauged minimal 6D supergravity,
where only one vector multiplet is coupled to the minimal ungauged supergravity (reviewed in section \ref{EH}), 
leads to a version of the Salam-Sezgin model.
It is straightforward to show that the supersymmetric ${\rm Minkowski}_4\times {\rm S}^2$ solution
is preserved also when any of the three curvature-squared invariants of section \ref{main-curvature-squared},
including the Gauss-Bonnet one, is added to the two-derivative gauged model but
it would be interesting to study the stability properties of this background in the presence of the higher-derivative
terms. For instance, the perturbative stability of the Salam-Sezgin model extended by the Riemann squared invariant 
was 
studied in \cite{Pang:2012xs}, where it was found that the inclusion of the supersymmetric Riemann squared term 
introduces 
tachyonic modes around the supersymmetric ${\rm Minkowski}_4\times {\rm S}^2$ solution. We expect that this will 
not be 
the case if the Gauss-Bonnet invariant instead of the Riemann squared invariant 
is added to the original Salam-Sezgin model.

More generally, it would be of interest to extend the analysis of our paper by considering the higher-derivative 
dynamics of general systems of vector multiplets, hypermultiplets, and tensor multiplets coupled to
conformal supergravity. For instance, a fully $\alpha^\prime$-corrected description of the 
gauged supergravity of the Salam-Sezgin model might include many others curvature-squared terms including
 $F^4$ interactions. It is then natural to seek off-shell $(1,0)$ supersymmetric extensions 
 of four-derivative terms for vector multiplets by using our techniques and extending the 5D analysis of
\cite{Ozkan:2016csy}.
To study general couplings, including higher-derivative ones, of off-shell hypermultiplets 
to conformal supergravity one could use the projective superspace approach of \cite{LT-M12},
based on the formalism developed in the last decade for supergravity theories with eight supercharges in 
$2\leq {\rm D}\leq 5$, see, e.g., 
\cite{ProjectiveSugra5D,Butter:2014xxa,ProjectiveSugra4D,KLRT-M2008,ProjectiveSugra2D,ProjectiveSugra3D,Projective-Dan-1,Projective-Dan-2,Butter:2012xg}
for relevant papers on the subject.
On the other hand, the off-shell description of a general system of $(1,0)$ tensor multiplets is less developed
and would be a very interesting avenue of research.
There have been proposals for an off-shell extension of the tensor multiplets which, 
similarly to the off-shell 6D charged hypermultiplets of \cite{LT-M12},
includes an infinite number of auxiliary fields, see \cite{SokatchevAA} and more recently \cite{LT-M12}, 
which could be used to construct general higher-derivative interactions for the tensor multiplets in a standard-Weyl 
multiplet background for conformal supergravity. 
It is natural to then wonder if and how one could use these multiplets to describe general higher-derivative
interactions, including the curvature-squared ones, based solely on a standard-Weyl multiplet and not the
 dilaton-Weyl used
for the invariants in section \ref{main-curvature-squared}.
In this framework it would be intriguing to verify whether the mechanisms of consistency of the dynamics of $D$ 
and for dynamical generation of a cosmological constant described in section \ref{new-R2} 
remain general features of curvature-squared models in the standard-Weyl multiplet.

Compactifications of the 6D GB invariant to 5 and 4 dimensions are also potentially interesting for various purposes. 
First of the all, the dimensional reduction leads to lower dimensional GB invariants coupled to matter multiplets,
which can be viewed as supersymmetrizaton of a particular class 
of the Horndeski scalar-tensor model widely studied in cosmology. 
The matter multiplets can be consistently truncated out preserving off-shell supersymmetry and the resulting 
invariants based 
solely on the off-shell supergravity multiplet can be compared to the existing supersymmetrization of GB invariants 
in 5D 
\cite{OP132} and 4D \cite{Butter:2013lta}. In the 4D STU model, \cite{DLR} proposed a string-string-string duality 
based on 
the properties of the 2-derivative Lagrangian. 
In the context of string theory, this duality should persist to all order in 
the $\alpha^\prime$ expansion. 
A first nontrivial check can be performed using the 2-torus reduction of the 6D Einstein-Gauss-Bonnet invariant. 

Finally, we recall that the 6D two-derivative supergravity admits supersymmetric black strings and black rings with 
AdS$_3$ 
near horizon geometry. Using the results presented in this paper, one can extend those results to include the leading 
higher-derivative corrections. 
Upon circle reductions, the 6D solutions give rise to black holes, black strings and black rings in 5D supergravities. 
The recent work \cite{Bonetti:2018lfb} 
has made an attempt to systematically analyze the geometry of black hole solutions in the 
presence of supersymmetric curvature squared terms. Their work utilized off-shell curvature squared
 invariants based on the 
standard-Weyl multiplet which are different from the curvature squared
 invariants coming from the reduction of the 6D invariants. 
However, on physical grounds, the different formulations of the curvature squared
 invariants should yield the same physical quantities for 
the same solution. Therefore, it might be worthwhile to carry out this comparison in more detail.  Due to the close 
relation 
between 6D and 5D, one can also look for the 6D analog of the non-renormalisation theorem for 6D black string 
entropy. For 
5D black string with ${\rm AdS}_3$ near horizon geometry, 
the non-renormalisation theorem ensures that the entropy of such objects 
does not receive corrections from terms with more than 4-derivatives 
(see \cite{Castro:2008ne} for a review  and references therein). 
Eventually, for a better understanding of the microstates underlying the black strings, it is indispensable to embed
the 6D solution into 10D string theory. In fact,  such an embedding is not unique and may lead to interesting physics 
regarding the apparent different 10D descriptions of the same 6D solutions. 
We leave these interesting problems for future investigation.

\acknowledgments
\noindent
We are grateful to Sylverster James Gates Jr., Fridrik Freyr Gautason, 
Pietro Antonio Grassi, Stefanos Katmadas,
William D. Linch III, Ergin Sezgin, Antoine Van Proeyen, and Thomas Van Riet
for discussions.
The work of DB was partially   supported by NSF under grants
PHY-1521099 and PHY-1620742 and the Mitchell Institute for Fundamental Physics and
Astronomy at Texas A\&M University.
JN was supported by the GIF, German-Israeli Foundation, for Scientific Research and Development 
and a Humboldt research fellowship of the Alexander von Humboldt Foundation.
The work of MO is supported in part by TUBITAK grant 118F091.
YP was supported by an Alexander von Humboldt fellowship and
by a Newton International Fellowship of the UK Royal Society.
The work of GT-M was supported by the
Interuniversity Attraction Poles Programme initiated by the Belgian Science Policy (P7/37),
 by the COST Action MP1210,
by the KU Leuven C1 grant ZKD1118 C16/16/005,
by the Albert Einstein Center for Fundamental Physics, University of Bern,
and by the Australian Research Council (ARC) Future Fellowship FT180100353.
GT-M thanks the Arnold-Regge Center and the Theory Group of the University of Turin,
the Mitchell Institute at Texas A\&M, the High Energy Theory Group at Brown University,
the High Energy Theory Group at the University of Vienna,
and the Theory Group of the University of Milan
for kind hospitality and support during part of this project.


\appendix


\section{Conformal superspace identities}
\label{conformal-identities}

In this appendix we collect results about conformal superspace in the traceless frame of \cite{BNT-M17}
that are of importance for this paper.

The Lorentz generators act on the superspace covariant derivatives $\de_A=(\de_a,\de_\a^i)$ as\bsubeq \label{SCA}
\begin{align}
[M_{ab} , M_{cd}] &= 2 \eta_{c[a} M_{b] d} - 2 \eta_{d [a} M_{b] c} \ , \\
[M_{ab} ,  {\nabla}_c ] &= 2 \eta_{c [a}  {\nabla}_{b]} \ , \\
 [M_\a{}^\b , \nabla_\g^k] &= - \d_\g^\b \nabla_\a^k + \frac{1}{4} \d^\b_\a \nabla_\g^k
~,
\end{align}
where $M_\a{}^\b=-\frac{1}{4}(\g^{ab})_\a{}^\b M_{ab}$.
The ${\rm SU(2)}_R$ and dilatation generators obey
\begin{align}
[J^{ij}, J^{kl}] &= \eps^{k(i} J^{j)l} + \eps^{l(i} J^{j)k} \ , \quad [J^{ij} , \nabla_\a^k] = \eps^{k(i} \nabla_\a^{j)} \ ,  \\
[\mathbb D ,  {\nabla}_a] &=  {\nabla}_a \ , \quad [\mathbb D , \nabla_\a^i] = \hf \nabla_\a^i \ .
\end{align}
The Lorentz and ${\rm SU(2)}_R$
generators
act
on the special conformal generators $K^A=(K^a,S^\a_i)$ as
\begin{align}
[M_{ab} , K^c] = 2 \d^c_{[a} K_{b]} \ , \quad
  [M_\a{}^\b , S^\g_k] = \d^\g_\a S^\b_k - \frac{1}{4} \d^\b_\a S^\g_k
\ , \quad
[J^{ij} , S^\g_k] = \d_k^{(i} S^{\g j)} \ ,
\end{align}
while the dilatation generator acts on  $K^A$ as
\begin{align}
[\mathbb D , K^a] = - K^a \ , \quad [\mathbb D, S^\a_i] &= - \hf S^\a_i \ .
\end{align}
Among themselves, the generators $K_A$ obey the only nontrivial anti-commutation relation
\begin{align}
\{ S^\a_i , S^\b_j \} = - 2 \ri \eps_{ij} (\tilde{\g}_c)^{\a\b} K^c \ .
\end{align}
The algebra of $K^A$ with ${\nabla}_A$ is given by
\bea
{[}K_a ,  {\nabla}_b{]} &=& 2 \eta_{ab} \mathbb D + 2 M_{ab} \ , \\
{[}K^a , \nabla_\a^i {]} &=& - \ri (\g^a)_{\a\b} S^{\b i} \ , \\
\{ S^\a_i , \nabla_\b^j \} &=& 2 \d^\a_\b \d^j_i \mathbb D - 4 \d^j_i M_\b{}^\a + 8 \d^\a_\b J_i{}^j \ , \\
{[}S^\a_i ,  {\nabla}_b{]}
&=&
 -\ri (\tilde{\g}_b)^{\a\b}  {\de}_{\b i}
+\frac{1}{10}W_{bcd}(\tilde{\g}^{cd})^\a{}_\g
S^\g_i
-\frac{1}{4}X^{\a}_{ i}K_b
\non\\
&&
+\Big{[}
\frac{1}{4}(\tilde{\g}_{bc})^{\a}{}_{\b} X^{\b}_{ i}
+\frac{1}{2}({\g}_{bc})_\b{}^{\g}X_{\g i}{}^{\b\a} \Big{]}K^c
~.
\label{Snabla}
\eea
\esubeq

The anticommutator of two spinor derivatives,
$\{{\de}_\a^i,{\de}_\b^j\}$, has the following non-zero
 torsion and curvatures\bsubeq\label{new-frame_spinor-spinor}
\bea
 {T}_{\a}^i{}_\b^j{}^c
&=&
2\ri\ve^{ij}(\g^c)_{\a\b}
~,
\\
 {R}(M)_{\a}^i{}_\b^j{}^{cd}
&=&
4\ri\ve^{ij}(\g_a)_{\a\b} W^{acd}
~,
\\
 {R}(S)_{\a}^i{}_\b^j{}_\g^k
&=&
-\frac{3}{2}\ve^{ij}\ve_{\a\b\g\d} X^{\d k}
~,
\\
 {R}(K)_{\a}^i{}_\b^j{}_c
&=&
\ri\ve^{ij}(\g^a)_{\a\b}\left(
\frac{1}{4} \eta_{ac}Y
- {\de}^b W_{a b c}
+ W_{a}{}^{ef} W_{cef}
\right)
~.
\eea
\esubeq
The non-zero torsion and curvatures in the commutator ${[} {\de}_a,{\de}_\b^{j}{]}$ are:
\bsubeq\label{new-frame_vector-spinor}
\bea
 {T}_{a}{}_\b^j{}^\g_k
&=&
-\frac{1}{2} (\g_{a})_{\b \d} W^{\d \g}\d^j_k
~,
\\
 {R}(\mathbb D)_a{}_\b^j
&=&
- \frac{\ri}{2} (\gamma_{a})_{\b\g} X^{\g j}
~,
\\
 {R}(M)_a{}_\b^j{}^{cd}
&=&
\ri \delta_{a}^{[c} (\g^{d]})_{\b\g} X^{\g j}
-\ri(\gamma_{a}{}^{ c d})_{\g \d} X_{\b}^j{}^{\g \d}
+2\ri(\gamma_{a})_{\b\g} (\gamma^{c d})_{\d}{}^{\r}X_{\r}^j{}^{\gamma \delta }
~,
\\
 {R}(J)_a{}_\b^j{}^{kl}
&=&
2 \ri (\gamma_{a})_{\b\g} X^{\g (k}\ve^{l)j}
~,
\\
 {R}(S)_a{}_\b^j{}_\g^k
&=&
- \frac{\ri}{4} (\gamma_{a})_{ \b \d} \, Y_{\g}{}^{\d}{}^{ jk}
+\frac{3\ri}{20}  (\gamma_{a})_{\g\d} Y_{\b}{}^{\d}{}^{ jk}
- \frac{\ri}{8} (\gamma_{a})_{\b \d}  {\de}_{\g \r}{W^{\d \r}}  \veps^{jk}
\non\\
&&
+ \frac{\ri}{40}(\gamma_{a})_{\g\d}  {\de}_{\b\r}{W^{\d \r}}\veps^{jk}
- \frac{\ri}{8} (\gamma_{a})_{\d \epsilon}\, \veps_{\b \r \t \g}\,W^{\d \r} W^{\epsilon \t} \veps^{jk}
~, \\
 {R}(K)_a{}_\b^j{}_c
&=&
\frac{\ri}{4} (\gamma_{c})_{\b \g}  {\de}_{a}{X^{\g j}}
- \frac{\ri}{4} (\gamma_{a c d})_{\g \d}  {\de}^{d}{X_{\b}^j{}^{\g \d }}
+ \frac{\ri}{3} (\gamma_{a})_{\b \d} (\gamma_{c d})_{\r}{}^{\g} {\de}^{d}{X_{\g}^j{}^{\d \r }}
\non\\
&&
- \frac{\ri}{8} (\gamma_{a})_{\b \g} (\gamma_{c})_{\d \r}W^{\g \d} X^{\r j}
+ \frac{5\ri}{12} (\gamma_{a})_{\b\r}(\gamma_{c})_{\g \epsilon} W^{\g \d} X_{\d}^j{}^{\r \epsilon }
\non\\
&&
+ \frac{\ri}{4} ( \gamma_{a})_{\g \r} (\gamma_{c})_{\b \epsilon} 	W^{\g \d} X_{\d}^j{}^{\r \epsilon }
- \frac{\ri}{2}(\gamma_{a})_{\g \r} (\gamma_{c})_{ \d \epsilon}W^{\g\d}X_{\b}^j{}^{\r\epsilon }
~.
\eea
\esubeq
The commutator of two vector derivatives,
$[ {\de}_a, {\de}_b]$,
has the following non-vanishing torsion and curvatures:
\bsubeq\label{new-frame_vector-vector}
\bea
 {T}_{ab}{}^\gamma_k
&=&
(\g_{ab})_\b{}^\a X_{\a k}{}^{\b\g}
~,
\label{new-frame_ALL-T-c}
\\
 {R}(M)_{a b}{}^{c d} &=&
Y_{a b}{}^{cd}
=\frac{1}{4}(\g_{ab})_\g{}^\a(\g^{cd})_\d{}^\b Y_{\a\b}{}^{\g\d}
~,
\label{new-frame_ALL-R-i}
\\
 {R}(J)_{ab}{}^{kl}
&=&
\frac{1}{2}(\g_{ab})_\d{}^\g Y_\g{}^\d{}^{kl} = Y_{a b}{}^{kl}
~,
\label{new-frame_ALL-R-J}
\\
 {R}(S)_{ab}{}_\gamma^k
&=&
- \frac{\ri}{3} (\gamma_{a b})_{\delta}{}^{\alpha}  {\de}_{\gamma \beta} X_{\alpha}^k{}^{\beta \delta }
- \frac{\ri}{6} (\gamma_{a b c})_{\alpha\beta}  {\de}^c X_{\gamma}^k{}^{\alpha \beta }
- \frac{\ri}{6} \veps_{\gamma \beta \epsilon \rho} (\gamma_{a b})_{\delta}{}^\rho
W^{\alpha \beta} X_{\alpha}^k{}^{\delta \epsilon}
~,~~~~~~~~~~~~
\label{new-frame_ALL-R-k}
\\
 {R}(K)_{ab}{}_{c}
&=&
\frac{1}{4}  {\de}^d Y_{a b c d}
+\frac{\ri}{3} X_{\alpha}^k{}^{\beta \gamma } X_{\beta k}{}^{\alpha \delta}(\gamma_{a b c})_{\gamma\delta}
+\ri (\gamma_{a b})_{\epsilon}{}^\alpha
(\gamma_{c})_{\gamma\delta} X_{\alpha}^k{}^{\beta \gamma } X_{\beta k}{}^{\delta \epsilon}
\non\\
&&
+\frac{\ri}{4}X^{\alpha k} X_{\beta k}{}^{\gamma \delta}
(\gamma_{a b})_{\gamma}{}^{\beta} (\gamma_{c})_{\alpha\delta}
~.
\label{new-frame_ALL-R-l}
\eea
\esubeq
Remember that the descendant superfields
$X^{\a i}$, $X_{\a i}{}^{\b\g}$, $Y$, $Y_\a{}^\b{}^{kl}$ (and equivalently $Y_{ab}{}^{kl}$),
$Y_{\a\b}{}^{\g\d}$ (and equivalently $Y_{ab}{}^{cd}$), were defined in
\eqref{Xfields} and \eqref{YfielDs}.

By using \eqref{WBI} and the previous definitions,
one can derive the following relations for the descendant superfields of the super-Weyl tensor:
\bsubeq \label{eq:Wdervs}
\bea
\nabla_\a^i X^{\b j}
&=&
- \frac{2}{5} Y_\a{}^\b{}^{ij}
- \frac{2 }{5} \eps^{ij}  {\nabla}_{\a \g} W^{\g\b}
- \hf \eps^{ij} \d_\a^\b Y
\ , \\
\nabla_\a^i X_\b^j{}^{\g\d}
&=&
\hf \d^{(\g}_\a Y_\b{}^{\d)}{}^{ij} - \frac{1}{10} \d_\b^{(\g} Y_\a{}^{\d)}{}^{ij}
- \hf \eps^{ij} Y_{\a\b}{}^{\g\d}
- \frac{1}{4}\eps^{ij}  {\nabla}_{\a\b} W^{\g\d}
\non\\
&&
+ \frac{3}{20} \eps^{ij} \d_\b^{(\g}  {\nabla}_{\a \r} W^{\d) \r}
-\frac{1}{4}\eps^{ij} \d^{(\g}_\a  {\nabla}_{\b \r} W^{\d) \r}
\ , \\
\nabla_\a^i Y
&=&
- 2 \ri  {\nabla}_{\a \b} X^{\b i}
\ , \\
\nabla_\g^k Y_{\a}{}^\b{}^{ij}
&=&
\frac{2}{3} \eps^{k (i} \Big(
- 8 \ri\,\de_{\g\d}{X^{j)}_{\alpha}{}^{\beta \delta}}
- 4 \ri\, \de_{\a\d}{X^{j)}_{\gamma}{}^{\beta \delta}}
+ 3 \ri \,\de_{\g\a}{X^{\beta j)}}
+ 3 \ri\,\delta^{\beta}_{\gamma}\,\de_{\a\d}{X^{\delta j)}}
\non\\ &&~~~~~~~~~
- \frac{3\ri}{2} \delta_{\alpha}^{\beta} \,\de_{\g\d}{X^{\delta j)}}
- 3 \ri \,\veps_{\alpha \gamma \delta \epsilon} W^{\beta \delta} X^{\epsilon j)}
+ 4 \ri\, \veps_{\alpha \gamma \epsilon \rho} \,W^{\delta \epsilon} X^{j)}_{\delta}{}^{\beta \rho}
\Big)\ , \\
\nabla_\e^l Y_{\a\b}{}^{\g\d} &=&
- 4 \ri \,\de_{\e(\a}{X_{\beta)}^l{}^{\gamma\delta}}
+ \frac{4\ri}{3} \delta_{(\alpha}^{(\gamma} \de_{\b)\rho}{X_{\epsilon}^l{}^{\delta) \rho}}
+ \frac{8\ri}{3} \delta_{(\alpha}^{(\gamma} \de_{|\epsilon\rho|}{X_{\beta)}^l{}^{\delta) \rho}}
+ 8 \ri\, \delta_{\epsilon}^{(\gamma} \de_{\rho(\alpha}{X_{\beta)}^l{}^{\delta) \rho}}
\non\\ &&
- \frac{4\ri}{3} \, W^{\rho \sigma} \,
	\delta_{(\alpha}^{(\gamma} \veps_{\beta) \epsilon \sigma \tau} X_{\rho}^l{}^{\delta) \tau}
- 8 \ri\, \veps_{\epsilon \rho \sigma (\alpha} W^{\rho (\gamma} X_{\beta)}^l{}^{\delta) \sigma}~.
\eea
\esubeq
These relations define the $Q$-supersymmetry transformations of the descendant superfields of the super-Weyl
tensor.
Their $S$-supersymmetry transformations are instead given by the following relations \cite{BKNT16}:
\bsubeq\label{S-on-X_Y}
\bea
S^\a_i X^{\b j} &=& \frac{8\ri}{5} \d_i^j W^{\a\b} ~,
~~~~~~
S^\a_i X_\b^j{}{}^{\g\d}
=
-\ri\d^j_i \d^\a_\b W^{\g\d}
+\frac{2\ri}{5} \d^j_i \d^{(\g}_{\b} W^{\d) \a}
~,
\label{S-on-X_Y-a}
\\
S^\g_k Y_\a{}^{\b}{}^{ij}
&=&
-\d^{(i}_k \left(
16 X_\a^{j)}{}^{\g\b}
- 2 \d_\a^\b X^{\g j)}
+ 8 \d^\g_\a X^{\b j)} \right)
~,
\label{S-on-X_Y-b}
\\
S^\r_j Y_{\a\b}{}^{\g\d}
&=&
24 \left( \d^\r_{(\a} X_{\b) j}{}^{\g\d}
- \frac{1}{3} \d^{(\g}_{(\a} X_{\b) j}{}^{\d) \r} \right)
~ ,~~~~~~
S^\a_i Y = - 4 X^\a_i
~ .
\label{S-on-X_Y-c}
\eea
\esubeq
The descendant superfields also satisfy the following indentities
\bsubeq\label{deaBI}
\bea
 {\de}^{\delta (\alpha} X_{\delta}^i{}^{\beta \gamma)}
&=&
W^{\delta (\alpha} X_{\delta}^i{}^{\beta \gamma)}
~,
\\
 {\de}_{\gamma (\alpha} Y_{\beta)}{}^{\gamma i j}
&=&
0~, \qquad
 {\de}^{\gamma (\alpha} Y_{\gamma}{}^{\beta) i j}
=
8\ri X^{\gamma (i} X_{\gamma}^{ j)}{}^{\alpha \beta}
~,
\\
 {\de}_{\delta (\alpha} Y_{\beta \gamma)}{}^{\delta \epsilon}
&=&
0~,
\qquad
 {\de}^{\delta (\alpha} Y_{\delta \epsilon}{}^{\beta \gamma)} =
24\ri X_{\epsilon}^k{}^{\delta (\alpha } X_{\delta k}{}^{\beta \gamma)}
- 8\ri X_{\rho}^k{}^{\delta (\alpha} \delta_{\epsilon}^{\beta} X_{\delta k}{}^{\gamma) \rho}
~.
\eea
\esubeq

We conclude by underlining that, compared to the frame chosen in  \cite{BKNT16},
 the superspace geometry in the traceless frame described in this appendix
is simply given by the following redefinition of the vector derivative
\bea
\de_a
&\to&
 \nabla_a
-  W_{a}{}^{b c} M_{bc}
+ \frac{3\ri}{8} (\gamma_a)_{\alpha \beta} X^{\alpha j} S^{\beta}_j
- \frac{1}{8} YK_a
+\hf \de^b W_{a b c} K^c
-\hf W_{a}{}^{ e f} W_{e f c} K^c
~.~~~~~~
\label{new-vector-derivatives}
\eea
Here on the left hand side $\de_a$ is the vector derivative of
\cite{BKNT16} while
the operator on the right hand side
is the vector covariant derivatives in the traceless frame of \cite{BNT-M17}, which we used everywhere in the paper,
defined in terms of the vector derivative of \cite{BKNT16}.


\section{Useful descendant components of the composite gauge 3-form \eqref{C2gauge3form}}
\label{descendants-C2gauge3form}

For the composite gauge 3-form \eqref{C2gauge3form}, in this appendix we collect the descendant components
relevant for the derivation of the invariant \eqref{newInvariant}. They are
\bsubeq\label{descendants-C2gauge3form-equations}
\bea
&&\Lambda_{\alpha a}{}^i
=
\tfrac{4}{3}(\gamma_{a})_{ \alpha \beta}
\big( \tfrac{2}{15} Y_\gamma{}^{\beta i j} {\chi}^\gamma_{j}
 - X_{\gamma j}{}^{\delta \beta}  Y_{\delta}{}^{\gamma i j} \big)
+4 (\gamma_{a})_{\beta \gamma} \big(
X_{\alpha j}{}^{\delta \beta}  Y_{\delta}{}^{\gamma i j}
- \tfrac{1}{15}  Y_{\alpha}{}^{\beta i j} {\chi}^\gamma_{j}
\big)
\non\\
&&~~~
+ \tfrac{2}{45}(\gamma_{a})_{ \alpha \beta}  D {\chi}^{\beta i}
+ 8(\gamma_{a})_{ \beta \gamma}   X_{\delta}^{i}{}^{\epsilon \beta}  Y_{\alpha \epsilon}{}^{\delta \gamma}
+2(\gamma_{a})_{\beta \gamma}
\big(
\de_{\delta \epsilon}{W^{\beta \delta}}   X_{\alpha}^{i}{}^{\gamma \epsilon}
-\tfrac{4}{3} W^{\beta \delta} \de_{\delta \epsilon}{ X_{\alpha}^{i}{}^{\gamma \epsilon}}
\big)
\non\\
&&~~~
+2 (\gamma_{a})_{\alpha \beta}
\big(
\de_{\gamma \delta}{W^{\epsilon \gamma}}   X_{\epsilon}^{i}{}^{\beta \delta}
 - \tfrac{2}{15}W^{\beta \gamma} \de_{\gamma \delta}{{\chi}^{\delta}{}^{i}}
 \big)
\non\\
&&~~~
+4 (\gamma_{a})_{\beta \gamma}
\big(
 \tfrac{1}{15} \de_{\alpha \delta}({W^{\beta \delta}}  {\chi}^\gamma{}^{i})
- \tfrac{4}{3} W^{\epsilon \beta} \de_{\alpha \delta}{ X_{\epsilon}^{i}{}^{\gamma \delta}}
-  \de_{\alpha \delta}{W^{\epsilon \beta}}   X_{\epsilon}^{i}{}^{\gamma \delta}
\big)
\non\\
&&~~~
+ \tfrac{4}{3} (\gamma_{a})_{\beta \gamma} \veps_{\alpha \delta \epsilon \rho}
\big(
2W^{\sigma \delta} W^{\beta \epsilon}  X_{\sigma}^{i}{}^{\gamma \rho}
+ \tfrac{1}{5} W^{\beta \delta} W^{\gamma \epsilon} {\chi}^{\rho}{}^{i}
\big)
~,
\eea
and
\bea
&&C_{a b}
=
(\gamma_{b})_{\alpha \beta} \big(
\tfrac{1}{8}( \gamma_{a c d})_{ \gamma \delta}\de^{c}W^{\alpha \gamma} \de^{d}W^{\beta \delta}
- \tfrac{2}{3}(\gamma_{a c})_{ \epsilon}{}^{\delta}  W^{\gamma \alpha} \de^{c} Y_{\gamma \delta}{}^{\beta \epsilon}
-(\gamma_{a c})_{\epsilon}{}^{\delta} \de^{c}W^{\gamma \alpha} Y_{\gamma \delta}{}^{\beta \epsilon}
 \big)
 \non\\
&&~~~
+\tfrac{1}{2} (\gamma_{c})_{\alpha \beta} \big(
(\gamma_{a b})_{ \epsilon}{}^{\delta}\de^{c}W^{\gamma \alpha}Y_{\gamma \delta}{}^{\beta \epsilon}
 -(\gamma_{a b d})_{\gamma \delta}W^{\alpha \gamma} \de^{c}\de^{d}W^{\beta \delta}
 + \tfrac{3}{4}(\gamma_{a b d})_{\gamma \delta}  \de^{c}W^{\alpha \gamma} \de^{d}W^{\beta \delta}
 \big)
 \non\\
&&~~~
+4\ri( \gamma_{a b c})_{ \gamma \delta} \de^{c} X_{\alpha i}{}^{\beta \gamma} X_{\beta}^{i}{}^{\alpha \delta}
- \tfrac{16\ri}{45}(\gamma_{a b c})_{ \beta \gamma}  \de^{c} X_{\alpha i}{}^{\beta \gamma}\chi^{\alpha}{}^{i}
- \tfrac{4\ri}{15}(\gamma_{a b c})_{\beta \gamma} \de^{c}{\chi}^{\alpha}_{i}   X_{\alpha}^i{}^{\beta \gamma}
\non\\
&&~~~
+ \tfrac{14\ri}{225}(\gamma_{a b c})_{ \alpha \beta}\de^{c}{\chi}^{\alpha}_{i}\chi^{\beta}{}^{i}
+ \tfrac{1}{45}\eta_{ab} D^2
+ \tfrac{1}{15} (\gamma_{a b c})_{ \alpha \beta} \de^{c}(D W^{\alpha \beta})
\non\\
&&~~~
+\tfrac{1}{2}\eta_{ab}
\big(
 \tfrac{1}{4} \de_{\alpha \beta}{W^{\alpha \gamma}} \de_{\gamma \delta}{W^{\beta \delta}}
 -  W^{\alpha \gamma} \de_{\alpha \beta}{\de_{\gamma \delta}{W^{\beta \delta}}}
  \big)
  \non\\
&&~~~
+(\gamma_{(a})_{ \alpha \beta}
\big(
W^{\alpha \gamma} \de_{b)}{\de_{\gamma \delta}{W^{\beta \delta}}}
 + \tfrac{1}{4}\de_{b)}{W^{\alpha \gamma}}\de_{\gamma \delta}{W^{\beta \delta}}
\big)
\non\\
&&~~~
+(\gamma_{a})_{\alpha \beta} (\gamma_{b})_{\gamma \delta}
\big(
 - \tfrac{1}{2} W^{\alpha \gamma} \Box {W^{\beta \delta}}
 - \tfrac{5}{8}\de^{c}{W^{\alpha \gamma}}  \de_{c}{W^{\beta \delta}}
+ \tfrac{2}{15} D W^{\alpha \gamma} W^{\beta \delta}
+ W^{\epsilon \rho} W^{\alpha \gamma}  Y_{\epsilon \rho}{}^{\beta \delta}
\big)
\non\\
&&~~~
- (\tilde{\gamma}_{a})^{\gamma \delta}(\gamma_{b})_{\alpha \beta}
Y_{\epsilon \gamma}{}^{\rho \alpha}  Y_{\rho \delta}{}^{\epsilon \beta}
+(\gamma_{a})_{ \alpha \beta}
\big(
4\ri\de_{b}{ X_{\gamma i}{}^{\delta \alpha}}   X_{\delta}^{i}{}^{\gamma \beta}
+ \tfrac{2\ri}{75}\de_{b}{{\chi}^{\alpha}_{i}}  {\chi}^{\beta}{}^{i}
\big)
\non\\
&&~~~
+(\gamma^{c})_{\alpha \beta}
\big(
 - 8\ri\eta_{ab}\de_{c}{ X_{\gamma i}{}^{\delta \alpha}}   X_{\delta}^{i}{}^{\gamma \beta}
- 4\ri(\gamma_{a b})_{\epsilon}{}^\gamma \de_{c}{ X_{\gamma i}{}^{\delta \alpha}}X_{\delta}^{i}{}^{\beta \epsilon}
+ \tfrac{8\ri}{45}(\gamma_{ab})_{ \delta}{}^\gamma \de_{c}{ X_{\gamma i}{}^{\alpha \delta}}  {\chi}^{\beta}{}^{i}
\non\\
&&~~~~~~~~~~~~~~~~
- \tfrac{4\ri}{15}(\gamma_{a b})_{ \delta}{}^\gamma\de_{c}{{\chi}^{\alpha}_{i}}   X_{\gamma}^{i}{}^{\beta \delta}
 - \tfrac{22\ri}{225}\eta_{ab}\de_{c}{{\chi}^{\alpha}_{i}}  {\chi}^{\beta}{}^{i}
 \big)
 \non\\
&&~~~
+(\gamma_{b})_{ \alpha \beta}
\big(
 - \tfrac{4\ri}{3}\de_{a}{ X_{\gamma i}{}^{\delta \alpha}}   X_{\delta}^{i}{}^{\gamma \beta}
+ \tfrac{2\ri}{75}\de_{a}{{\chi}^{\alpha}_{i}}  {\chi}^{\beta}{}^{i}
 - \tfrac{28\ri}{3}(\gamma_{a c})_{ \delta}{}^{\epsilon}
 \de^{c}{ X_{\gamma i}{}^{\alpha \delta}}   X_{\epsilon}^{i}{}^{\gamma \beta}
 \non\\
&&~~~~~~~~~~~~~~~~
 + 4\ri(\gamma_{a c})_{ \epsilon}{}^\gamma\de^{c}{ X_{\gamma i}{}^{\delta \alpha}} X_{\delta}^{i}{}^{\beta \epsilon}
- \tfrac{16\ri}{45}(\gamma_{a c})_{\delta}{}^\gamma \de^{c}{ X_{\gamma i}{}^{\alpha \delta}}  {\chi}^{\beta}{}^{i}
\non\\
&&~~~~~~~~~~~~~~~~
 +\tfrac{8\ri}{15}(\gamma_{a c})_{ \delta}{}^\gamma\de^{c}{{\chi}^{\alpha}_{i}}   X_{\gamma}^{i}{}^{\beta \delta}
  \big)
  \non\\
&&~~~
- \tfrac{1}{6} (\gamma_{a b})_{ \sigma}{}^{\rho} \veps_{\gamma \delta \epsilon \rho}
W^{\alpha \gamma} W^{\sigma \delta} \de_{\alpha \beta}{W^{\beta \epsilon}}
+\tfrac{5}{12} \eta_{ab} Y_{\alpha}{}^{\beta i j}  Y_{\beta}{}^{\alpha}{}_{ij}
+ \tfrac{1}{4} (\gamma_{a b})_{\gamma}{}^{\alpha}  Y_{\alpha}{}^{\beta i j}  Y_{\beta}{}^{\gamma}{}_{ ij}
\non\\
&&~~~
+ (\gamma_{a c})_{ \sigma}{}^{\rho} ( \gamma_{b})_{ \alpha \beta} \veps_{\gamma \delta \epsilon \rho}
\big(
\tfrac{1}{2} W^{\alpha \gamma} W^{\beta \delta} \de^{c}{W^{\sigma \epsilon}}
- \tfrac{1}{3} W^{\alpha \gamma} W^{\sigma \delta} \de^{c}{W^{\beta \epsilon}}
\big)
\non\\
&&~~~
+ \veps_{\alpha \beta \gamma \delta}
\big(
\tfrac{4\ri}{3}(\gamma_{a b})_{ \sigma}{}^{\delta}
W^{\epsilon \alpha}  X_{\epsilon i}{}^{\rho \beta}  X_{\rho}^{i}{}^{\sigma \gamma}
+\tfrac{2\ri}{5}( \gamma_{a b})_{\rho}{}^{\delta}W^{\epsilon \alpha}  X_{\epsilon i}{}^{\rho \beta} {\chi}^\gamma{}^{i}
\big)
\non\\
&&~~~
+\tfrac{1}{6} \veps_{\alpha \beta \gamma \delta} \veps_{\epsilon \rho \sigma \tau}
\big(
\eta_{ab}W^{\alpha \epsilon} W^{\beta \rho} W^{\gamma \sigma} W^{\delta \tau}
-(\gamma_{a b})_{ \l}{}^{\tau}
W^{\l \alpha} W^{\beta \epsilon} W^{\gamma \rho} W^{\delta \sigma}
\big)
\non\\
&&~~~
+ \tfrac{1}{3} (\gamma_{a})_{ \alpha \beta} (\gamma_{b})_{ \gamma \delta}
\big(
2W^{\alpha \epsilon} W^{\gamma \rho} \de_{\epsilon \rho}{W^{\beta \delta}}
+ W^{\alpha \gamma} W^{\beta \epsilon} \de_{\epsilon \rho}{W^{\delta \rho}}
-2 W^{\alpha \gamma} W^{\delta \epsilon} \de_{\epsilon \rho}{W^{\beta \rho}}
\big)
\non\\
&&~~~
+( \gamma_{a})_{ \alpha \beta} (\gamma_{b})_{ \gamma \delta}
\big(
\tfrac{16\ri}{3}W^{\alpha \gamma}  X_{\epsilon i}{}^{\rho \beta}  X_{\rho}^{i}{}^{\epsilon \delta}
- 12\ri W^{\epsilon \alpha}  X_{\epsilon i}{}^{\rho \gamma}  X_{\rho}^{i}{}^{\beta \delta}
+ \tfrac{44\ri}{3}  W^{\epsilon \gamma}  X_{\epsilon i}{}^{\rho \alpha}  X_{\rho }^{i}{}^{\beta \delta}
\non\\
&&~~~~~~~~~~~~~~~~~~~~~~~~
- \tfrac{20\ri}{3}W^{\epsilon \rho}  X_{\epsilon i}{}^{\alpha \gamma}  X_{\rho}^{i}{}^{\beta \delta}
- \tfrac{8\ri}{5} W^{\alpha \gamma}  X_{\epsilon i}{}^{\beta \delta} {\chi}^{\epsilon}{}^{i}
- \tfrac{22\ri}{45}W^{\epsilon \alpha}  X_{\epsilon i}{}^{\beta \gamma} {\chi}^{\delta}{}^{i}
\non\\
&&~~~~~~~~~~~~~~~~~~~~~~~~
+ \tfrac{14\ri}{45} W^{\epsilon \gamma}  X_{\epsilon i}{}^{\alpha \delta} {\chi}^{\beta}{}^{i}
+ \tfrac{2\ri}{75} W^{\alpha \gamma} {\chi}^{\beta}_{i} {\chi}^{\delta}{}^{i}
\big)
\non\\
&&~~~
+ \tfrac{1}{2}(\gamma_{a})_{ \alpha \beta}( \gamma_{b})_{ \gamma \delta} \veps_{\epsilon \rho \sigma \tau}
W^{\alpha \epsilon} W^{\beta \rho} W^{\gamma \sigma} W^{\delta \tau}
+ \tfrac{20\ri}{3}(\tilde{\gamma}_{a})^{\epsilon \rho}( \gamma_{b})_{ \alpha \beta}
\de_{\gamma \delta}{ X_{\epsilon i}{}^{\alpha \gamma}}   X_{\rho}^{ i}{}^{\beta \delta}
\non\\
&&~~~
 - \tfrac{1}{4} (\tilde{\gamma}_{a})^{\gamma \delta}(\gamma_{b})_{ \alpha \beta}
 Y_\gamma{}^{\alpha i j}  Y_{\delta}{}^{\beta}{}_{ ij}
~.
\eea
\esubeq
Inserting the expression for $B_a{}^{ij}$, eq.\,\eqref{C2gauge3form},
 and its corresponding descendants $\L_{\a a}{}^i$ and $C_{ab}$ presented in this appendix
  into the action principle \eqref{BH-action}, 
  one can obtain the action \eqref{newInvariant} together with all the fermionic terms that complete
   the new supersymmetric invariant described in subsection  \ref{new-invariant-subsection}.


\section{Full bosonic terms of the invariant \eqref{new_Ricci+scalar_invariant}}
\label{full-nasty-new-invariant}

In this appendix we present the full bosonic contribution to the new higher-derivative locally superconformal
invariant constructed in section \ref{new-R2}.
In a general gauge and conformal supergravity background,
up to fermionic terms,
the completion of the Lagrangian \eqref{new_Ricci+scalar_invariant} reads
 \bea
&&
e^{-1}\cL_{\rm new-linear}=
 -\frac{1}{12}\, L^{\frac{1}{2}}  { {{\Box}}} D
 - \frac{1}{20}L^{\frac{1}{2}}D^2
 + \frac{1}{64} L^{-\frac{3}{2}}  {\rm E}^{a} {\rm E}_{a}  D
 -  \frac{1}{24} L^{-\frac{3}{2}} D  L^{i j}  {\Box}L_{i j}
\non\\
&&~~~
  +\frac{1}{192}L^{-\frac{3}{2}}D  (\de^{a}L^{i j})  \de_{a}L_{ij}
 +  \frac{1}{128}L^{-\frac{7}{2}}D  L^{i j} L^{k l} (\de^{a}L_{i j})\de_{a}L_{k l}
 \non\\
&&~~~
 +\frac{15}{32} L^{-\frac{3}{2}} L^{i j} L^{k l}  Y^{ab}{}_{ij} Y_{ab}{}_{ k l}
 - \frac{5}{16} L^{-\frac{3}{2}}  L^{i j}  {\Box}^2L_{i j}
   + \frac{5}{32}L^{-\frac{7}{2}} L^{i j} L^{k l} ( {\Box}L_{i j})  {\Box}L_{k l}
   \non\\
 &&~~~
 + \frac{5}{32}L^{-\frac{3}{2}}( {\Box}L^{i j}) {\Box}L_{ij}
 + \frac{5}{32}L^{-\frac{7}{2}}  L^{i j} L^{k l}( \nabla^{a}\nabla^{b}L_{ij})  \nabla_{a}\nabla_{b}L_{k l}
  - \frac{15}{64}L^{-\frac{3}{2}}   ( \nabla^{a}\nabla^{b}L^{ij})\nabla_{a}\nabla_{b}L_{ij}
 \non\\
 &&~~~
 +\frac{15}{256}L^{-\frac{7}{2}} {\rm E}^{a} L^{i j} (\de_{a}L_i{}^{k}) {\Box}L_{jk}
-\frac{15}{64}L^{-\frac{7}{2}} {\rm E}^{a} L^{i j} (\de^{b}L_i{}^{k})\nabla_{a}\nabla_{b}L_{jk}
\non\\
&&~~~
 - \frac{105}{256}L^{-\frac{7}{2}}L^{i j} (\de^{a}{\rm E}^{b}) ( \de_{a}L_i{}^{k})\de_{b}L_{jk}
 - \frac{105}{256}L^{-\frac{11}{2}}{\rm E}^aL^{i j} L^{k l} (\de_aL_i{}^{p})(\de^bL_{jp})\de_bL_{kl}
 \non\\
 &&~~~
 + \frac{5}{64} L^{-\frac{3}{2}}  {\rm E}^{a}  {\Box}{\rm E}_{a}
 + \frac{25}{128}L^{-\frac{3}{2}}  (\de^{a}{\rm E}^{b})\de_{a}{{\rm E}_{b}}
 - \frac{35}{1024}L^{-\frac{7}{2}} {\rm E}^{a} {\rm E}_{a} L^{i j} {\Box}L_{i j}
 \non\\
 &&~~~
 - \frac{15}{512}L^{-\frac{7}{2}} {\rm E}^{a} {\rm E}^{b} L^{i j} \nabla_{a}\nabla_{b}L_{i j}
- \frac{15}{512}L^{-\frac{7}{2}}{\rm E}^{a} (\de_{a}{\rm E}^{b})L^{i j}\de_{b}L_{i j}
   \non\\
 &&~~~
 - \frac{155}{512}L^{-\frac{7}{2}}{\rm E}^{a}(\de^{b}{{\rm E}_{a}}) L^{i j} \de_{b}L_{i j}
 + \frac{35}{256}L^{-\frac{11}{2}} {\rm E}^{a} {\rm E}_{a} L^{i j} L^{k l} (\de^{b}L_{i j}) \de_{b}L_{k l}
   \non\\
 &&~~~
 - \frac{275}{4096}L^{-\frac{7}{2}} {\rm E}^a {\rm E}_{a} (\de^{b}L^{ij})\de_{b}L_{ij}
 + \frac{105}{2048}L^{-\frac{11}{2}} {\rm E}^{a} {\rm E}^{b} L^{i j} L^{k l} (\de_{a}L_{i j})  \de_{b}L_{k l}
  \non\\
 &&~~~
 - \frac{105}{2048}L^{-\frac{7}{2}} {\rm E}^{a} {\rm E}^{b}(\de_{a}L^{ij})\de_{b}L_{ij}
 - \frac{45}{16384} L^{-\frac{7}{2}}  ({\rm E}^{a} {\rm E}_{a})^2
\non\\
&&~~~
 + \frac{35}{64}L^{-\frac{7}{2}} L^{i j} L^{k l} (\de^{a}L_{i j})\nabla_{a} {\Box}L_{k l}
-\frac{5}{32}L^{-\frac{3}{2}} (\de^{a}L^{ij})\nabla_{a} {\Box}L_{ij}
  \non\\
 &&~~~
-\frac{25}{64}L^{-\frac{11}{2}}L^{i j}L^{k l}L^{p q}(\de^{a}L_{ij})(\de_{a}L_{kl}) {\Box}L_{p q}
- \frac{25}{256}L^{-\frac{7}{2}}L^{i j} (\de^{a}L_{ij})(\de_{a}L^{kl})  {\Box}L_{kl}
\non\\
&&~~~
+ \frac{155}{512} L^{-\frac{7}{2}}L^{ij}(\de^{a}L^{kl})(\de_{a}L_{kl})  {\Box}L_{ij}
- \frac{65}{128}
 L^{-\frac{11}{2}}L^{i j} L^{k l} L^{p q} (\de^{a}L_{i j})(\de^{b}L_{k l})\nabla_{a}\nabla_{b}L_{p q}
\non\\
&&~~~
+\frac{85}{128}L^{-\frac{7}{2}}L^{i j} (\de^{a}L_{i j})(\de^{b}L^{kl}) \nabla_{a}\nabla_{b}L_{kl}
 +\frac{5}{256}L^{-\frac{7}{2}}L^{ij}(\de^{a}L^{kl})(\de^{b}L_{kl}) \nabla_{a}\nabla_{b}L_{ij}
 \non\\
 &&~~~
+ \frac{15}{64}L^{-\frac{3}{2}}T^-{}^{abc} {\rm E}_{a} \de_{b}{{\rm E}_{c}}
  + \frac{15}{32}L^{-\frac{3}{2}} L^{i j}  Y^{ab}{}_{i j}\de_{a}{\rm E}_{b}
 + \frac{15}{16}L^{-\frac{3}{2}}L^{i j} (\de_{a}L_i{}^{k})\de_{b} Y^{ab}{}_{jk}
\non\\
&&~~~
 + \frac{105}{256}L^{-\frac{7}{2}}{\rm E}^{a} Y_{ab}{}_{ i j}L^{i j} L^{k l}  \de^{b}L_{k l}
 - \frac{45}{128}L^{-\frac{3}{2}} Y^{ab}{}^{i j}{\rm E}_{a}\de_{b}L_{ij}
   \non\\
 &&~~~
  -\frac{45}{256}L^{-\frac{7}{2}}T^-{}^{abc}{\rm E}_{a} L^{ij}(\de_{b}L_i{}^{k})\de_{c}L_{jk}
-\frac{165}{256}L^{-\frac{7}{2}}L^{i j} L^{k l} Y^{ab}{}_{ ij}(\de_{a}L_{k}{}^p)\de_{b}L_{lp}
  \non\\
 &&~~~
-\frac{45}{128}L^{-\frac{3}{2}} Y^{ab}{}^{ij}(\de_{a}L_{i}{}^{k})\de_{b}L_{jk}
 -\frac{15}{8}L^{\frac{1}{2}} (\de_{a}T^-{}^{acd})  \de^{b}T^-_{bcd}
+\frac{15}{32}L^{-\frac{3}{2}}{\rm E}^{a} L^{i j}T^-_{abc} Y^{bc}{}_{ij}
   \non\\
 &&~~~
 +\frac{15}{256}L^{-\frac{3}{2}} {\rm E}_{a}{\rm E}^{b}T^-{}^{acd} T^-_{bcd}
 -\frac{15}{32}L^{-\frac{3}{2}} T^-{}^{acd}T^-{}^b{}_{cd} L^{ij}\nabla_{a}\nabla_{b}L_{ij}
  \non\\
 &&~~~
 + \frac{45}{128}L^{-\frac{7}{2}}T^-{}^{acd}T^-_{bcd}L^{i j} L^{k l}(\de_{a}L_{i j}) \de^{b}L_{k l}
-\frac{45}{128}L^{-\frac{3}{2}}T^-{}^{acd}T^-_{bcd}(\de_{a}L^{ij}) \de^{b}L_{ij}
\non\\
&&~~~
+\frac{15}{32}L^{-\frac{3}{2}}L^{i j} T^-{}^{abc} Y_{ab}{}_{i}{}^{k} \de_{c}L_{jk}
-\frac{15}{16}L^{-\frac{3}{2}}L^{i j}T^-{}^{acd}(\de_{a}L_{ij})\de^bT^-_{bcd}
  \non\\
 &&~~~
 -\frac{15}{32}L^{\frac{1}{2}}T^-{}^{acd}T^-_{bcd}T^-_{aef}T^-{}^{bef}
+ \frac{165}{512}
L^{-\frac{15}{2}} L^{i j} L^{k l} L^{p q} L^{r s}
  (\de^{a}L_{i j})( \de_{a}L_{k l}) (\de^{b}L_{p q})  \de_{b}L_{r s}
  \non\\
  &&~~~
-\frac{135}{256}
L^{-\frac{11}{2}}L^{i j} L^{k l}(\de^{a}L_{i j})(\de_{a}L_{k l})(\de^{b}L^{pq}) \de_{b}L_{pq}
\non\\
&&~~~
+\frac{5}{1024} L^{-\frac{11}{2}}L^{i j} L^{kl}
(\de^{a}L_{i j})(\de_{a}L^{pq}) (\de^{b}L_{pq})\de_{b}L_{kl}
  \non\\
 &&~~~
+\frac{1005}{4096}L^{-\frac{7}{2}}(\de^{a}L^{ij})(\de_{a}L_{ij})(\de^{b}L^{kl})\de_{b}L_{kl}
\non\\
&&~~~
 -\frac{325}{2048}
  L^{-\frac{7}{2}}(\de^{a}L^{i j})(\de_{a}L^{k l})(\de^{b}L_{kl})\de_{b}L_{ij}
  \non\\
&&~~~
+{\rm fermions}
  ~.~~~~~~
  \label{new_Ricci+scalar_invariant-full}
\eea


\section{Different 6D $\cN=(1,0)$ notation and conventions}
\label{notation}

In this appendix we describe the differences between
 our notation and conventions, based on \cite{BKNT16,BNT-M17},
and the ones used in \cite{BSVanP} and \cite{CVanP}.
By using the map between notations described here the reader can rewrite the results of our paper
in the notation of \cite{BSVanP} and \cite{CVanP}.

Throughout our paper we have used chiral four-component spinor notation
while in \cite{BSVanP,CVanP} eight-component spinor notation is used. 
To translate our results, the reader should first reinterpret  our formulae using eight component  spinors.
According to  appendix A of \cite{BKNT16},
our $8\times 8$ Dirac spinors $\Psi$ and matrices $\G^a$ are
\begin{align}
\Psi =
\begin{pmatrix}
\psi^\alpha \\
\chi_\alpha
\end{pmatrix}
~, \qquad
\Gamma^a =
\begin{pmatrix}
0 & (\tilde\gamma^a)^{\alpha\beta} \\
(\gamma^a)_{\alpha\beta} & 0 
\end{pmatrix}
~,\qquad
\Gamma_* =
\begin{pmatrix}
\delta^\a_\b & 0 \\
0 & -\delta_\a^\b
\end{pmatrix}
~,
\end{align}
where $\G_*$ obeys
$\Gamma_{[a} \Gamma_b \Gamma_c \Gamma_d \Gamma_e \Gamma_{f]} = \veps_{abcdef} \Gamma_*$.
Similarly, there is a direct relation between ${\g}^{a_1\cdots a_n}$,
$\tilde{\g}^{a_1\cdots a_n}$ and $\G^{a_1\cdots a_n}:=\G^{[a_1}\G^{a_2}\cdots\G^{a_n]}$
since a product of chiral $\g$s are straightforwardly lifted to a product of Dirac $\G$s.
For example, the eight component spinor generators of the 6D $\cN=(1,0)$ superconformal algebra are 
$
Q^i=\left(
\begin{array}{c}
0\\ 
Q_{\a}^i
\end{array}
\right)
$
and
$S_{i}= \left(
\begin{array}{c}
S^\a_{ i}
\\
0
\end{array}
\right)
$.
Similarly, all the (anti)chiral spinor fields are straightforwardly lifted to eight components, such that, e.g.
$\psi_m{}^{\a}_{i}Q_{\a}^{i}\to\bar{\psi}_m{}_{i}Q^{i}$,
${{\f}}_m{}_\a^{i}S^\a_{i}\to\bar{{\f}}_m{}^{i}S_{i}$.
To map our results to the notation of \cite{BSVanP} and \cite{CVanP},
the reader should then use the table \ref{tab:ConvComp-1}.
\begin{table*}[!t]
\centering
\renewcommand{\arraystretch}{1.4}
\resizebox{15.1cm}{!}{
\begin{tabular}{@{}ccc@{}} \toprule
Our notation 
&  Bergshoeff et al. \cite{BSVanP} 
&  Coomans \& Van Proeyen \cite{CVanP}
\\ \midrule
$\eta^{ab}$, 
$\veps^{abcdef}$, $\G^a$, $\G^{a_1\cdots a_n}$, $\cdots$
&
$\d^{ab}$,
$\ri\veps^{abcdef}$, $-\ri \gamma^a$, $(-\ri)^n\g^{a_1\cdots a_n}$, $\cdots$
&
$\eta^{ab}$,
$-\veps^{abcdef}$, $-\ri \gamma^a$, $(-\ri)^n\g^{a_1\cdots a_n}$, $\cdots$
\\
\midrule
$P_a$, $K^a$, $Q^i$,  $S_i$ 
&
$P_a$, $K^a$, $-2Q^i$,  $2S_i$ 
&
$P_a$, $K^a$, $-2Q^i$,  $2S_i$ 
 \\
$M_{ab}$,
$J_{ij}$,
$\mathbb D$
&
$-2M_{ab}$,
$2U_{ij}$,
$D$
&
$-M_{ab}$,
$-U_{ij}$,
$D$
\\ \midrule
$e_m{}^a$,
$\psi_m{}^i$,
$\cV_m{}^{ij}$, 
$b_m$
& 
 $e_\mu{}^a$,
 $\psi_\mu{}^i$,
 $\hf V_\mu{}^{ij}$,
$b_\mu$,
 &
 $e_\mu{}^a$,
 $\psi_\mu{}^i$,
$-\cV_\mu{}^{ij}$,
$b_\mu$,
 \\
$\omega_m{}^{ab}$,
 $\phi_m{}^i$
&
 $-\omega_\mu{}^{ab}$,
 $\phi_\mu{}^i +\tfrac{1}{60}\g_\mu\chi^i$
&
 $-\omega_\mu{}^{ab}$,
 $\phi_\mu{}^i +\tfrac{1}{60}\g_\mu\chi^i$
 \\
${\frak{f}}_m{}_a$
&
$f_\mu{}_a-e_\mu{}^a\big(\tfrac{1}{240}\d_{ab}D-\tfrac{1}{8}T^{-}_a{}^{cd}T^-_{cdb}\big)$
&
$f_\mu{}_a-e_\mu{}^a\big(\tfrac{1}{240}\eta_{ab}D-\tfrac{1}{8}T^{-}_a{}^{cd}T^-_{cdb}\big)$
\\
$\de_a$
&
$
\hat{\mathscr{D}}_a
+\frac{1}{60} \chi^{i}\g_aS_i
+\Big(
\frac{1}{240}\d_{ab}D
-\frac{1}{8}T^-_{a}{}^{cd} T^-_{cdb} 
\Big)K^b
$
&
$
{\cD}_a
+\frac{1}{60} \chi^{i}\g_aS_i
+\Big(
\frac{1}{240}\eta_{ab}D
-\frac{1}{8}T^-_{a}{}^{cd} T^-_{cdb} 
\Big)K^b
$
\\ \midrule
$T^-_{abc}$,
$\chi^i$, 
$D$
&
$T^-_{abc}$,
$\chi^i$, 
$D$
&
$T^-_{abc}$,
$\chi^i$, 
$D$
\\ \midrule
$R(P)_{ab}{}^{c}=R(\mathbb D)_{ab}=0$
&
$\hat R(P)_{ab}{}^{c}=\hat R(D)_{ab}=0$
&
$\hat R(P)_{ab}{}^{c}=\hat R(D)_{ab}=0$
\\
$R(J)_{ab}{}^{ij}$,
$R(Q)_{ab}{}^i$
&
$\hf\hat R(U)_{ab}{}^{ij}$,
$\tfrac{1}{2}\hat{R}(Q)_{ab}{}_k
+\tfrac{1}{60}\g_{ab}\chi_k $
&
$-\hat R(U)_{ab}{}^{ij}$,
$\tfrac{1}{2}\hat{R}(Q)_{ab}{}_k
+\tfrac{1}{60}\g_{ab}\chi_k $
 \\
$R(M)_{ab}{}^{cd}$
&
$-\hat{R}(M)_{a b}{}^{c d}
-\tfrac{1}{30} D \d_{a}^{[c}\d_b^{d]}
-T^{-}_{a b f} T^{-}{}^{fcd} $
&
$-\hat{R}(M)_{a b}{}^{c d}
-\tfrac{1}{30} D \d_{a}^{[c}\d_b^{d]}
-T^{-}_{a b f} T^{-}{}^{fcd} $
\\
$ R(S)_{ab}{}_i$ 
&
$\tfrac{1}{2}\hat{R}(S)_{ab}{}^k
-\tfrac{1}{60}  \g_{[a}\hat{\mathscr{D}}_{b]}\chi^{k}
+\tfrac{19}{600}T^{-}_{abc}\g^{c}\chi^{k}$
&
$\tfrac{1}{2}\hat{R}(S)_{ab}{}^k
-\tfrac{1}{60}  \g_{[a}{\cD}_{b]}\chi^{k}
+\tfrac{19}{600}T^{-}_{abc}\g^{c}\chi^{k}$
\\
$R(K)_{ab}{}_{c}$
&
$\Big{[}\hat{R}(K)_{ab}{}_{c}
+\frac{1}{120}\eta_{c[a}\hat{\mathscr{D}}_{b]} D
+2 T^-_{c}{}^{de}\hat{\mathscr{D}}_{[a} T^-_{b]de}
$
&
$\Big{[}\hat{R}(K)_{ab}{}_{c}
+\frac{1}{120}\eta_{c[a}\cD_{b]} D
+2 T^-_{c}{}^{de}\cD_{[a} T^-_{b]de}
$
\\
&
$+\tfrac{1}{1440}\chi^i\big(
\g_{[a}\hat{R}(Q)_{b]c}{}_i
-7\g_{c}\hat{R}(Q)_{ab}{}_i
\big)~~~~~
$
&
$+\tfrac{1}{1440}\chi^i\big(
\g_{[a}\hat{R}(Q)_{b]c}{}_i
-7\g_{c}\hat{R}(Q)_{ab}{}_i
\big)~~~~~
$
\\
&
$+\tfrac{1}{900}\chi^i \g_{abc}\chi_i\Big{]}$
~~~~~~~~~~~~~~~~~~~~~~~~~~~~~~
&
$+\tfrac{1}{900}\chi^i \g_{abc}\chi_i\Big{]}$
~~~~~~~~~~~~~~~~~~~~~~~~~~~~~~
\\ \midrule
$b_{mn}$,
$\s$,
$\psi_i$,
$H_{abc}$
&
$B_{\mu\nu}$,
$\s$,
$-2\psi_i$,
$(F^+_{abc}+2\s T^-_{abc})$
&
$B_{\mu\nu}$,
$\s$,
$-2\psi_i$,
$\hat{F}(B)_{abc}$
\\ \midrule
$v_m$,
$\L^i$,
$X^{ij}$,
$F_{ab}$
&
$-\tfrac{1}{2} W_\mu$,
$-\ri\O^i$,
$-\tfrac{1}{2} Y^{ij}$,
$-\tfrac{1}{2}\hat{F}(W)_{ab}$
&
$-\tfrac{1}{2} W_\mu$,
$-\ri\O^i$,
$-\tfrac{1}{2} Y^{ij}$,
$-\tfrac{1}{2}\hat{F}(W)_{ab}$
\\ \midrule
$b_{mnpq}$
$L^{ij}$,
$\varphi^i$,
$H_a$,
$L$
&
$E_{\mu\nu\rho\tau}$
$L^{ij}$,
$\varphi^i$,
$E_a$,
$\tfrac{1}{\sqrt{2}}L$
&
$E_{\mu\nu\rho\tau}$
$L^{ij}$,
$\varphi^i$,
$E_a$,
$\tfrac{1}{\sqrt{2}}L$
\\
\bottomrule
\end{tabular}}
\caption{Different notation and conventions}
\label{tab:ConvComp-1}
\end{table*}
This table shows how our gamma matrices, generators of the superconformal group, 
connections, curvatures,  fields of matter multiplets, etc, 
described in the first column,
should be replaced with the terms in the second and third columns
to match the notation and conventions of \cite{BSVanP} and \cite{CVanP}, respectively.
This table can be obtained by 
using results of appendix A of \cite{BNT-M17} 
which describes in general how to change frames for the local gauging of the superconformal algebra.

Note also that our definitions of local superconformal transformations are
different from the ones of  \cite{BSVanP} and \cite{CVanP}.
If we restrict only to $\delta_Q + \delta_S + \delta_K$ transformations, 
it can be then shown that
the differences between our notation and the ones of
\cite{BSVanP} and \cite{CVanP}
only amount to a simple rescaling of the local parameters $\x_i$,
$\eta^i$ and $\l^a$
 as described in 
table \ref{tab:ConvComp-3}.
\begin{table*}[!t]
\centering
\renewcommand{\arraystretch}{1.4}
\resizebox{11cm}{!}{
\begin{tabular}{@{}ccc@{}} \toprule
Traceless frame 
&  Bergshoeff et al. \cite{BSVanP} 
&  Coomans \& Van Proeyen \cite{CVanP}
\\ \midrule
$\x_i$,
$\eta^i$,
$\l^a$
&~~~~~~
$\hf\ve_i$,
$\hf\eta^i$,
$\L_K^a$
&~~~~~~
$\hf\ve_i$,
$\hf\eta^i$,
$\l_K^a$
\\
\bottomrule
\end{tabular}}
\caption{Transformation parameters for $\delta_Q + \delta_S + \delta_K$}
\label{tab:ConvComp-3}
\end{table*}


\begin{footnotesize}

\end{footnotesize}

\end{document}